\documentclass[authoryear, preprint, 12pts]{article}

\usepackage{amsfonts}
\usepackage{amssymb}
\usepackage{amsmath}
\usepackage{amsthm}
\usepackage{dsfont}
\usepackage{appendix}
\usepackage{tabulary}
\usepackage{setspace}
\usepackage{comment}
\usepackage[bottom]{footmisc}
\usepackage{lipsum}
\usepackage[hmargin=1.5in,vmargin=1.5in]{geometry}
\usepackage{pdflscape}
\usepackage{ccaption}
\usepackage{afterpage}
\usepackage[colorlinks=true]{hyperref}
\hypersetup{citecolor=blue}
\usepackage{authblk}
\usepackage{graphicx}
\usepackage{rotating}
\usepackage{color}
\usepackage{longtable}
\usepackage{booktabs}
\usepackage{multirow}
\usepackage{cellspace}
\usepackage{hyperref}
\usepackage{float}
\usepackage[caption = false]{subfig}
\usepackage[table,xcdraw]{xcolor}
\usepackage[section]{placeins}

\usepackage[style=authoryear-comp,maxcitenames=2,uniquelist=false,backend=biber,natbib,maxbibnames=99,doi=false,isbn=false,url=false]{biblatex}
 \renewbibmacro{in:}{%
\ifentrytype{article}{}{\printtext{\bibstring{in}\intitlepunct}}}
 
\renewbibmacro*{volume+number+eid}{%
 \setunit{\addcomma\space}
 \printfield{volume}%
 \printfield{number}%
 \setunit{\addcomma\space}%
\printfield{eid}}
\DeclareFieldFormat[article]{number}{\mkbibparens{#1}}
\DeclareFieldFormat[article]{pages}{#1}
\DeclareFieldFormat[article,inbook,incollection]{title}{#1}

\DeclareCiteCommand{\cite}
  {\usebibmacro{prenote}}
  {\usebibmacro{citeindex}%
   \printtext[bibhyperref]{\usebibmacro{cite}}}
  {\multicitedelim}
  {\usebibmacro{postnote}}

\DeclareCiteCommand*{\cite}
  {\usebibmacro{prenote}}
  {\usebibmacro{citeindex}%
   \printtext[bibhyperref]{\usebibmacro{citeyear}}}
  {\multicitedelim}
  {\usebibmacro{postnote}}

\DeclareCiteCommand{\parencite}[\mkbibparens]
  {\usebibmacro{prenote}}
  {\usebibmacro{citeindex}%
    \printtext[bibhyperref]{\usebibmacro{cite}}}
  {\multicitedelim}
  {\usebibmacro{postnote}}

\DeclareCiteCommand*{\parencite}[\mkbibparens]
  {\usebibmacro{prenote}}
  {\usebibmacro{citeindex}%
    \printtext[bibhyperref]{\usebibmacro{citeyear}}}
  {\multicitedelim}
  {\usebibmacro{postnote}}

\DeclareCiteCommand{\footcite}[\mkbibfootnote]
  {\usebibmacro{prenote}}
  {\usebibmacro{citeindex}%
  \printtext[bibhyperref]{ \usebibmacro{cite}}}
  {\multicitedelim}
  {\usebibmacro{postnote}}

\DeclareCiteCommand{\footcitetext}[\mkbibfootnotetext]
  {\usebibmacro{prenote}}
  {\usebibmacro{citeindex}%
   \printtext[bibhyperref]{\usebibmacro{cite}}}
  {\multicitedelim}
  {\usebibmacro{postnote}}

\DeclareCiteCommand{\textcite}
 {\boolfalse{cbx:parens}}
  {\usebibmacro{citeindex}%
    \printtext[bibhyperref]{\printnames{labelname}%
      \printtext{ (\printfield{year}\printtext{)}}}}
 {\ifbool{cbx:parens}
  {\bibcloseparen\global\boolfalse{cbx:parens}}
  {}%
 \multicitedelim}
{\usebibmacro{textcite:postnote}}

\newcommand{\columnname}[1]
{\makebox[\tempwidth][c]{\textbf{#1}}}

\AtBeginEnvironment{table}{\singlespacing}
\AtBeginEnvironment{figure}{\singlespacing}

\begin{document}
\begin{titlepage}
\title{\singlespacing \textbf{Gentrification, displacement, and income trajectory of incumbents}\thanks{I am especially indebted to David Green, for invaluable guidance, advising, and support. I am grateful to Victor Couture, Réka Juhász, Florian Mayneris, and Pablo Valenzuela, all participants at the VSE-Sauder Spatial brown bag seminar,  and the VSE PhD labour group for their insightful comments and suggestions.  I gratefully acknowledge the financial support of the Fonds de Recherche du Québec – Société et Culture (FRQSC). The analysis is based on Statistics Canada's Census of Population and Longitudinal Administrative Databank (LAD). This research was conducted at the Québec Interuniversity Centre for Social Statistics (QICSS) and at the UBC RDC, part of the Canadian Research Data Centre Network (CRDCN). This service is provided through the support of QICSS’ Member Universities, the University of British Columbia, the provinces of Québec and British Columbia, the Canadian Foundation for Innovation, the Canadian Institutes of Health Research, the Social Science and Humanity Research Council, the Fonds de Recherche du Québec, and Statistics Canada. All errors remain my own. The views expressed in this paper are those of the author.}\\}
\author{
Pierre-Loup Beauregard\thanks{Vancouver School of Economics, UBC, Email: pierreloup.beauregard@gmail.com} \\ 
\vspace*{-0.75cm} \textit{University of British Columbia} \\ \vspace*{-0.7cm}}
\date{February 2024\\
\href{https://plbeauregard.github.io/document/beauregard_gentrification.pdf}{Click Here for Latest Version}}
\maketitle
\thispagestyle{empty}
\vspace*{-1cm}
\begin{abstract}
\singlespacing
Gentrification is associated with rapid demographic changes within inner-city neighborhoods. While many fear that gentrification drives low-income people from their homes and communities, there is limited evidence of the consequences of these changes. I use Canadian administrative tax files to track the movements of incumbent workers and their income trajectory as their neighborhood gentrifies. I exploit the timing at which neighborhoods gentrify in a matched staggered event-study framework. I find no evidence of displacement effects, even for low socioeconomic status households. In fact, families living in gentrifying neighborhoods are more likely to stay longer. I suggest that this might be related to tenant rights protection laws. When they endogenously decide to leave, low-income families do not relocate to worse neighborhoods. Finally, I find no adverse effects on their income trajectory, suggesting no repercussions on their labor market outcomes.   
\end{abstract}

\end{titlepage}



\section{Introduction}\label{chap:intro}

Since the 1990s, a growing number of higher socioeconomic status (SES) households, who once fled the city for the suburbs, are choosing to migrate back to the city \parencite{hwang2016have, couture2020urban, baum2020accounting}; a phenomenon known as gentrification. This is a significant reversal from the socioeconomic decline of inner-city neighborhoods in the 70s and 80s. These changes have sparked interest among urban economists, contributing to a growing literature on the relationship between income distribution and spatial sorting patterns \parencite{couture2021income, diamond2016determinants}. Concurrently, many prominent studies in labor economics show that neighborhoods have significant impacts on socioeconomic and labor market outcomes \parencite{chetty2016effects,galster2012mechanism,oreopoulos2003long, laliberte2021long}. Thus far, the literature on neighborhood effects has focused on the high- and low-poverty neighborhood differences. This overlooks an important aspect of residential stratification in urban areas: not all disadvantaged neighborhoods stay that way. Gentrification thus has the potential to reshape the way we think of neighborhood effects dramatically. 

It is unclear how those rapid neighborhood changes affect incumbent households. There is a widespread presumption that those transformations negatively affect incumbent families --- especially low-income ones. The main concern is that the influx of affluent residents might push original residents away from city-center and economic opportunities. When households are involuntarily displaced, they might be negatively affected if they lose strong local ties, face high moving costs, or find themselves in ``worse" neighborhoods\footnote{Neighborhoods with higher poverty, longer commute, fewer job opportunities and lower education level.}. Stayers could also end up worse if they feel uprooted in their own community \parencite{valli2015sense}, if it destroys their current jobs and they don't benefit from new job creation \parencite{meltzer2017does} or if they face higher rents. However, the net effect could be positive if the benefits of lower exposure to poverty and an increase in amenities offset the adverse effects.

In this paper, I exploit longitudinal administrative data on Canadian taxfilers to study the relationship between gentrification and incumbent workers' mobility and their income. Does gentrification push incumbent residents away? Where do low-income households reallocate when they leave gentrifying neighborhoods? How does this affect their income trajectory? 

I first identify neighborhoods that gentrified between 1996 and 2016 in the three largest Canadian cities: Montréal, Toronto, and Vancouver. I measure gentrification as an increase in the share of residents with a bachelor's degree in initially low-income central-city census tracts relative to that increase at the city level. Then, using a staggered event-study design, I study how those demographic changes affect incumbent residents' probability of leaving the neighborhood. For movers, I investigate their relocation choices. Finally, I estimate the impact of gentrification on their income trajectory. 

The main challenge in studying neighborhood changes is that they occur non-randomly. Areas that undergo gentrification might differ from those that don't, and people living there initially might also be different from those originally living in non-gentrifying neighborhoods. I mitigate those concerns by including baseline characteristics of neighborhoods along with pre-gentrification trends in those characteristics. Section \ref{chap:data} suggests evidence that gentrifying neighborhoods don't have significantly different trends than their non-gentrifying counterparts. To further minimize those concerns, I build a matched control group of households and workers used in the main analysis. 

An additional problem raised by past research is that using a measure combining income and education is essential as both capture different spatial and temporal patterns of change \parencite{behrens2022gentrification}. My analysis is robust to using the change in median income to measure gentrification instead of the growth in the share of residents with a bachelor's degree.

I find that incumbent households are not more likely to move out of their neighborhoods during the gentrification process. In fact, (high-) low-income households are (5 ppts) 2 ppts more likely to live in their origin neighborhood ten years later than their counterpart living in other initially poor neighborhoods. The effect on low-incumbent workers is entirely driven by Montréal, which as the strongest rent regulation and tenant protection law in my sample of cities. I briefly discuss this in Section \ref{chap:discussion}. Moreover, households who endogenously decide to move do not select into worse neighborhoods; in other words, when moving out, households relocate to similar areas than those moving out of non-gentrifying neighborhoods. 

In contrast to the effects on mobility, my results indicate no effects on labor force outcomes. Low-income workers who initially lived in gentrifying neighborhoods have similar income trajectories compared to those in not gentrifying neighborhoods. This suggests that local labor market structure changes induced by gentrification are not an important vector impacting incumbent workers. The documented industrial transformation in gentrifying areas documented by \citet{lester2014long} might not affect local residents, as the labor markets are much more integrated. 

My work builds on previous multidisciplinary literature on the effect of gentrification on incumbent households (see \textcite{ellen2012gentrification} and \textcite{hwang2016have} for exhaustive reviews of the literature). Early papers studying the effects of gentrification as we define it today primarily focused on the question of displacement \parencite{vigdor2002does, freeman2004gentrification, ellen2011low}. They typically find no evidence that households from gentrifying areas are more likely to move out of their housing units than similar households in non-gentrifying neighborhoods. \textcite{freeman2004gentrification} even suggest that low socioeconomic status households had a slower residential turnover in gentrifying neighborhoods during the 1990s in New York City. \citet{ellen2011low} finds that low-income households' exits are not the driving force of neighborhood changes but rather the entry of higher SES households, especially homeowners. Most of those studies usually have significant data limitations. For instance, they lack longitudinal data or study changes at a very large spatial level.\footnote{Sometimes using areas composed of more than 100,000 residents, while gentrification takes place at a much more local level.} I contribute to this literature by studying mobility at a significantly more granular level and exploiting longitudinal administrative data instead of surveys. My results are consistent with the previous literature, suggesting that there are no displacement effects. As in \citet{freeman2004gentrification}, I find that incumbents had a lower turnover in gentrifying neighborhoods than similar poor neighborhoods that did not gentrify. 

A recent strain of the literature uses higher-quality data\footnote{They often still suffer fairly small sample sizes and are often limited to specific regions or cities.} (e.g. Medicaid data, Credit data, linked Census) and focus on welfare effects \parencite{ brummet2021effects, baum2019long, ding2016gentrification, dragan2020does}. In addition to studying mobility, they find that children of less-educated homeowners exposed to gentrification had a higher probability of attending and completing college \parencite{brummet2021effects}, that movers tend to move to worse neighborhoods and farther from the CBD \parencite{ ding2016gentrification, dragan2020does}, that children exposed to higher gentrification had higher credit scores, credit limits and were more likely to have a mortgage 17 years later \parencite{baum2019long}. However, these papers are often limited to two points in time (before and after gentrification). One notable exception is \citet{french2024quantifying}, which uses administrative US Census data to track the earnings, workplaces, and residential addresses of low-income urban renters in the US. They suggest that gentrification does not directly affect incumbent outcomes but does so by changing the characteristics of other neighborhoods in their choice sets. In this vein, I follow incumbents every year before, during, and after gentrification occurs. Consequently, my second contribution is to explore dynamic effects on incumbents' mobility and earnings by observing their location and income annually as gentrification occurs (or not) in their neighborhood.

Finally, this paper also relates to the literature on urban revival, spatial sorting, and their relationship with welfare inequality. Increases in spatial sorting by socioeconomic status between 1980 and 2000 \parencite{diamond2016determinants} and since 2000 \parencite{couture2021income} increased welfare inequality. My results add to the growing evidence on the welfare consequences of re-urbanization on original residents by providing mid- to long-term effects on incumbent households.

The paper is structured as follows: Section 2 describes the data, sample of interest, and the gentrification measure used. Section 3 discusses the empirical strategy used to answer each question. Section 4 presents the results. Section 5 discusses tenant protection laws as a potential mechanism. Finally, Section 6 concludes.


\section{Data and Sample Characteristics} \label{chap:data}

\subsection{Census of population}

I use the Canadian Census of Population\footnote{In 2011, the long-form Census was discontinued and replaced by a non-mandatory National Household Survey (NHS). I use the NHS in an equivalent manner as I do for the other census years.} to construct neighborhood-level and city-level characteristics. The Census is conducted every five years and consists of a representative 25\% sample of the Canadian population designed to provide information about individuals' housing, demographic, social, and economic characteristics. Every census year, I obtain neighborhood and city indicators from which I build gentrification measures exploiting time variations.  

I define neighborhoods as census tracts (CT), small and relatively stable geographic units that usually have populations of between 2,500 and 8,000 persons. I define cities as census metropolitan areas (CMA), a geographic unit comprising one or several municipalities around a major urban core. In this sense, CMAs are similar in nature to Commuting Zones. To be classified as CMA, a metropolitan area must have a population of at least 100,000, of which 50,000 or more live in the urban core. 

I restrict the analysis to Canada's three largest metropolitan areas: Toronto, Montréal, and Vancouver. The smallest CMA is Vancouver, with a population of 2,642,825 in 2021, similar in size to the Orlando-Kissimmee-Sanford Metropolitan Statistical Area in Florida (2,691,925 people). Hence, the sample of cities used is comparable in size to the top 25 Metropolitan Statistical Areas in the US, excluding New York, LA, and Chicago, which have considerably larger populations.

\subsection{Longitudinal Administrative Databank}

To assess the impacts on incumbent workers, I use Statistic Canada's Longitudinal Administrative Databank (LAD). The LAD is a longitudinal linkage of family tax files (T1FF) of a random 20\% of Canadian workers. It allows me to follow taxpayers year to year from 1982 onward. The sampling of the dataset is such that if a worker is selected in a particular reference year, they are included in any other later (or earlier) years in which they are present in the T1FF database. Any individual with a social insurance number and completed a T1 tax return or who received Canada Child Benefit can be selected into the LAD.\footnote{This limits the longitudinal representativeness to individuals who have started filing income tax returns and their partners. According to Statistics Canada, the potentially sampled population covers around 75\% of the official population estimates.}  Once an individual is selected in the database, their non-tax-filing spouses and non-tax-filing children 19 years old or younger who previously filed tax forms will have a reliable identifier that can be followed across years. 

The longitudinal nature of the LAD allows me to follow workers in time geographically but also in their income trajectory. The primary income measure I use is Statistics Canada's definition of total pre-tax income. The total income is found on Line 150 of the T1 tax form, which refers to the sum of a tax filer's income for the Canada Revenue Agency's purpose. The total income includes the tax filer's income from taxable and non-taxable sources.\footnote{Hence, it includes earnings, dividends, interest	and investment income, rental income, self-employment net income, other income, pension income, Indian exempt employment income, and benefits such as employment insurance, family and child benefits, social assistance, and tax credits, but excludes capital gains or losses. Due to confidentiality rules, amounts are rounded to the nearest \$100.} The second outcome variable is the places of residence defined by the full six-character postal code\footnote{Canadian postal codes are small. In urban areas, they represent one block face or a single building.} of the family. A household with a different postal code from the previous year is flagged as moving. The postal code variable is then converted to census tracts using Statistics Canada's Postal Code Conversion File.

In addition to the income and place of residence, the LAD includes some limited characteristics of individuals and their families. Namely, the individual's age, gender, marital status, and immigration status. T1 files also provide an array of amounts received as government transfers and tax credits from various sources. At the family level, the T1FF contains the family structure, its size, the number of children, and their respective dates of birth. 

It must be noted that while the LAD is a random sample of the tax filers population, it might not be as representative to study low-income households in gentrifying neighborhoods. Indeed, the neighborhood changes studied here might lead individuals to leave the tax filer population if they lose formal employment links. While it is possible that some incumbents undergo long periods of unemployment, I believe that most of the incumbent workers will still have some formal link in each given year or receive some transfer. This, however, means that more extreme cases where individuals lose all formal employment links, stop receiving any transfers, or even stop having a residential address will be overlooked. 

\subsection{Measuring gentrification}
\label{sec:measure}

Gentrification is a multifaced phenomenon; hence, it is hard to quantify using a single measure that captures all the dimensions we usually associate with gentrification. The urban economics literature has yet to converge to a unified measure, but some neighborhood characteristics are systematically used \parencite{barton2016exploration}. \citet{behrens2022gentrification} reviewed 27 academic papers, and their bibliometric analysis shows the most prominent dimensions used to define gentrification are changes in income and education: ``Most of the papers further retain an eligibility criterion: a neighborhood gentrifies if, starting from an \textit{initially low}-income level, it experiences a \textit{substantial increase} in income and/or in the share of highly educated residents" \parencite[p.6]{behrens2022gentrification}.

My preferred measure --- and the most used in the recent literature on the effect of gentrification \parencite{baum2019long, brummet2021effects} and its causes \parencite{couture2020urban, baum2020accounting, diamond2016determinants} --- is the growth in the share of residents aged 25 or older with a bachelor's degree living in census tract $j$. I calculate this growth relative to that of the CMA $c$, over overlapping ten years periods:
\begin{align}\label{equation:gent_measure}
    gent_{jc,t} = \frac{grad25_{jc,t+10} - grad25_{jc,t}}{pop25_{jc,t}} - \frac{grad25_{c,t+10} - grad25_{c,t}}{pop25_{c,t}}
\end{align}

Where $grad25_{jc,t}$ is the number of bachelor's graduates in tract $j$ of CMA $c$ in time $t$, and $pop25_{jc,t}$ is the total population aged 25 or older. Variables without the $j$ subscripts are the CMA-level equivalent statistics.  

Using education level for the population above 25 limits the potential ``outcomes effect" in which, because of neighborhood change, younger people achieve higher education. Education level seems to be one of the first noticeable neighborhood changes, possibly leading to rent and income changes. Hence, using education helps finding neighborhoods earlier in the gentrification process. Using education rather than income also detects what is considered as ``early gentrifiers", such as artists, students, and young professionals who contribute to gentrification but who don't have significantly high income (yet). 

The city ``normalization" prevents from capturing the well-documented general education level and income increase in cities \parencite{couture2020urban, diamond2016determinants}. It also allows me to compare the measure across different cities. In appendix \ref{sec:robustness}, I show that the results are robust to an alternative measure of gentrification based on the median income instead of the share of individuals with a college degree. 

For a neighborhood to be considered gentrifiable, it must meet certain criteria. I define gentrifiable neighborhoods as census tracts (1) situated in the central city and (2) initially poor, based on their median income falling in the bottom half of the distribution within their Metropolitan area.

\begin{figure}[ht!]
\centering
\caption{Montréal central business district and 10 km radius\label{fig:radius_mtl}}
\includegraphics[scale=0.55]{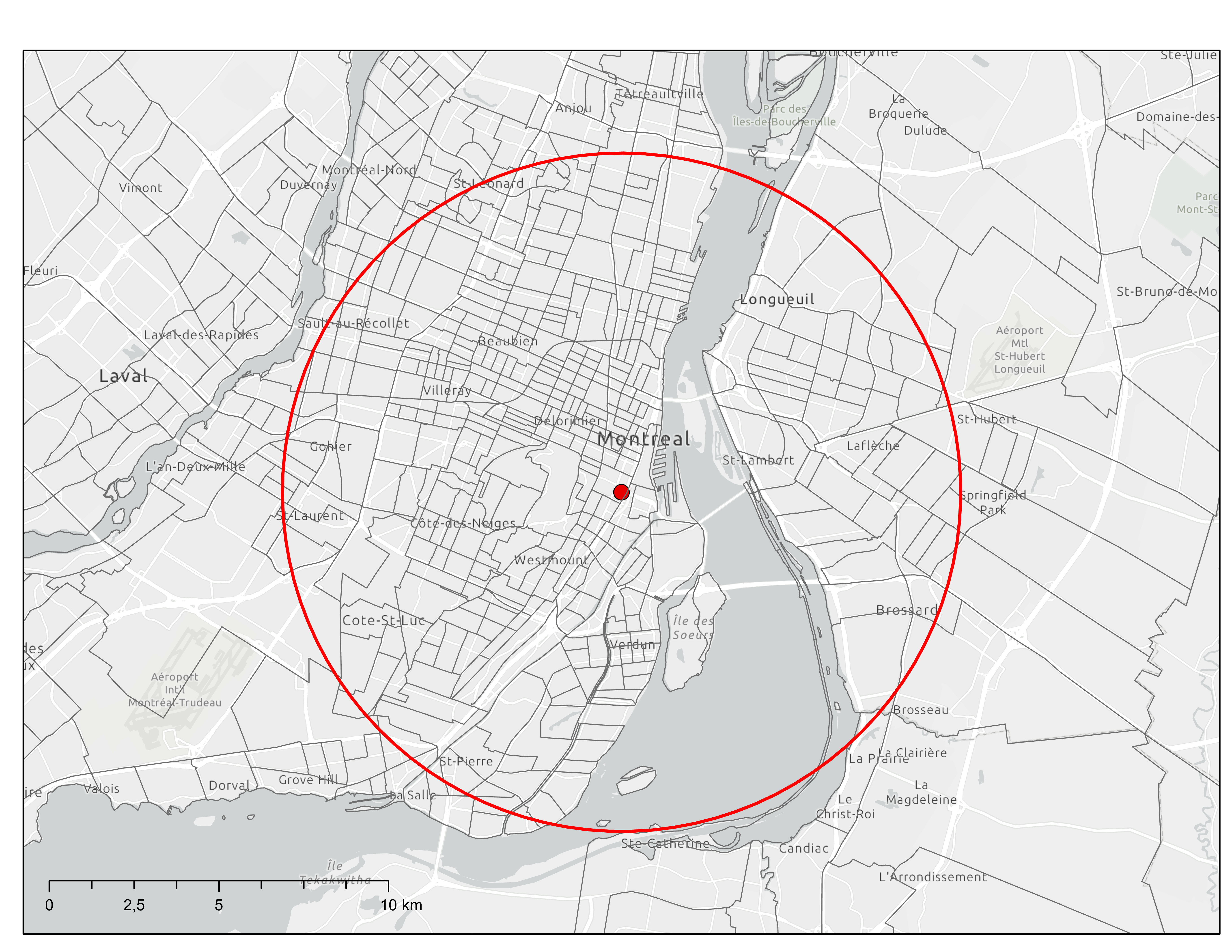}\\
\begin{minipage}{0.9\textwidth} 
{\footnotesize \vspace{0.1cm} Note: The red dot marks the central business district. The red circle represents a 10km radius circle around the CBD. Census tract boundaries are represented by black lines. \par}
\end{minipage}
\end{figure}

\begin{figure}[ht!]
\centering
\caption{Toronto central business district and 10 km radius\label{fig:radius_tor}}
\includegraphics[scale=0.55]{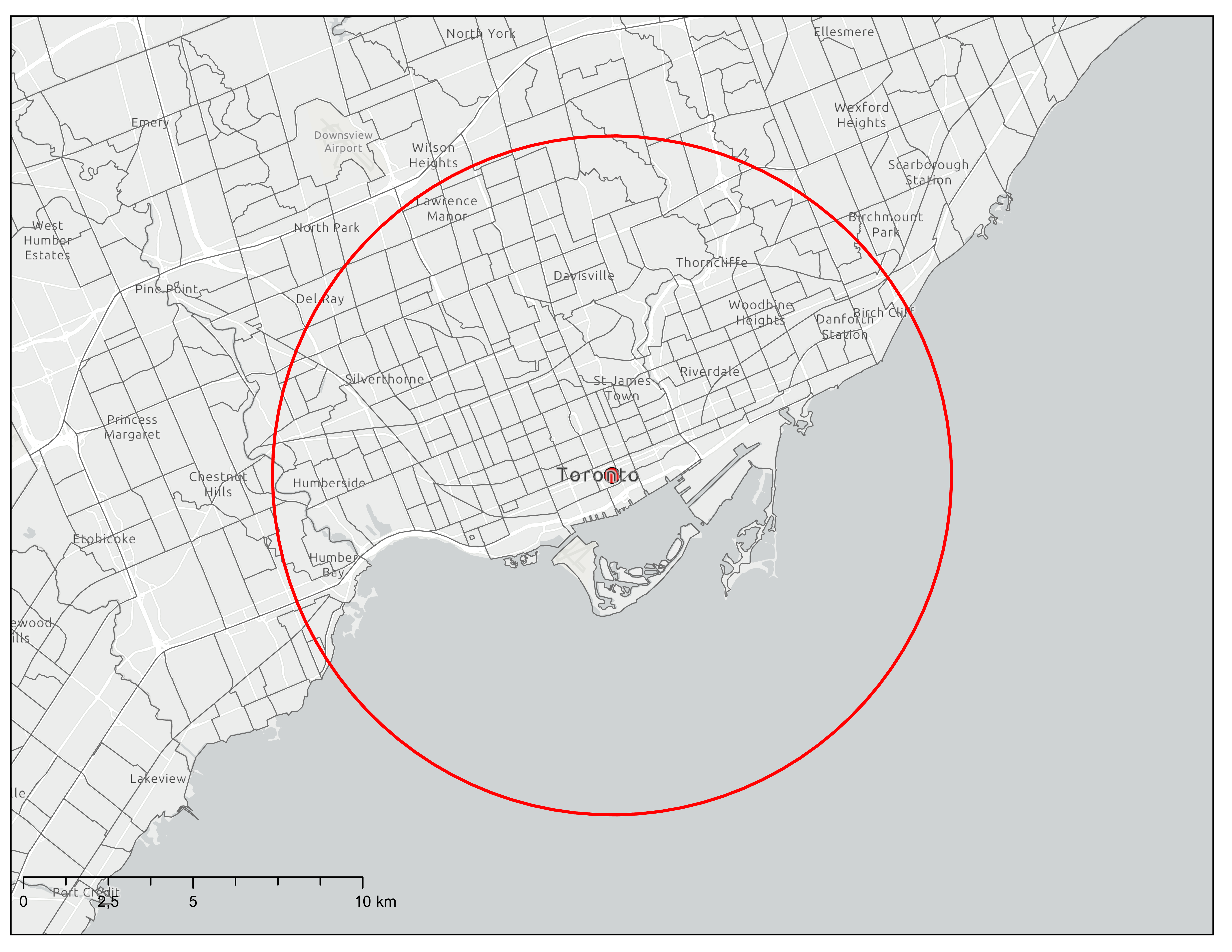}\\
\begin{minipage}{0.9\textwidth} 
{\footnotesize Note: The red dot marks the central business district. The red circle represents a 10km radius circle around the CBD. Census tract boundaries are represented by black lines. \par}
\end{minipage}
\end{figure}

\begin{figure}[ht!]
\centering
\caption{Vancouver central business district and 10 km radius\label{fig:radius_van}}
\includegraphics[scale=0.55]{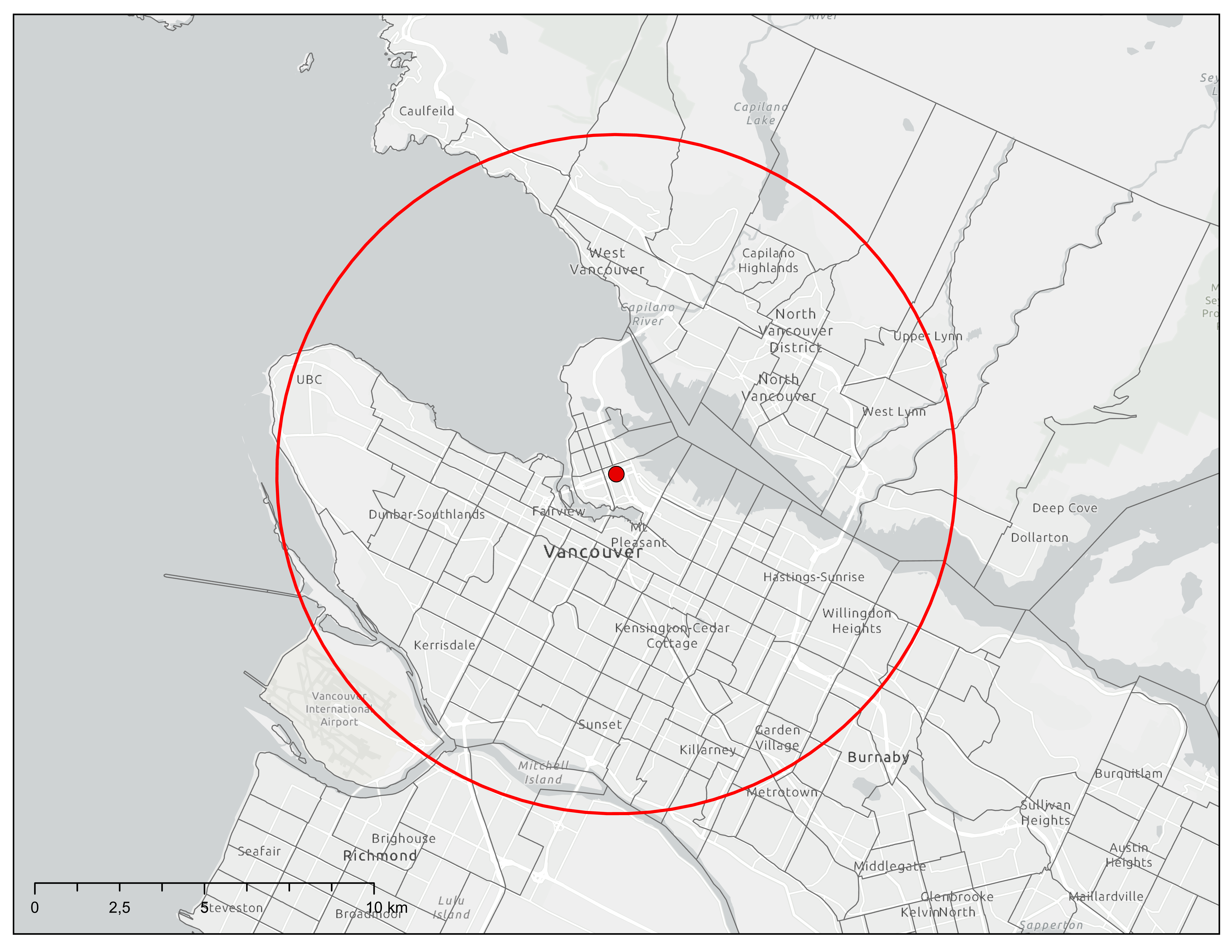}\\
\begin{minipage}{0.9\textwidth} 
{\footnotesize \vspace{0.1cm} Note: The red dot marks the central business district. The red circle represents a 10km radius circle around the CBD. Census tract boundaries are represented by black lines. \par}
\end{minipage}
\end{figure}

Determining what constitutes the central city can be a point of contention, with varying definitions proposed in the literature.\footnote{\citet{su2022rising} defines central city as all CTs located within 5 miles (about 8 km) from downtown. \citet{baum2020accounting} set the limit as 4 km from the CBD. Alternatively, \citet{couture2020urban} define a downtown in each CBSA as the set of tracts closest to the city center accounting for 5\% of a Core-Based Statistical Area's population (10\% in \citet{couture2021income}). On the other hand, \citet{behrens2022gentrification} capture 60\% of New York's population in their 30 km-radius central-city area.} To provide a middle-ground definition, I consider tracts that are within 10 km of the central business district as gentrifiable.\footnote{Central business district location are defined as follow; The intersection of Bay Street and King Street for Toronto, The \textit{Gare centrale} for Montréal, and The City Centre station for Vancouver.} This 10 km radius includes more than a third of the population in both Montréal and Vancouver, and a fifth of the population in Toronto. Formally, to calculate whether a census tract is within the 10 km radius, I measure the distance between the tract's centroid and the CBD coordinates. If the distance is within a 10 km radius, the tract is considered an inner-city area. Figures \ref{fig:radius_mtl}, \ref{fig:radius_tor}, and \ref{fig:radius_van} outline the 10 km radius from the CBD in each city.

\subsection{Sample characteristics}

This study focuses on incumbent households living in initially low-income, central city neighborhoods of the three most populous CMA before their CT starts gentrifying. I exclude individuals less than 20 years old when gentrification starts. Incumbent households are divided into high-income and low-income, depending on whether they are above or below the median income in the income distribution of their respective CMA. The low-income neighborhood sample comprises census tracts with median household income in the bottom half of the CT income distribution in their CMA and within 10km from downtown. 

Table \ref{tab:CTdist} shows the distribution of census tracts by their gentrification start date across cities. I split CT into six groups: \textit{Did not gentrify} (not gentrifiable, or gentrifiable that did not see a relatively large increase in the fraction of university graduates),  \textit{Improving before 1996} (neighborhoods that were already on an upward trend before the 1996-2006 period) and gentrifying neighborhood by their gentrification start date. The city of Montréal has the largest number of gentrifying neighborhoods over the study period, with 17\% of its tracts gentrifying between 1996 and 2016. In contrast, Toronto and Vancouver only saw around 5\% of their tract gentrify over the same period. It must be noted that in all three CMAs, a large portion of the CTs were already improving before 1996.  

For each 10-year period in the analysis, Tables \ref{tab:descstat_1996}, \ref{tab:descstat_2001} and \ref{tab:descstat_2006} describe some characteristics of neighborhoods by their gentrification status: not gentrifiable (outer-city), not gentrifiable (inner-city) and gentrifying (by quartile). Panel A shows the baseline characteristics of neighborhoods. Comparing the gentrifiable neighborhoods, CTs at the top and bottom quartiles look surprisingly similar in most aspects, except for the average distance from CBD and population. This highlights that neighborhoods closer to downtown and with lower density are gentrifying the most over the period. The similarity in average rent, median income, and the fraction of university graduates between the top and bottom quartiles is striking, especially when we compare them to the rich (not gentrifiable) inner-city neighborhoods. The only notable difference is in minority and immigration profiles: faster-gentrifying neighborhoods have fewer immigrants and visible minorities at baseline. 

Panel B pictures the changes in the fraction of college-educated, the fraction of low-income, and employment rates between 2006 and 2016. All those measures are strongly related to my measure of gentrification: comparing highly gentrifying neighborhoods (Q4) to not gentrifying neighborhoods (Q1) for the 2001-2011 period, the fraction of college-educated increased more (16.6 ppt vs 3.3 ppt), the employment rate increased more (5.4 ppt vs -1.2 ppt), and the fraction of low income decreased more (-5.9 ppt vs -1.7 ppt).


\section{Empirical Strategy} \label{chap:strategy}

\subsection{Gentrification induced displacement}
Guided by popular concerns, I start by studying the potential displacement effects of gentrification on low-income families. To evaluate whether gentrification leads to more out-migration of incumbent households, I estimate the following staggered event-study: 
\begin{align}\label{equation:reg_migration}
   Move_{hjc,t+b} = \beta_0 +  \sum_{b=-5}^{10} \delta_b  \widetilde{gent}_{jc,t} + \beta_2 X_{hjc,t} + \beta_3 A_{jc,t} + \beta_4 \Delta A_{jc,t-1} + \mu_c + \mu_t +\epsilon_{hjc,t}
\end{align}

Where $Move_{hjc,t+b}$ is a dummy variable equal to one if the household $h$ lives in a different census tract than their baseline census tract $j$, $b$ period after period $t$. $\widetilde{gent}_{jc,t}$ is a dummy variable whether the gentrification measure presented in Equation \ref{equation:gent_measure} is in the top half of the distribution. This captures the total effect of the inflow of university-educated households, including the effect on rent and endogenous amenities. 


$X_{hjc,t}$ is a vector of household characteristics. It includes age, family composition, including the number of children and whether one of the adults is an immigrant. $A_{jc,t}$ is a vector of pre-period tract characteristics such as initial median income, education levels, average rent and distance from the central business center. $\Delta A_{jc,t-1}$ is a vector of changes in neighborhood indicators during the previous five-year period, which includes income, education, rent, and minority rate.  Including this vector of change minimizes the concerns about individuals selecting into neighborhoods based on already existing trends. Finally, $\mu_c$ and $\mu_t$ are CMA and years fixed effects.

\subsection{Neighborhood exposure}

Beyond household displacement, gentrification may have an effect on relocation choices. Hence, I explore whether workers moving out of gentrifying neighborhoods end up in worse neighborhoods than those moving out of non-gentrifying neighborhoods. This highlights the displacement effect: involuntary movers might not select their new neighborhood as carefully as those who planned their move. I restrict the sample to households who do not currently live in their baseline neighborhood (i.e. movers). I run the following regression:
\begin{align}\label{equation:reg_exposure}
    Y_{hjc,t} = \beta_0 + \sum_{b=-5}^{10} \delta_b  \widetilde{gent}_{jc,t} + \beta_2 X_{hjc} + \beta_3 A_{jc,t} + \beta_4 \Delta A_{jc,t-1} + \mu_c + \mu_t + \epsilon_{hjc,t}
\end{align}

$Y_{hjc,t}$ is the exposure to a neighborhood characteristic for household $h$ in census tract $j$, in CMA $c$. The indicators are computed at baseline to avoid any endogeneity. Neighborhood characteristics exposure outcomes include the distance from the city center, the share of college-educated, the employment rate, the median income, and the poverty rate. It also includes some demographic composition of the neighborhoods, namely the share of visible minorities and the share of immigrants.

As in Equation \ref{equation:reg_migration}, $\widetilde{gent}_{jc,t}$ is an indicator of whether the gentrification measure presented in Equation \ref{equation:gent_measure} is in the top half of the distribution, $X_{hjc}$ is a vector of individual characteristics, $A_{jc,t}$ and $\Delta A_{jc,t-1}$ are neighborhood indicators and their changes in the previous period and $\mu_c$ and $\mu_t$ are a CMA and year fixed effects.

\subsection{Income trajectory}

I then turn to income effects. Gentrification can affect someone's labor outcomes in several ways. For stayers, it might result in a loss of employment if their current jobs are destroyed, and they don't benefit from the new job creation. It might also lead to mismatch in potential job-to-job transitions that are known to be beneficial for an individual's income trajectory. However, it could actually improve access to good jobs. Therefore, the long-term effects of gentrification on stayers' income are not clear-cut.

Conversely, for those who choose to move due to gentrification, quitting their current job may be necessary if they have to relocate too far away. As a result, they may end up in lower-paying jobs in their new location, or they may have to deal with longer commutes to their current jobs.

As I do not have access to specific employment data, I am limited to measuring the overall effect of gentrification on yearly earnings. While this approach may provide an incomplete picture of the employment situation, it is still significant in understanding the impact on individuals' welfare. Consequently, I shift my analysis from a household-level to an individual-level perspective to better capture the effects of gentrification on individuals' earnings.
\begin{align}\label{equation:reg_income}
  Y_{ijc,t+b} = \beta_0 + \sum_{b=-5}^{10} \delta_b  \widetilde{gent}_{jc,t} + \beta_2 X_{ijc} + \beta_3 A_{jc,t} + \beta_4 \Delta A_{jc,t-1} + \mu_c + \mu_t + \epsilon_{ijc,t}
\end{align}

$Y_{ijc,t}$ is the individual $i$'s income in year $t$. $\widetilde{gent}_{jc,t}$ is defined as in the previous analyses. $X_{ijc}$ contains individual characteristics, including age, age squared, sex, immigration and marital status. In addition to total gross income, I explore other income definitions to disentangle potential mechanisms. 

\subsection{Identification} \label{chap:indentification}

So far, I assumed gentrification to be solely the result of exogenous shocks in population composition. However, it is possible that population changes may also coincide with shifts in employment opportunities or other unobserved changes, which could potentially impact the results. Therefore, even after controlling for pre-trends, there may be concerns about potential selection bias based on unobserved or expected neighborhood changes. In the absence of a strong instrumental variable, the best approach I can take is to provide evidence that my control group is not significantly different in trends before the treatment.

Table \ref{tab:Bal_table_gent} shows how balanced the pre-trend in several neighborhood characteristics are. Although unconditional trends may show slight differences, once I control for a narrow set of controls, the pre-trends do not appear to be significant between neighborhoods that will undergo gentrification and those that won't. The only significant difference is in the pre-trend of the fraction of university-educated individuals, suggesting a small dip in my gentrification measure before treatment. 

Finally, I try to alleviate the remaining concerns by building a matched control group. The recent literature on heterogeneous treatment effects in event studies with timing in treatment raises concerns about using simple TWFE models. Here, I match each treated individual to one living in a never-treated neighborhood. I then stack each treatment-time group in a single regression. 

\subsection{Matched Samples} 
\label{sec:Matched_Samples}

For the moving analysis, I construct a comparison group of households initially living in neighborhoods that did not go through gentrification using a matching algorithm. For each treated unit, I take the set of untreated households living in the same city. A household is a potential control unit for household $h$ if: (1) the household never lived in a gentrifying neighborhood during our sample period, and (2) the household has the same family structure, the same tenure in the neighborhood (top codded at five years), and is less than half a standard deviation away from $h$ in the family income distribution the years prior to treatment. Potential controls are randomly matched to treated units without replacement.\footnote{I use the \textit{calipmatch} package from \citet{ stepner2017calipmatch}. For each case, \textit{calipmatch} searches for matching controls until it either finds a match or runs out of controls. While it is possible that some cases end up unmatched because all possible matching controls have already been matched with another case, the high match rate indicates that this isn’t a major problem here.}

This matching strategy finds a counterfactual household for about 97.6 percent of all treated units. Panel A of Table \ref{tab:matching_table} shows household characteristics for the moving analysis sample. The characteristics of the matched treated and untreated households are quite similar, even for characteristics that were not part of the matching algorithm, such as the number of kids. 

For the income trajectory analysis, I build the control group in a similar fashion. For each treated individual, I take the set of untreated individuals who live in the same city. An individual is a potential control unit for treated unit $i$ if: (1) the individual never lived in a gentrifying neighborhood during our sample period, and (2) the individual has the marital status, less than five years difference of age, and are less than half a standard deviation away from $i$ in the income distribution the years prior to treatment. 

The matching algorithm finds a counterfactual for 96.8 percent of the treated individuals. Panel B of Table \ref{tab:matching_table}  shows individual characteristics for the income trajectory analysis sample. The characteristics of the matched treated and untreated individuals are quite similar, even for characteristics beyond those used by the matching algorithm.

\section{Results} \label{chap:results}

In this section, I examine how gentrification affects incumbent residents in several dimensions. I first present the results from estimating the effect of gentrification on residents' mobility, corresponding to the estimate of equation \ref{equation:reg_migration}. I then focus on incumbents who endogenously decide to move and explore how gentrification affects their choice of area to relocate and how those choices impact their exposure to several neighborhood characteristics. Those estimates correspond to Equation \ref{equation:reg_exposure}. Finally, I present the estimate from Equation \ref{equation:reg_income} and picture the effect of gentrification on income trajectory, stratifying the sample by high/low income workers.

\subsection{Gentrification induced displacement} \label{chap:results_displacement}

\begin{figure}[ht!]
\caption{Effect of gentrification on displacement }
\centering
\subfloat[Low-income]{\includegraphics[width = 5in]{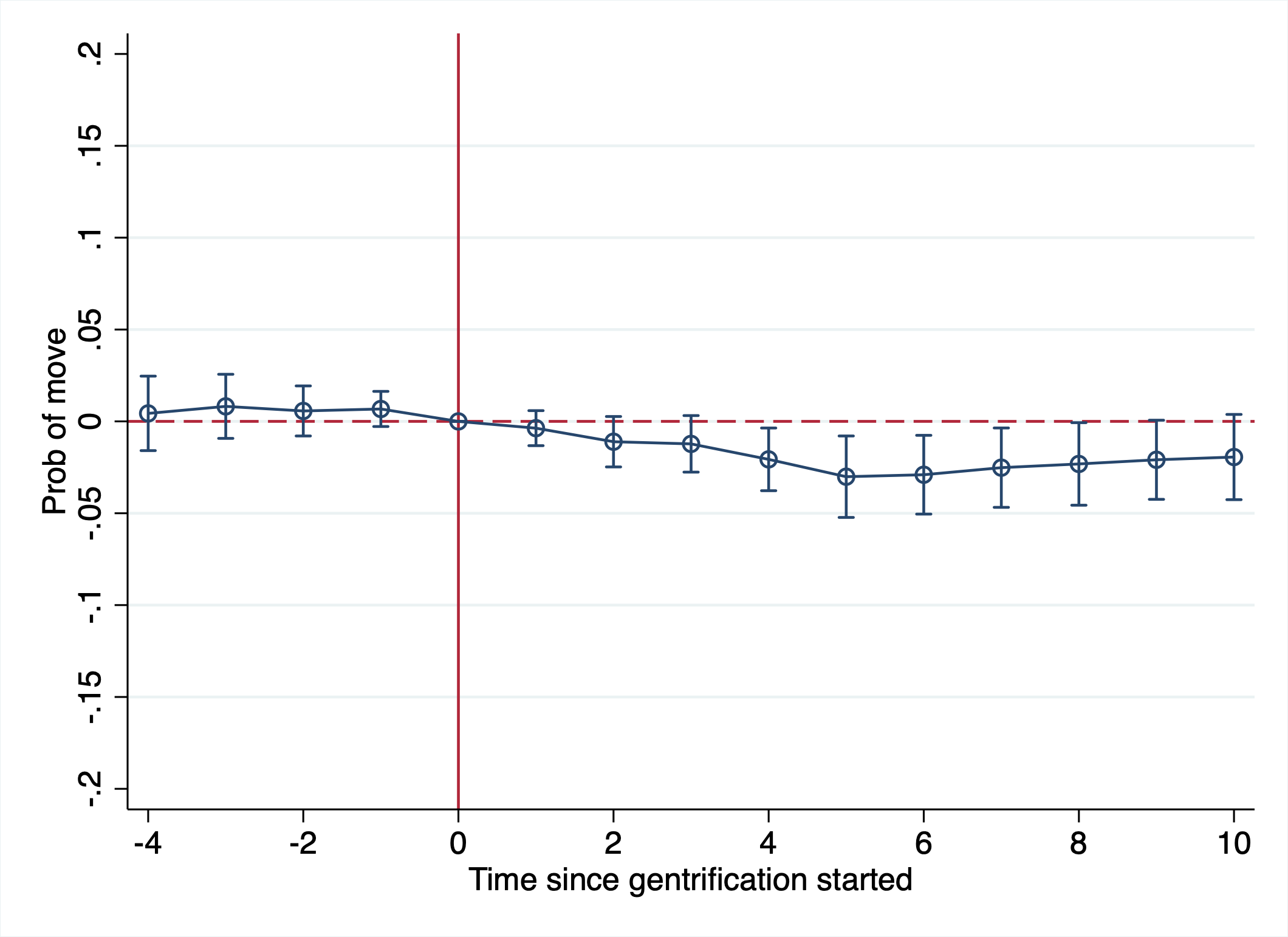}} \\
\subfloat[High-income]{\includegraphics[width = 5in]{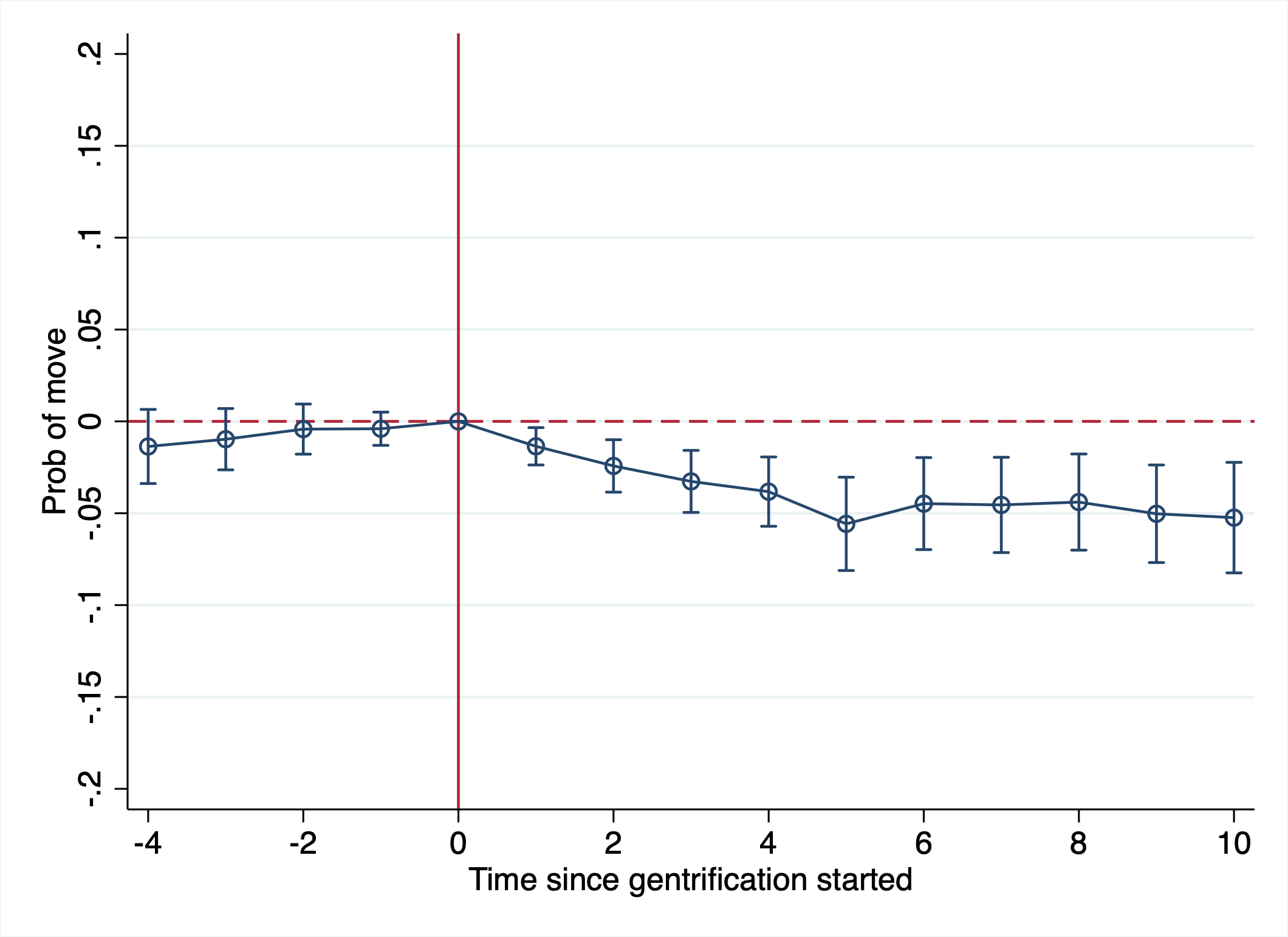}}\\
\label{fig:Displacement}
\begin{minipage}{0.9\textwidth} 
{\footnotesize \vspace{0.1cm} Source: Author's calculations using the Census and LAD.\\
Note: Dependant variables: dummy equal to one if the household lives in a different tract than at baseline. The sample is incumbent households living in gentrifiable neighborhoods (initially low-income and central city) prior to one of the baseline years. The control group is the matched sample discussed in section \ref{sec:Matched_Samples}. The regressions also include family-level socioeconomic control variables (age, family composition,  number of children, and immigrant indicator), baseline Census Tract controls (college-educated share, median income, share for low income, average rent, employment rate, visible minority share, the share of immigrant, distance from CBD), and pre-period variation controls (changes of college-educated share, median income, average rent, employment rate). Low-income number of observations: 991,125. High-income number of observations: 759,985.\par}
\end{minipage}
\end{figure}

Figure \ref{fig:Displacement} shows OLS estimates of the effect of gentrification on incumbent mobility. Gentrification can have a ``first order" effect by displacing the incumbent population, directly affecting their welfare. We see that gentrification decreases the probability that incumbent residents leave their neighbourhood. Low-income households are 2.0 percentage points less likely to have moved out of their baseline neighborhood ten years into the gentrification process than their counterpart in non-gentrifying neighborhoods. High-income households are also more likely to stay longer, 5.2 percentage points more likely to live in their origin neighborhoods ten years later. 

While the popular belief is that gentrification leads low-income people to leave their historical neighborhoods, my results suggest it does the opposite. Several mechanisms can explain this phenomenon. For renters, rent stickiness might allow them to stay longer than one could expect. By staying, they benefit from improving amenities, such as reduced crime, lower poverty, and improved infrastructure. This is likely the case for low-income households. 

For high-income households, who are more likely to be homeowners, economic incentives might lead them to stay longer in their original neighborhood. As gentrification occurs, they benefit from the transformations happening in the neighborhood, such as the inflow of new businesses, new infrastructure, and improved transit. Incumbent residents who own property in the neighborhood may see their property values rise, which could provide a financial incentive to stay and benefit from the increased value of their homes. 

While those effects do not look very large, they represent a 6.7\% to 10.9\% decrease in the probability of moving within 10 years from the control group's mobility rate of 55\%.

\subsection{Neighborhood exposure}

While my results show no displacement effects, it does not mean it has no adverse effect on location choice. As incumbent residents try to stay in their neighborhood, forces might push them away, such as large rent increases and evictions. If moves from gentrifying neighborhoods are more involuntary, they might lead to worse location choices if individuals have to make rushed decisions. 

I explore how gentrification might affect relocation choices. Looking at movers only, I study different neighborhood characteristics of CTs households after leaving their baseline neighborhood. Dependant variables are pre-period measures to avoid endogeneity between the tract characteristics and the individual's location choice. 

\begin{figure}[ht!]
\caption{Effect of gentrification on location choice: Distance from CBD}
\centering
\subfloat[Low-income]{\includegraphics[width = 5in]{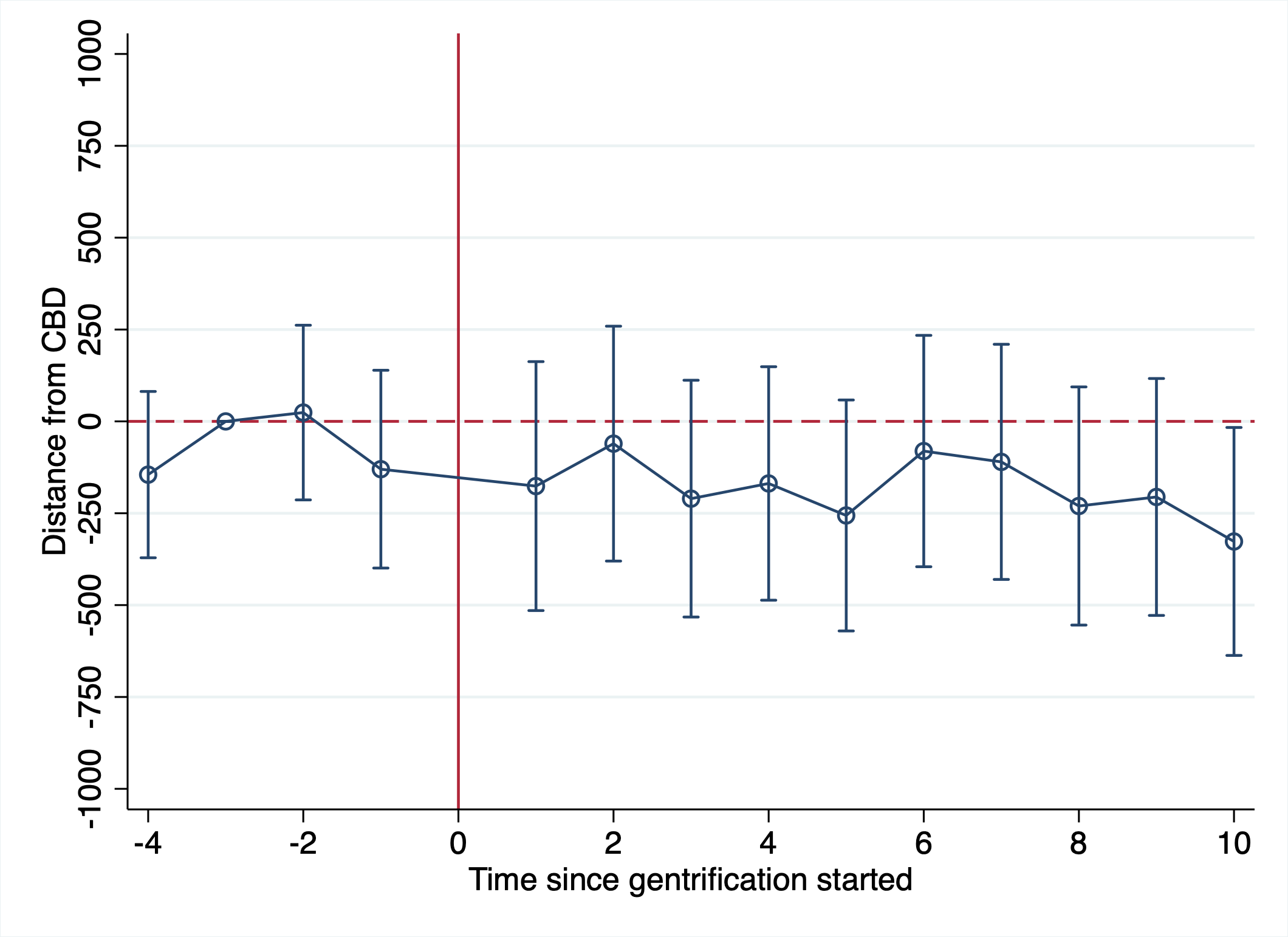}} \\
\subfloat[High-income]{\includegraphics[width = 5in]{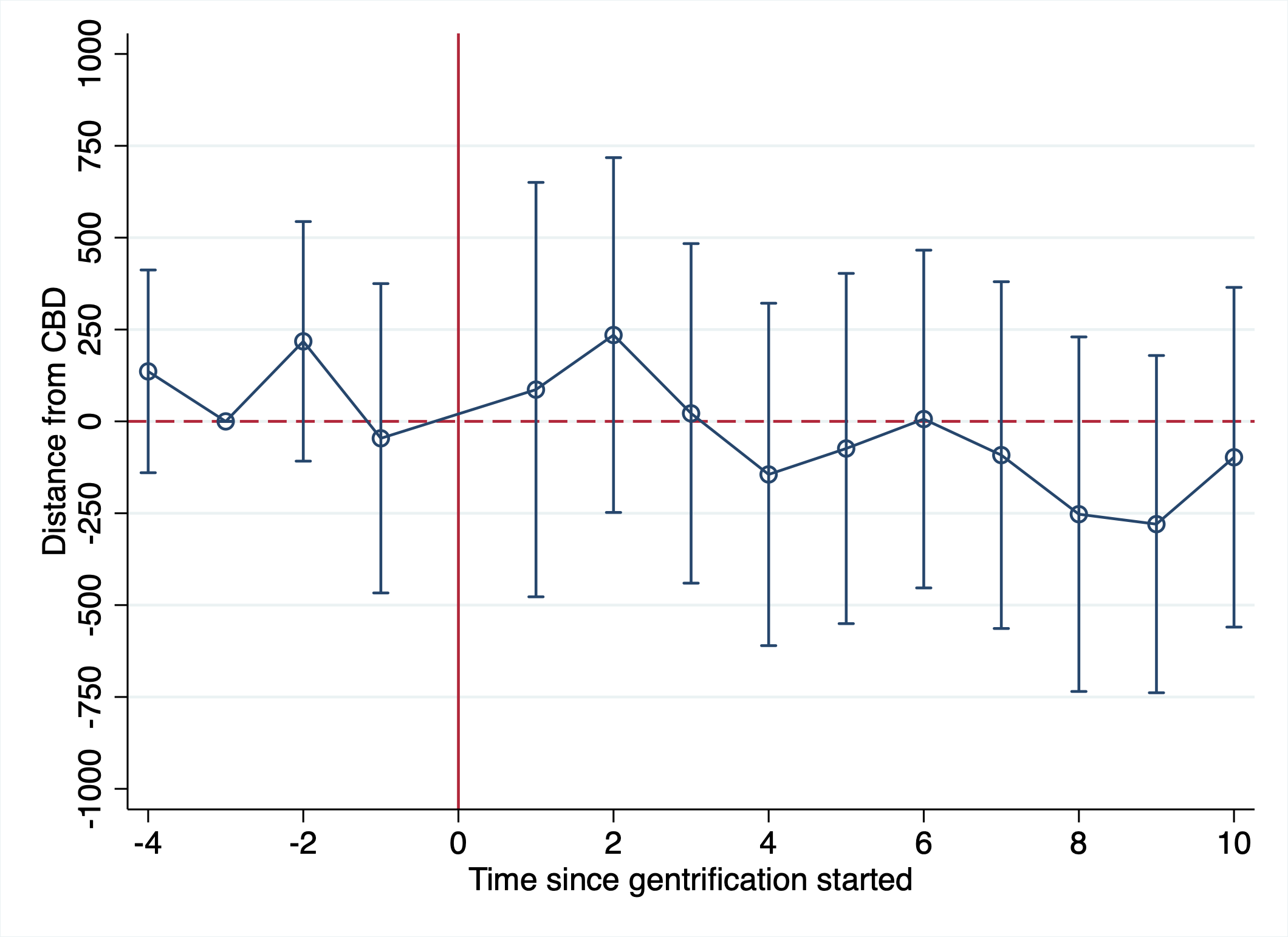}}\\
\label{fig:exp_distance}
\begin{minipage}{0.9\textwidth} 
{\footnotesize \vspace{0.1cm} Source: Author's calculations using the Census and LAD.\\
Note: Dependant variables: distance from central business district. The sample is incumbent households living in gentrifiable neighborhoods (initially low-income and central city) in one of the baseline years, but do not currently live in that census tract. The control group is the matched sample discussed in section \ref{sec:Matched_Samples}. The regressions also include family-level socioeconomic control variables (age, family composition,  number of children and immigrant indicator), baseline Census Tract controls (college-educated share, median income, share for low income, average rent, employment rate, visible minority share, the share of immigrant, distance from CBD), and pre-period variation controls (changes of college-educated share, median income, average rent, employment rate). Low-income number of observations: 356,900. High-income number of observations: 202,530. \par}
\end{minipage}
\end{figure}

\begin{figure}[ht!]
\caption{Effect of gentrification on location choice: Share of low-income}
\centering
\subfloat[Low-income]{\includegraphics[width = 5in]{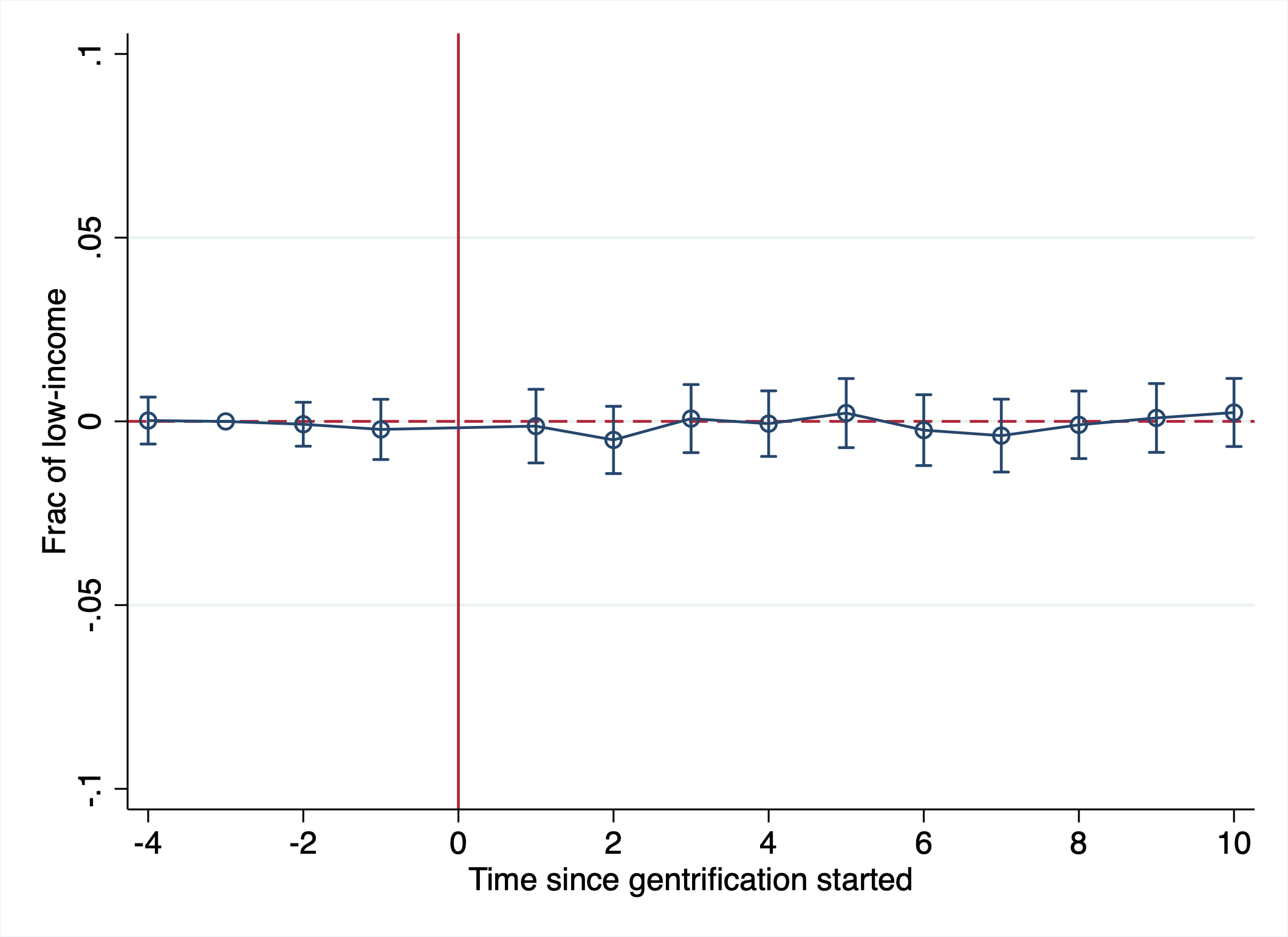}} \\
\subfloat[High-income]{\includegraphics[width = 5in]{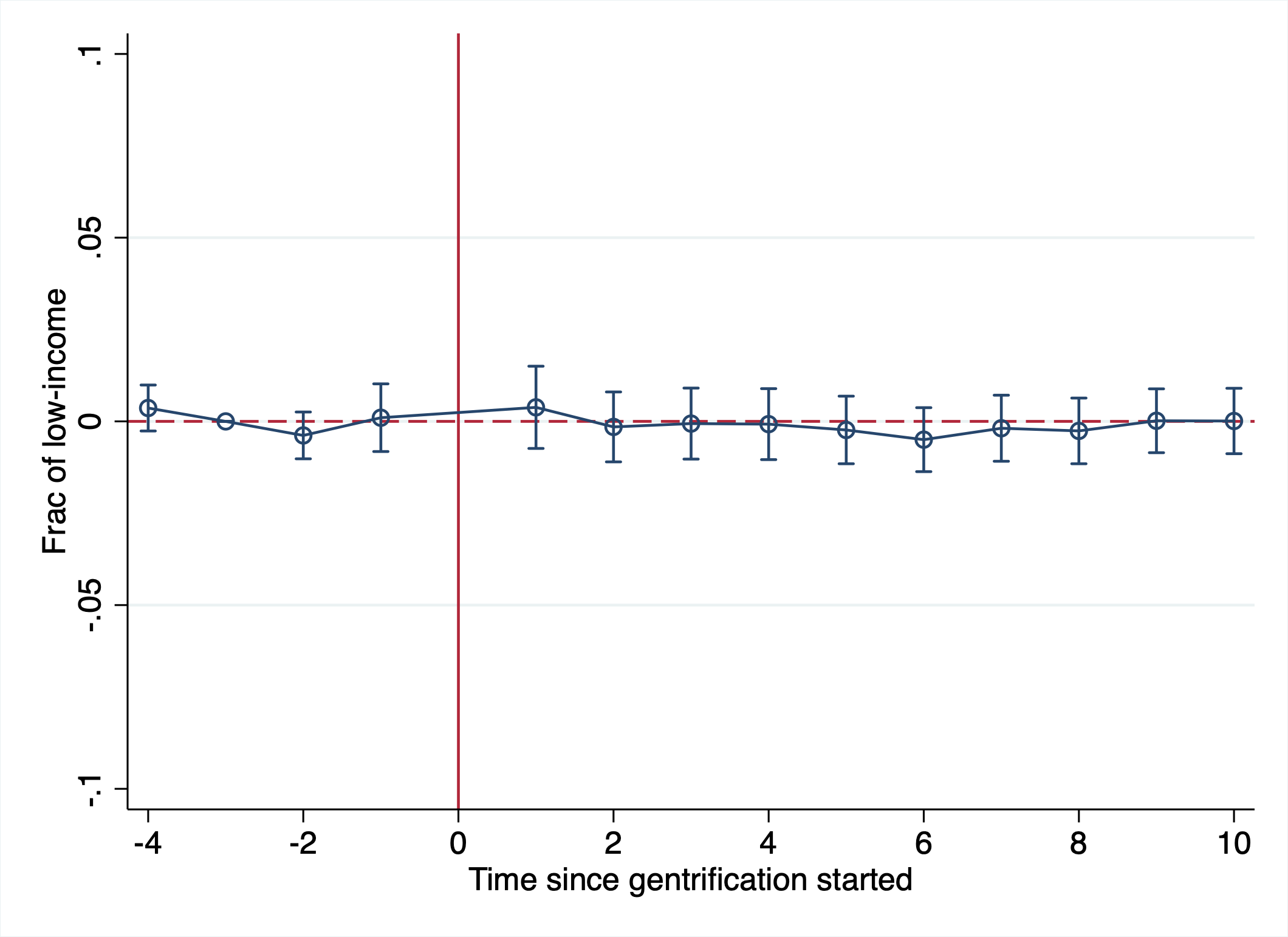}}\\
\label{fig:exp_lowinc}
\begin{minipage}{0.9\textwidth} 
{\footnotesize \vspace{0.1cm} Source: Author's calculations using the Census and LAD.\\
Note: Dependant variables: share of low-income households in the CT. The sample is incumbent households living in gentrifiable neighborhoods (initially low-income and central city) in one of the baseline years, but do not currently live in that census tract. The control group is the matched sample discussed in section \ref{sec:Matched_Samples}. The regressions also include family-level socioeconomic control variables (age, family composition,  number of children and immigrant indicator), baseline Census Tract controls (college-educated share, median income, share for low income, average rent, employment rate, visible minority share, the share of immigrant, distance from CBD), and pre-period variation controls (changes of college-educated share, median income, average rent, employment rate). Low-income number of observations: 356,900. High-income number of observations: 202,530. \par}
\end{minipage}
\end{figure}

\begin{figure}[ht!]
\caption{Effect of gentrification on location choice: Share of university-educated}
\centering
\subfloat[Low-income]{\includegraphics[width = 5in]{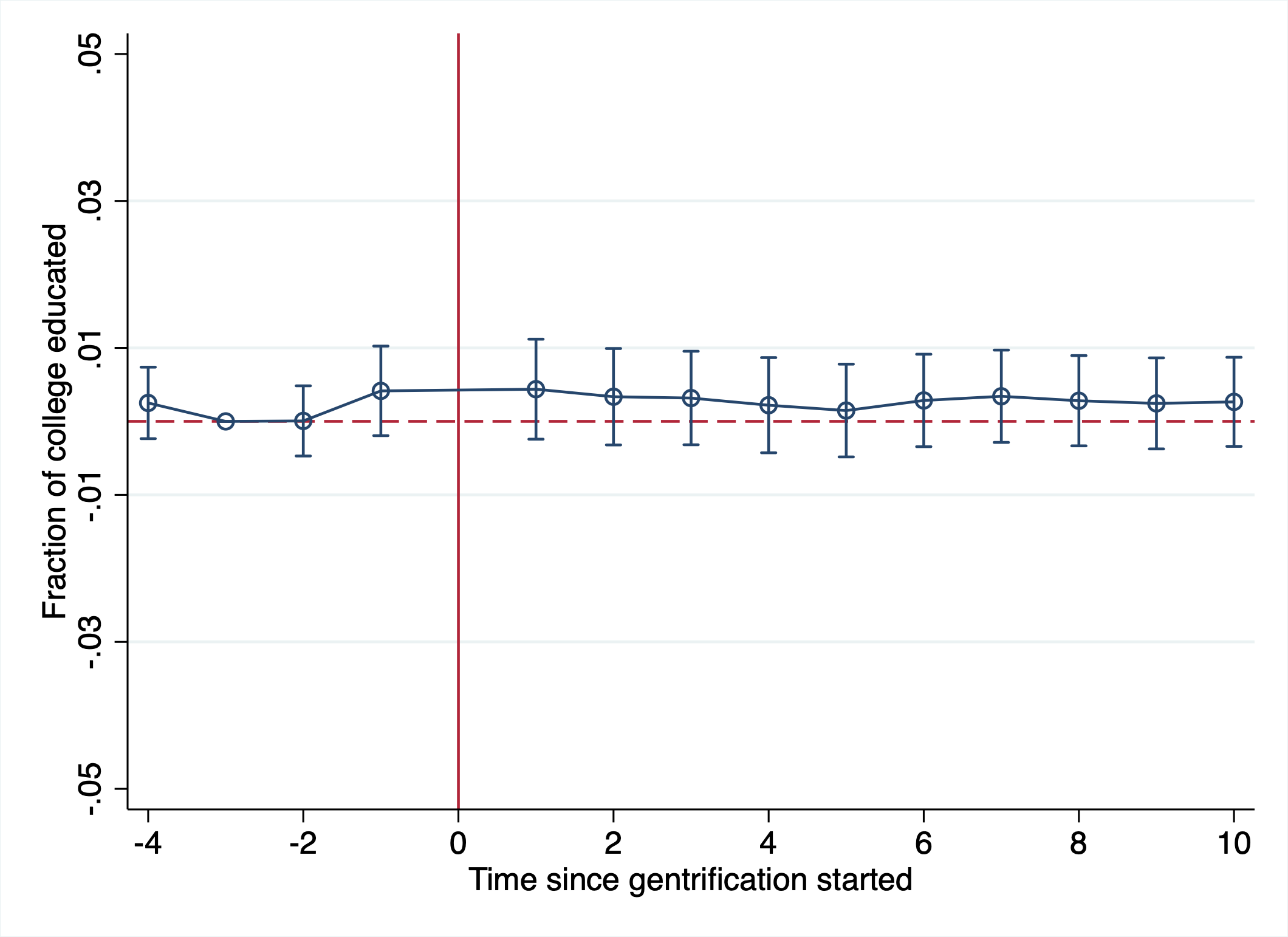}} \\
\subfloat[High-income]{\includegraphics[width = 5in]{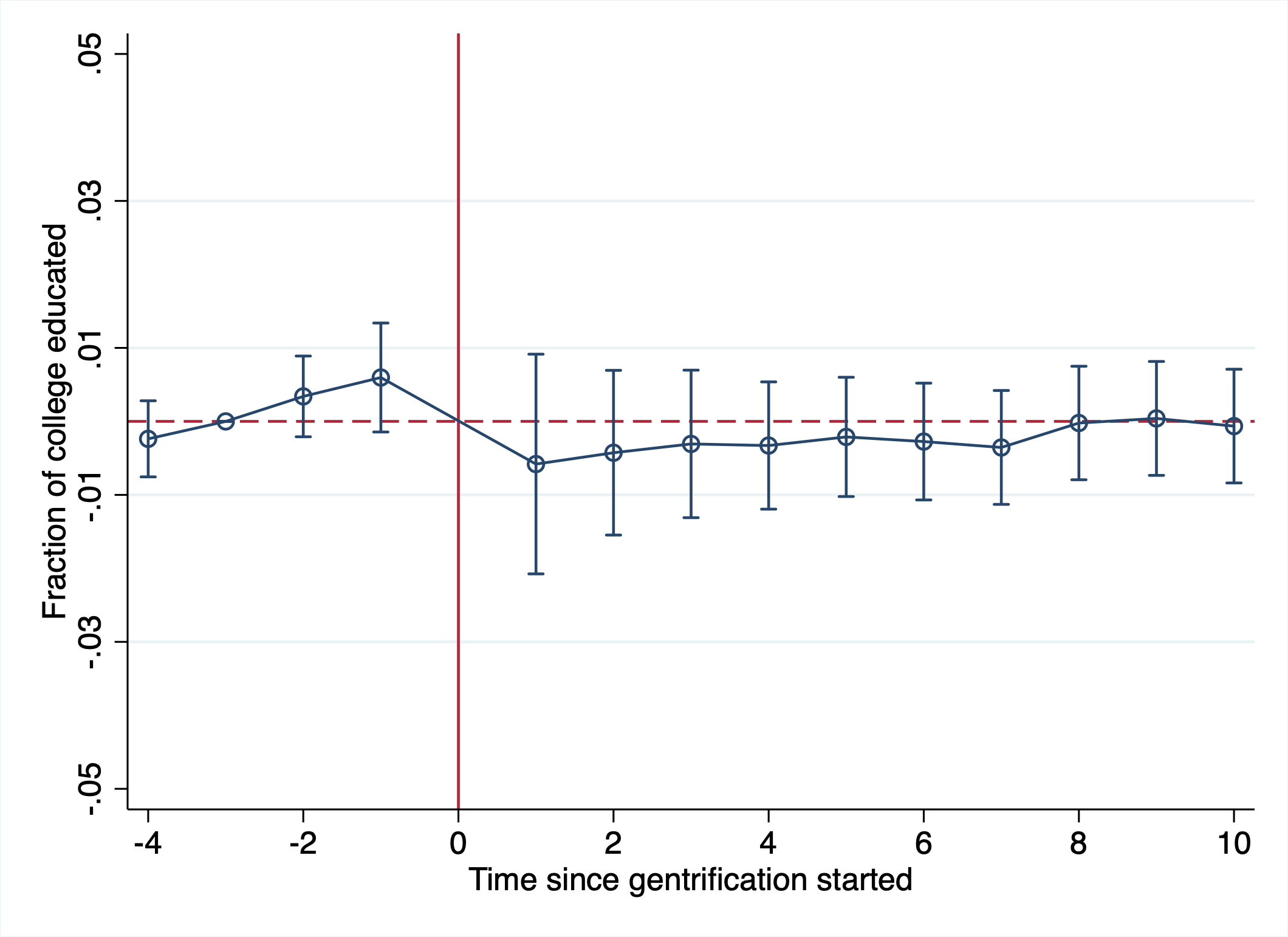}}\\
\label{fig:exp_colgrad}
\begin{minipage}{0.9\textwidth} 
{\footnotesize \vspace{0.1cm} Source: Author's calculations using the Census and LAD.\\
Note: Dependant variables: share of university-educated in the CT. The sample is incumbent households living in gentrifiable neighborhoods (initially low-income and central city) in one of the baseline years, but do not currently live in that census tract. The control group is the matched sample discussed in section \ref{sec:Matched_Samples}. The regressions also include family-level socioeconomic control variables (age, family composition,  number of children and immigrant indicator), baseline Census Tract controls (college-educated share, median income, share for low income, average rent, employment rate, visible minority share, the share of immigrant, distance from CBD), and pre-period variation controls (changes of college-educated share, median income, average rent, employment rate). Low-income number of observations: 356,900. High-income number of observations: 202,530. \par}
\end{minipage}
\end{figure}

\begin{figure}[ht!]
\caption{Effect of gentrification on location choice: Employment rate}
\centering
\subfloat[Low-income]{\includegraphics[width = 5in]{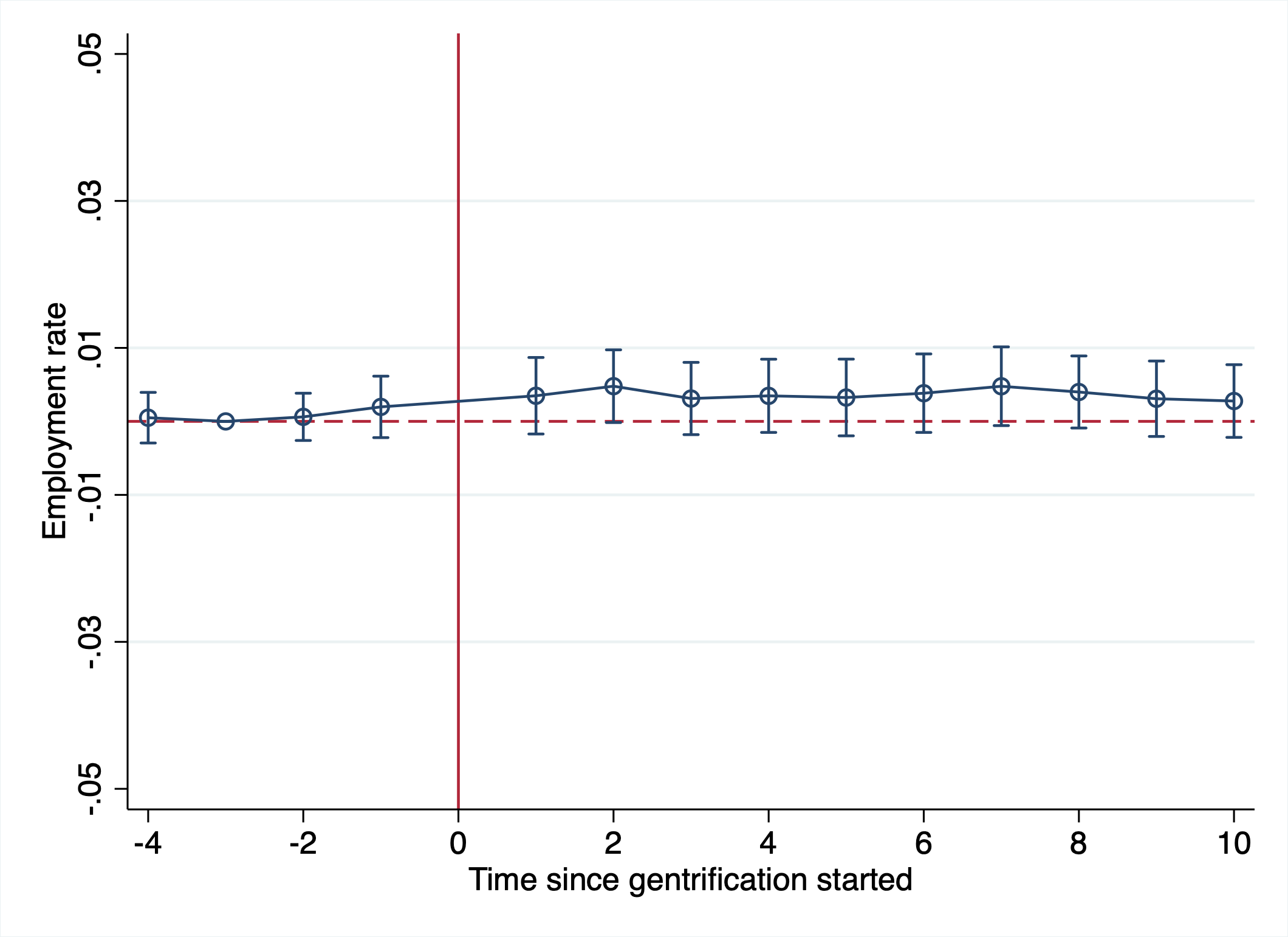}} \\
\subfloat[High-income]{\includegraphics[width = 5in]{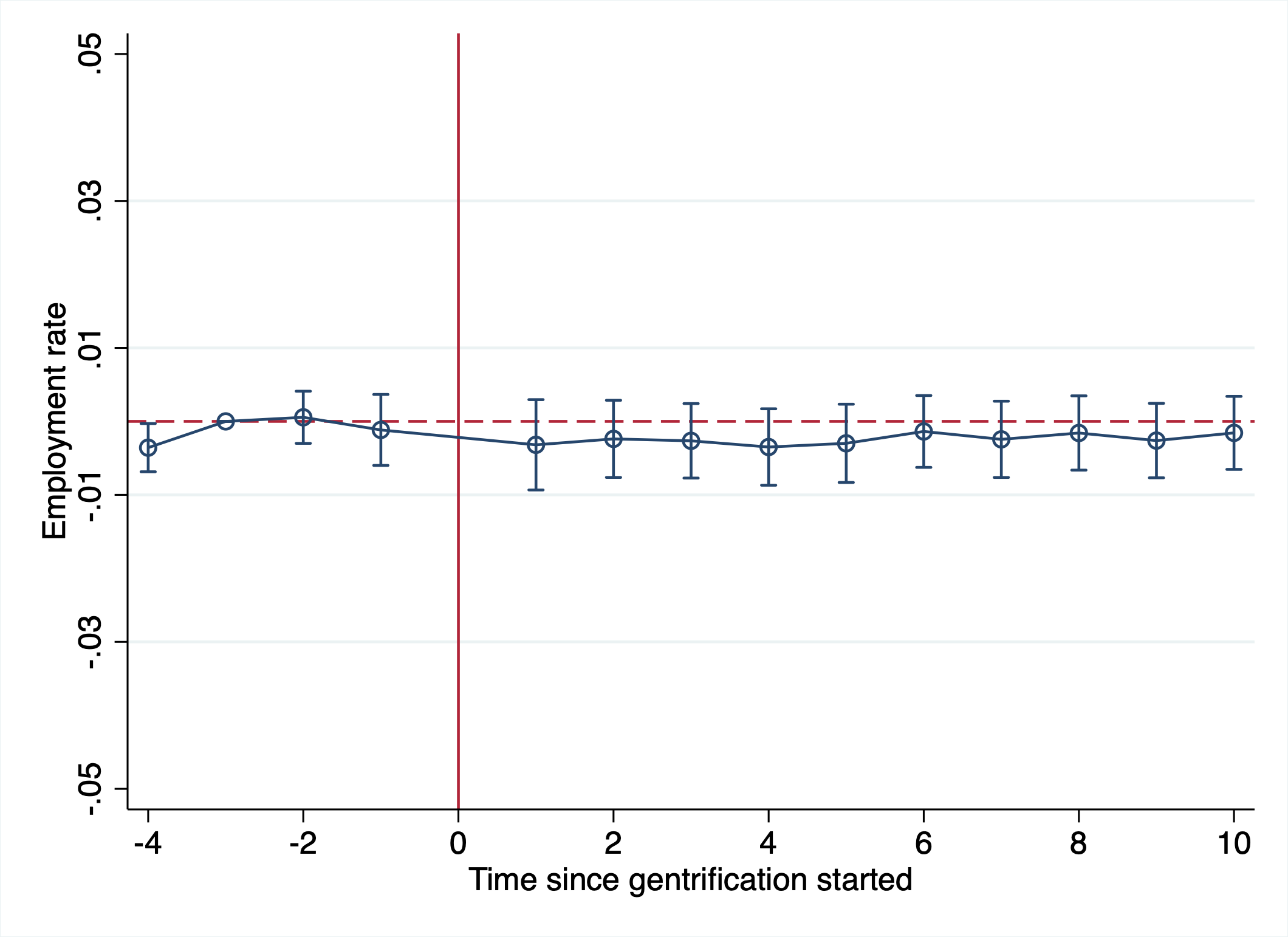}}\\
\label{fig:exp_unempl}
\begin{minipage}{0.9\textwidth} 
{\footnotesize \vspace{0.1cm} Source: Author's calculations using the Census and LAD.\\
Note: Dependant variables: employment rate in the CT. The sample is incumbent households living in gentrifiable neighborhoods (initially low-income and central city) in one of the baseline years, but do not currently live in that census tract. The control group is the matched sample discussed in section \ref{sec:Matched_Samples}. The regressions also include family-level socioeconomic control variables (age, family composition,  number of children and immigrant indicator), baseline Census Tract controls (college-educated share, median income, share for low income, average rent, employment rate, visible minority share, the share of immigrant, distance from CBD), and pre-period variation controls (changes of college-educated share, median income, average rent, employment rate). Low-income number of observations: 356,900. High-income number of observations: 202,530. \par}
\end{minipage}
\end{figure}

\begin{figure}[ht!]
\caption{Effect of gentrification on location choice: Median income}
\centering
\subfloat[Low-income]{\includegraphics[width = 5in]{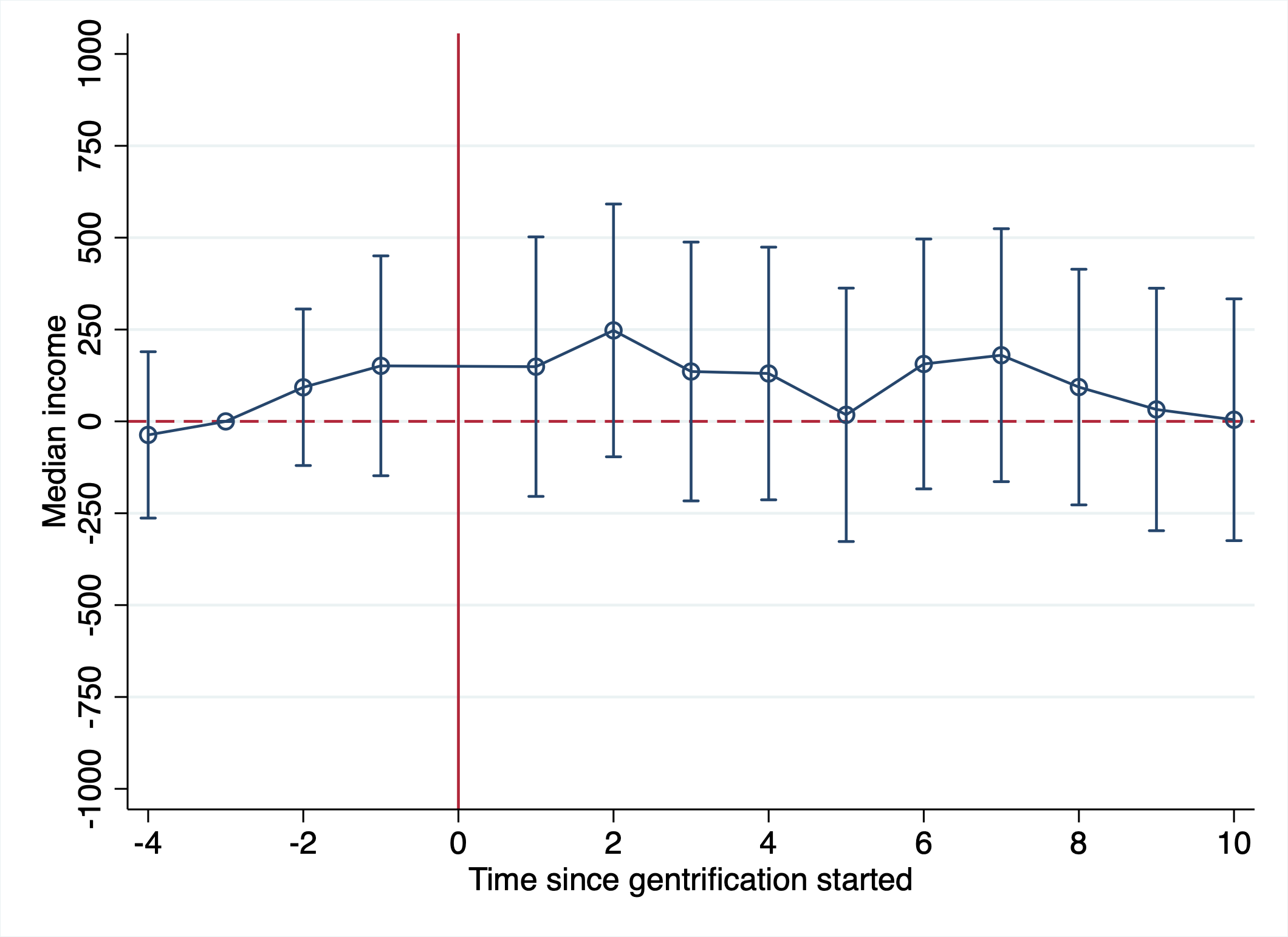}} \\
\subfloat[High-income]{\includegraphics[width = 5in]{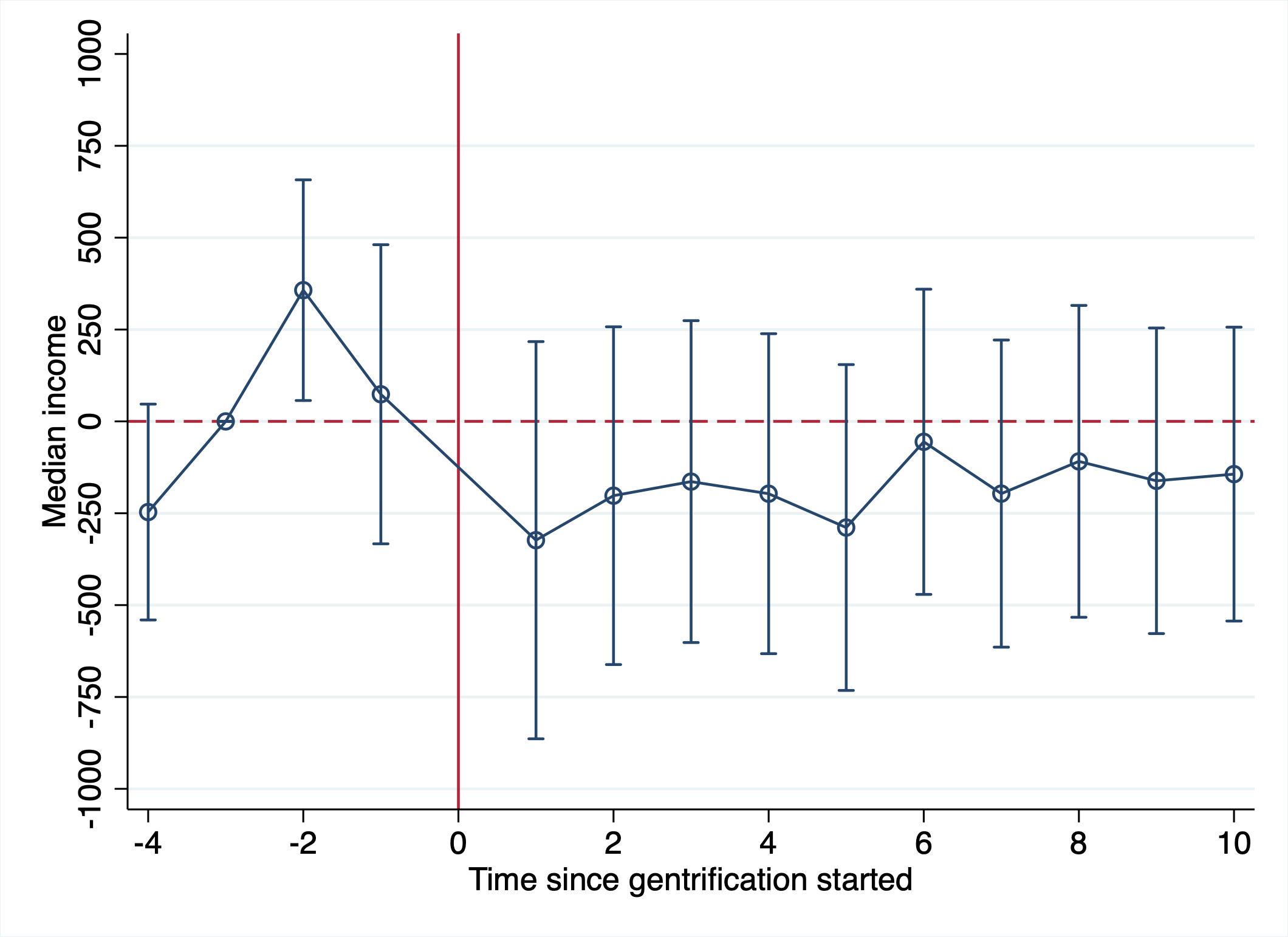}}\\
\label{fig:exp_meadian_inc}
\begin{minipage}{0.9\textwidth} 
{\footnotesize \vspace{0.1cm} Source: Author's calculations using the Census and LAD.\\
Note: Dependant variables: median household income in the CT. The sample is incumbent households living in gentrifiable neighborhoods (initially low-income and central city) in one of the baseline years, but do not currently live in that census tract. The control group is the matched sample discussed in section \ref{sec:Matched_Samples}. The regressions also include family-level socioeconomic control variables (age, family composition,  number of children and immigrant indicator), baseline Census Tract controls (college-educated share, median income, share for low income, average rent, employment rate, visible minority share, the share of immigrant, distance from CBD), and pre-period variation controls (changes of college-educated share, median income, average rent, employment rate). Low-income number of observations: 356,900. High-income number of observations: 202,530. \par}
\end{minipage}
\end{figure}

The results show gentrification has no adverse effect on location choice for individuals who endogenously decide to leave. Households are not relocating to neighborhoods with more low-income households (Figure \ref{fig:exp_lowinc}), lower employment rate (Figure \ref{fig:exp_unempl}), lower median income (Figure \ref{fig:exp_meadian_inc}) or higher fraction of university-educated (Figure \ref{fig:exp_colgrad}). The only significant difference in trend is the distance from the Central business district (Figure \ref{fig:exp_distance}). Households leaving gentrifying neighborhoods relocate closer to downtown than their counterparts in non-gentrifying neighborhoods. This might reflect different general location preferences not influenced by neighborhood changes. 

While results suggest that gentrification leads to lower household turnover, once households endogenously decide to move, they do not relocate to worse locations than those moving out of non-gentrifying neighborhoods. A potential explanation is that once low-income households decide to leave, they face the same housing market as similar households. It also indicates that neighborhood changes do not rush residents into a moving decision, potentially leading to poorer location choices.

\subsection{Income trajectory} \label{chap:income_results}
\begin{figure}[ht!]
\caption{Effect of gentrification on total income}
\centering
\subfloat[Low-income]{\includegraphics[width = 5in]{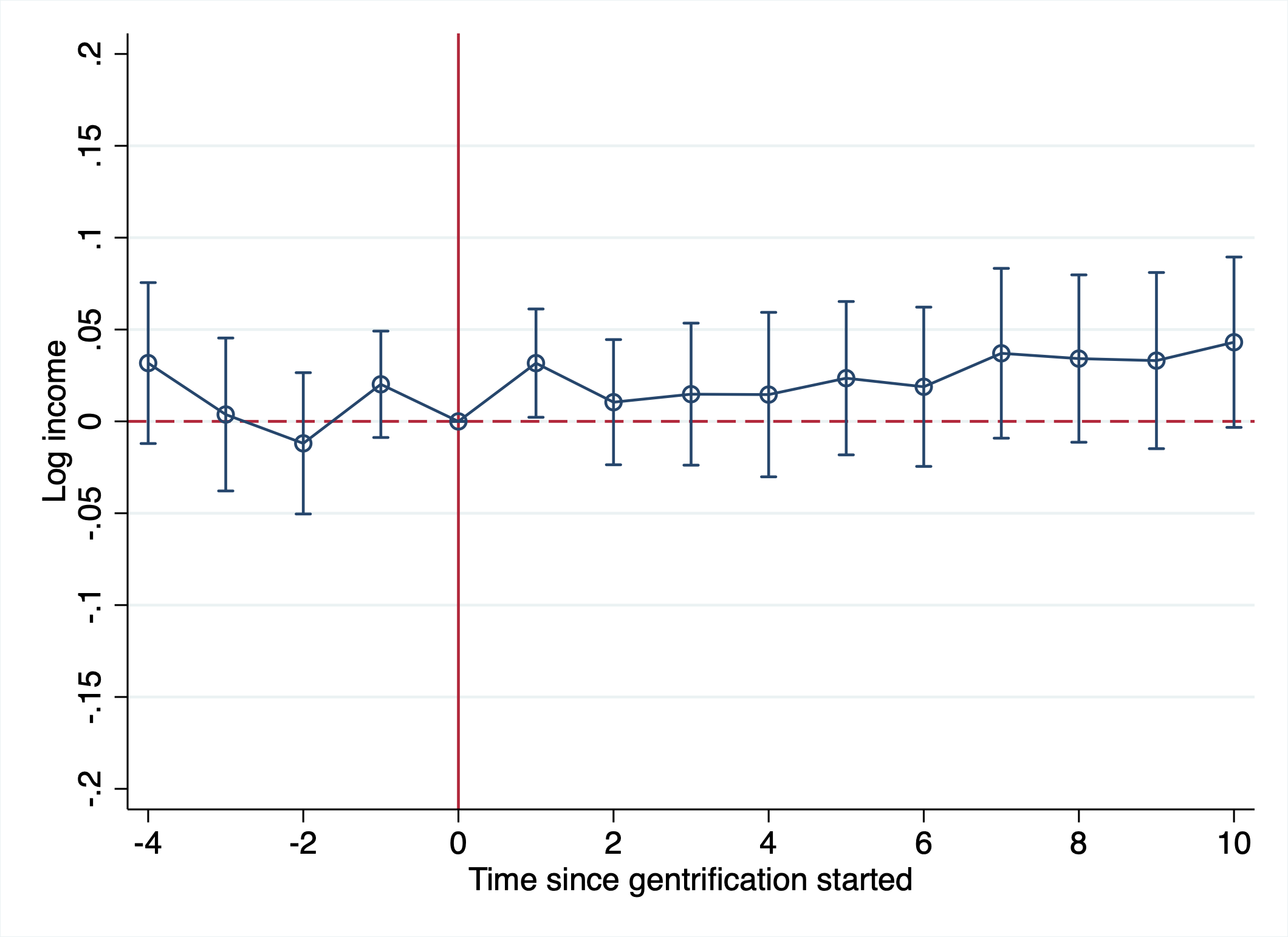}} \\
\subfloat[High-income]{\includegraphics[width = 5in]{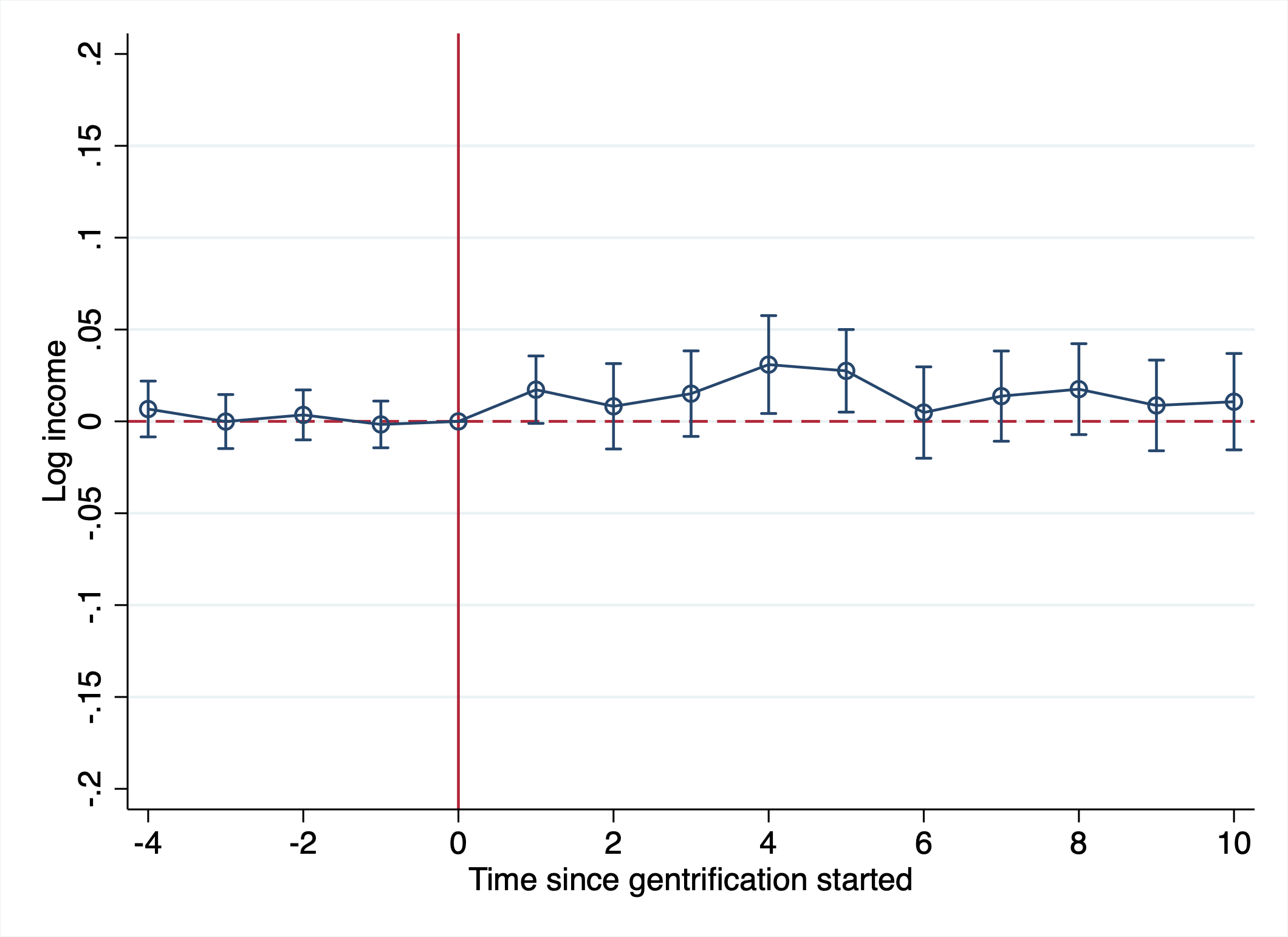}}\\
\label{fig:ES_total_earnings}
\begin{minipage}{0.9\textwidth} 
{\footnotesize \vspace{0.1cm} Source: Author's calculations using the Census and LAD.\\
Note: Dependant variables: total gross income. The sample is incumbent households living in gentrifiable neighborhoods (initially low-income and central city) in one of the baseline years. The control group is the matched sample discussed in section \ref{sec:Matched_Samples}.  The regressions also include individual-level control variables (age, age squared, gender, family composition,  number of children and immigrant indicator), baseline Census Tract controls (college-educated share, median income, share for low income, average rent, employment rate, visible minority share, the share of immigrant, distance from CBD), and pre-period variation controls (changes of college-educated share, median income, average rent, employment rate). Low-income number of observations: 4,698,305. High-income number of observations: 2,559,580. \par}
\end{minipage}
\end{figure}

Location and neighborhood characteristics might directly affect individuals' welfare but could also impact their labor force outcomes, influencing their welfare. Figure \ref{fig:ES_total_earnings} shows the matched estimates of Equation \ref{equation:reg_income} for total income. We see no effect on total income.

\begin{figure}[ht!]
\caption{Effect of gentrification on employment earnings}
\centering
\subfloat[Low-income]{\includegraphics[width = 5in]{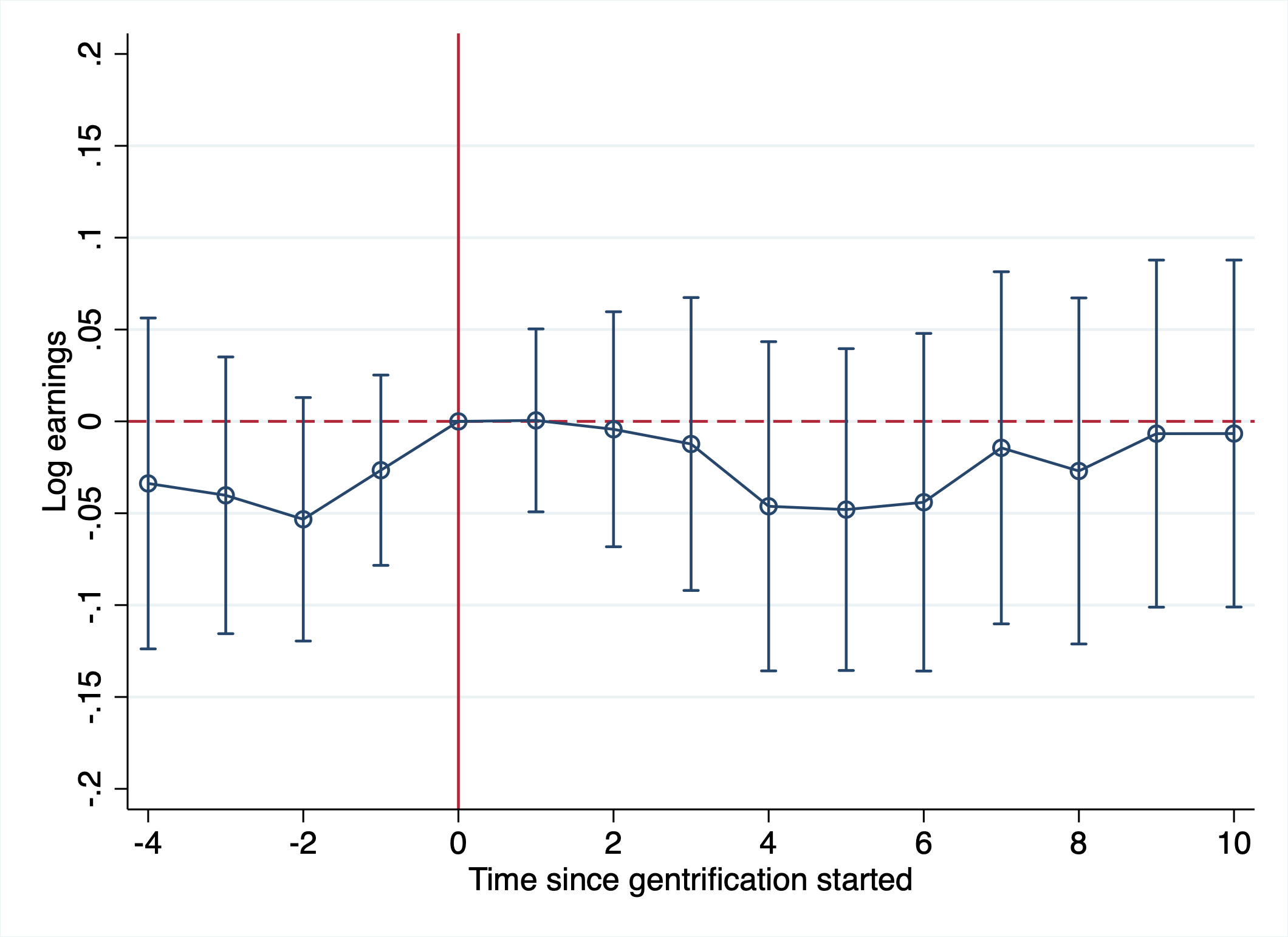}} \\
\subfloat[High-income]{\includegraphics[width = 5in]{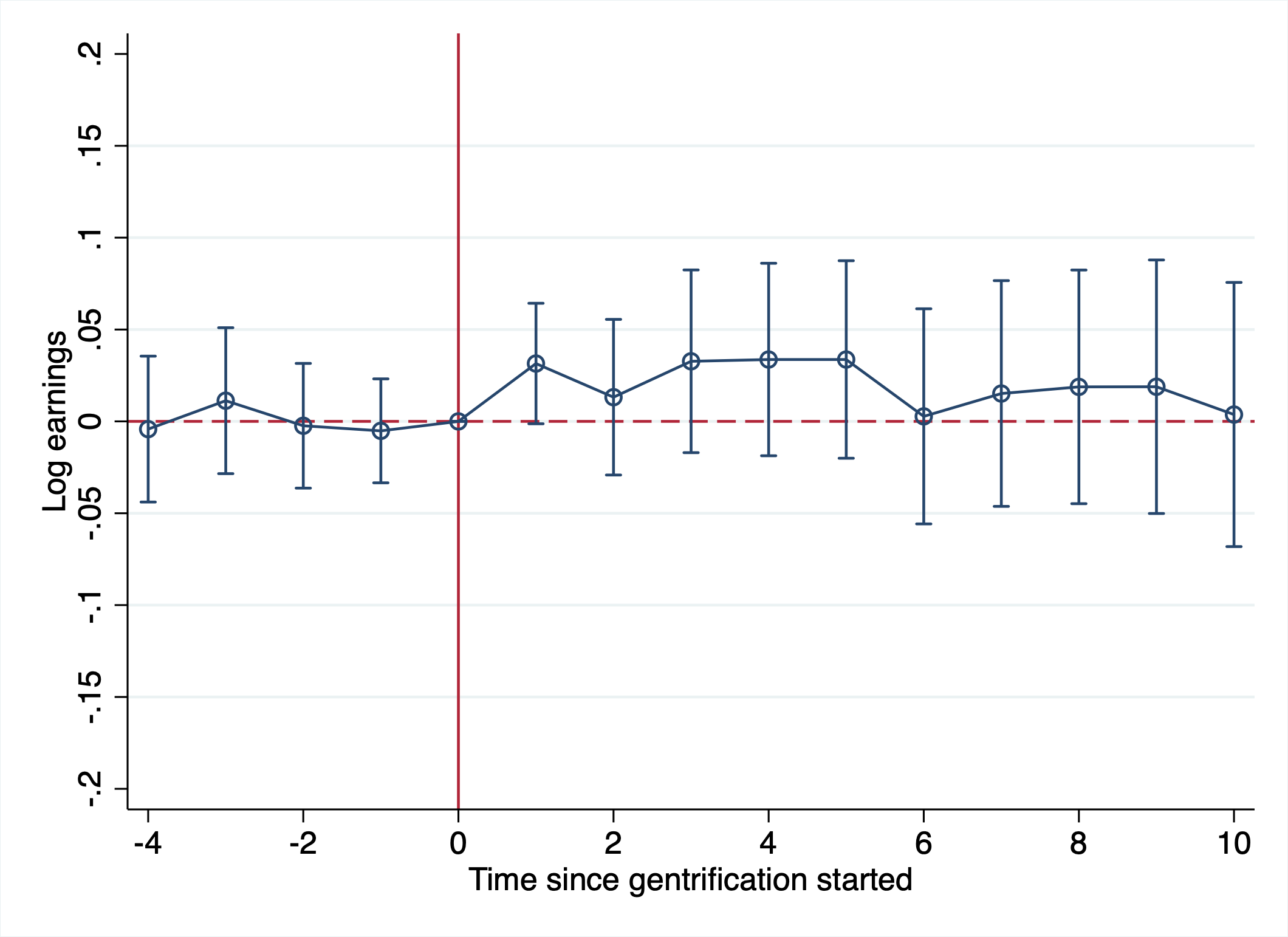}}\\
\label{fig:ES_empl_earnings}
\begin{minipage}{0.9\textwidth} 
{\footnotesize \vspace{0.1cm} Source: Author's calculations using the Census and LAD.\\
Note: Dependant variables: employment earnings. The sample is incumbent households living in gentrifiable neighborhoods (initially low-income and central city) in one of the baseline years. The control group is the matched sample discussed in section \ref{sec:Matched_Samples}.  The regressions also include individual-level control variables (age, age squared, gender, family composition,  number of children and immigrant indicator), baseline Census Tract controls (college-educated share, median income, share for low income, average rent, employment rate, visible minority share, the share of immigrant, distance from CBD), and pre-period variation controls (changes of college-educated share, median income, average rent, employment rate). Low-income number of observations: 4,698,305. High-income number of observations: 2,559,580. \par}
\end{minipage}
\end{figure}

\begin{figure}[ht!]
\caption{Effect of gentrification on other earnings}
\centering
\subfloat[Low-income]{\includegraphics[width = 5in]{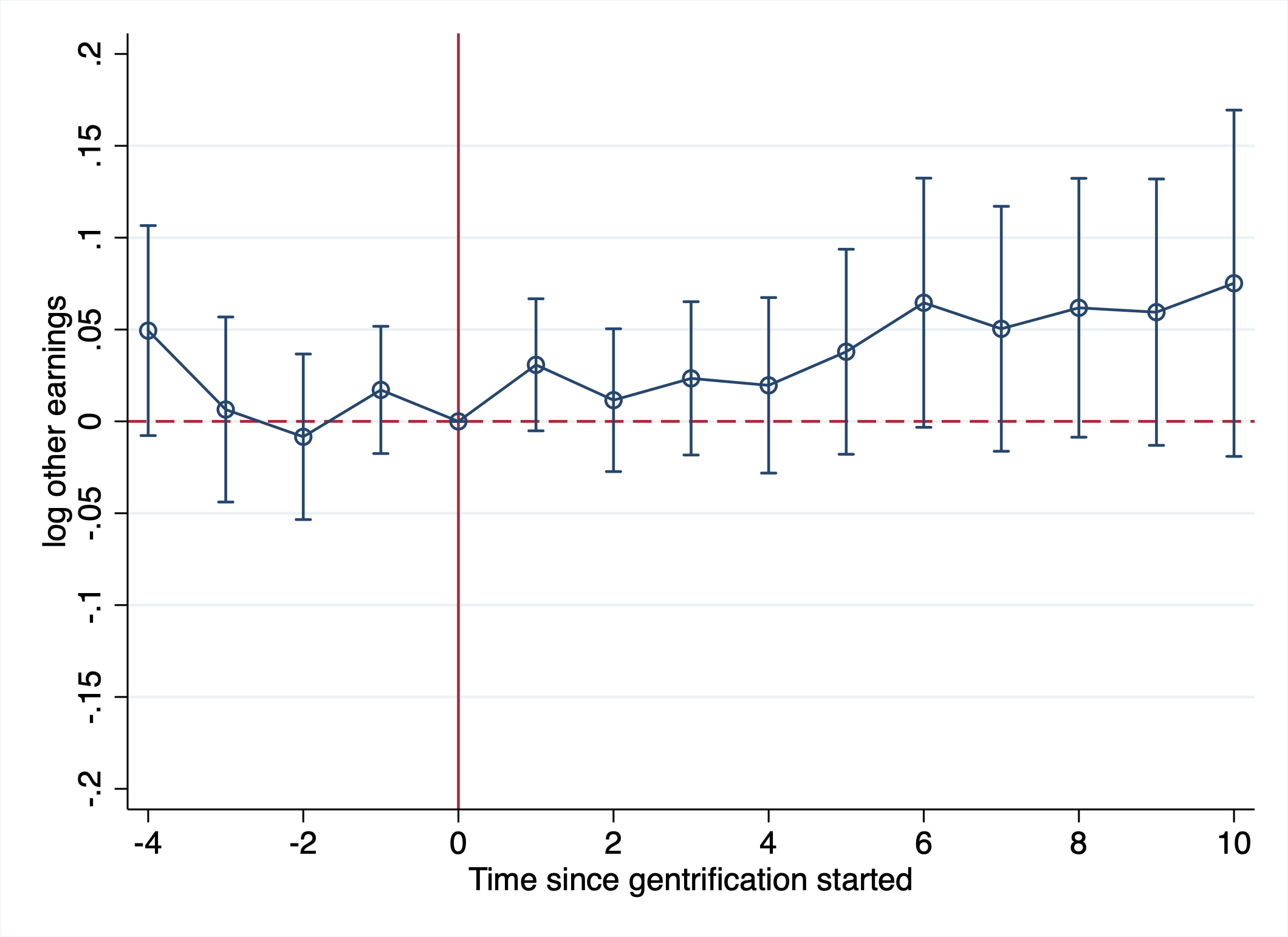}} \\
\subfloat[High-income]{\includegraphics[width = 5in]{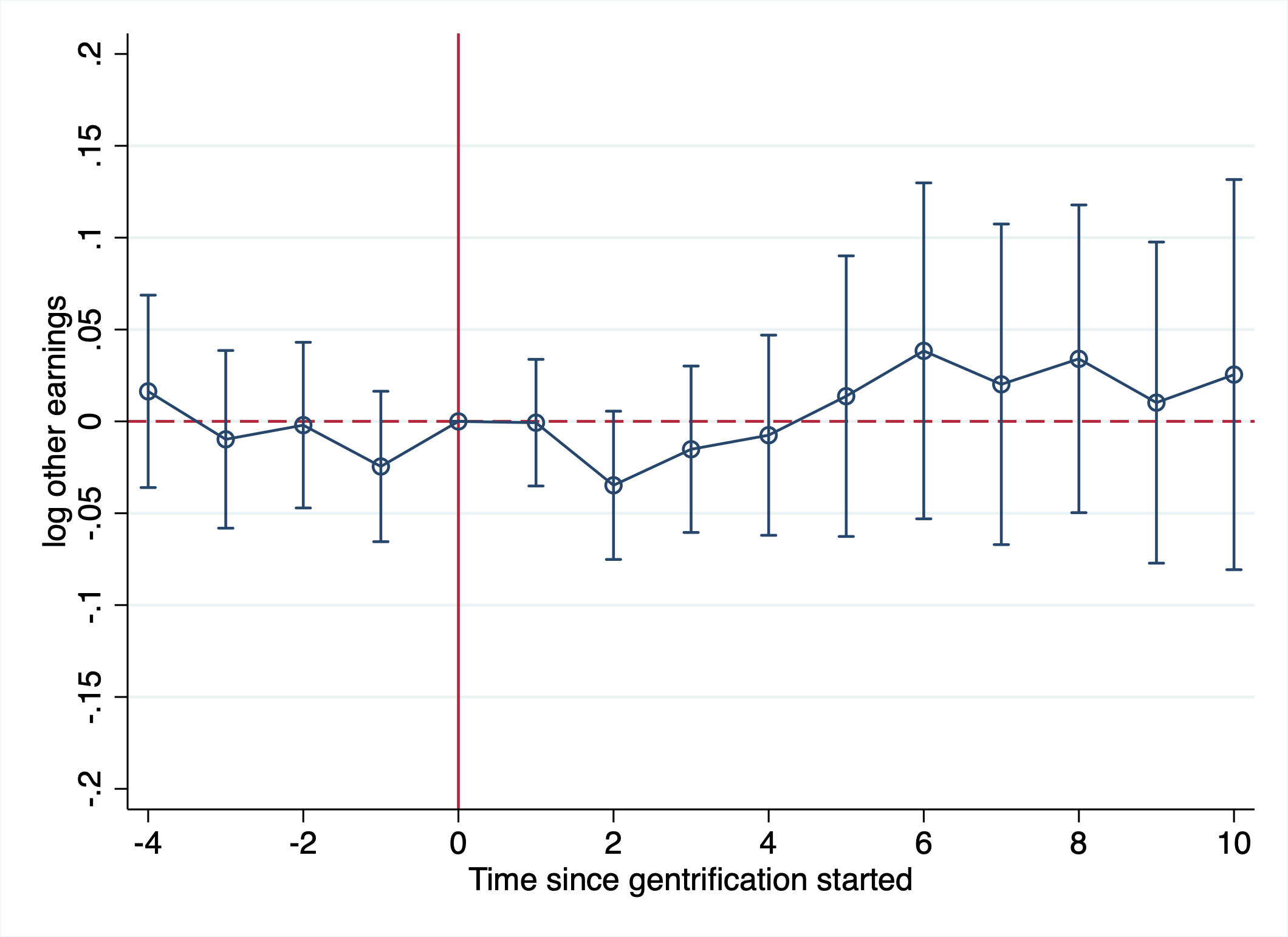}}\\
\label{fig:ES_other_earnings}
\begin{minipage}{0.9\textwidth} 
{\footnotesize \vspace{0.1cm} Source: Author's calculations using the Census and LAD.\\
Note: Dependant variables: total gross income minus employment earnings. The sample is incumbent households living in gentrifiable neighborhoods (initially low-income and central city) in one of the baseline years. The control group is the matched sample discussed in section \ref{sec:Matched_Samples}. The regressions also include individual-level control variables (age, age squared, gender, family composition,  number of children and immigrant indicator), baseline Census Tract controls (college-educated share, median income, share for low income, average rent, employment rate, visible minority share, the share of immigrant, distance from CBD), and pre-period variation controls (changes of college-educated share, median income, average rent, employment rate). Low-income number of observations: 4,698,305. High-income number of observations: 2,559,580. \par}
\end{minipage}
\end{figure}

I decompose total income into two components: employment earnings and other income. Figure \ref{fig:ES_empl_earnings} shows that there is no effect on employment income. Hence, I rule out that gentrification significantly impacts the labor market outcomes of incumbent households. Beyond employment, gentrification could affect other sources of earnings, e.g., rental income or government transfer. Figure \ref{fig:ES_other_earnings} shows a slight positive effect for low-income households' other income sources. This increase is not significant at the 5\% level but could indicate some compensation for negative shocks. 

Overall, income is not affected by gentrification. A simple explanation is that the labor market access of workers goes far beyond their census tract. Micro-geographic transformations do not significantly change the labor market opportunities for workers.


\section{Discussion on tenant protection laws} \label{chap:discussion}

In section \ref{chap:results_displacement}, I show that, in the Canadian context, gentrification had a retention effect rather than a displacement effect. I discuss this surprising result here. I should point out that the no-displacement result is in line with previous literature \parencite{vigdor2002does, ellen2011low}, especially with papers suggesting that gentrification slows down the turnover of incumbent \parencite{freeman2004gentrification}.   

The absence of evidence supporting displacement poses a perplexing situation for many observers. \citet{dragan2020does} suggest two explanations. Firstly, displacement is simply more salient in gentrifying areas. People may be less likely to notice the evictions and forced moves in other neighborhoods because new entrants in those areas more closely resemble the local population. Secondly, researchers have failed to capture the phenomenon because of inadequate measures or data. 

I propose an alternative explanation to this paradox: behind gentrification's null (or negative) effects on incumbents' mobility lies a large heterogeneity in effects across cities. This heterogeneity might be related to how strong tenant-protection laws are.   

The null results found in the literature might result from balancing opposite forces. Households benefit from improving their neighborhood's urban amenities and increasing their willingness to stay. On the other hand, increasing rents and potential pressure from landlords decrease their stay capacity. Stronger rent control and tenant protection laws would weaken the later driver. The three Canadian cities in my analysis have stronger rent control laws than the average American city previously studied in the literature, weakening the vector that pushes residents away. 

In Canada, rent regulation is provincial jurisdiction; hence, there are 13 variants in place. The literature on state rent regulation and its application in the different provinces is limited. However, it is commonly accepted that Québec is likely positioned at one end of the spectrum, boasting a firmly established rent control system that significantly influences the market, accompanied by robust tenant protection and rights. In contrast, Alberta represents the opposite extreme, providing safeguards against eviction and specifying landlord obligations yet lacking any restrictions on annual rent hikes \parencite{mykkanen2018right}. British Columbia and Ontario fall in the middle of the spectrum, with Ontario tending towards a weaker system and BC leaning towards a stricter one. Furthermore, Québec has some form of vacancy control, a feature absent in the other two provinces. Vacancy control decreases the financial incentive for landlords to evict tenants frequently. 

The significant impact of rent control on tenant turnover has been documented in several contexts \parencite{diamond2019effects, autor2014housing, munch2002rent}. However, there's little evidence on how those laws interact with the gentrification process. If tenant protection law plays a role in the turnover rate of renters in gentrifying neighborhoods, we'd expect the effect for high-income households (likely to be homeowners) to be similar across cities, but the effect for low-income households (likely to be renters) to be larger in Montréal (Québec) than in the Vancouver (British Columbia) and Toronto (Ontario).

\begin{landscape}
\begin{figure}[ht!]
\caption{Effect of gentrification by cities}
\label{fig:ES_by_mtl_vs_rest}
\begin{sloppypar}
\hspace{7cm} \textbf{Montréal} \hspace{5cm}        
\textbf{Toronto \& Vancouver}   \\
\end{sloppypar}
\centering
\subfloat[Low-income]{\includegraphics[width = 3.1in]{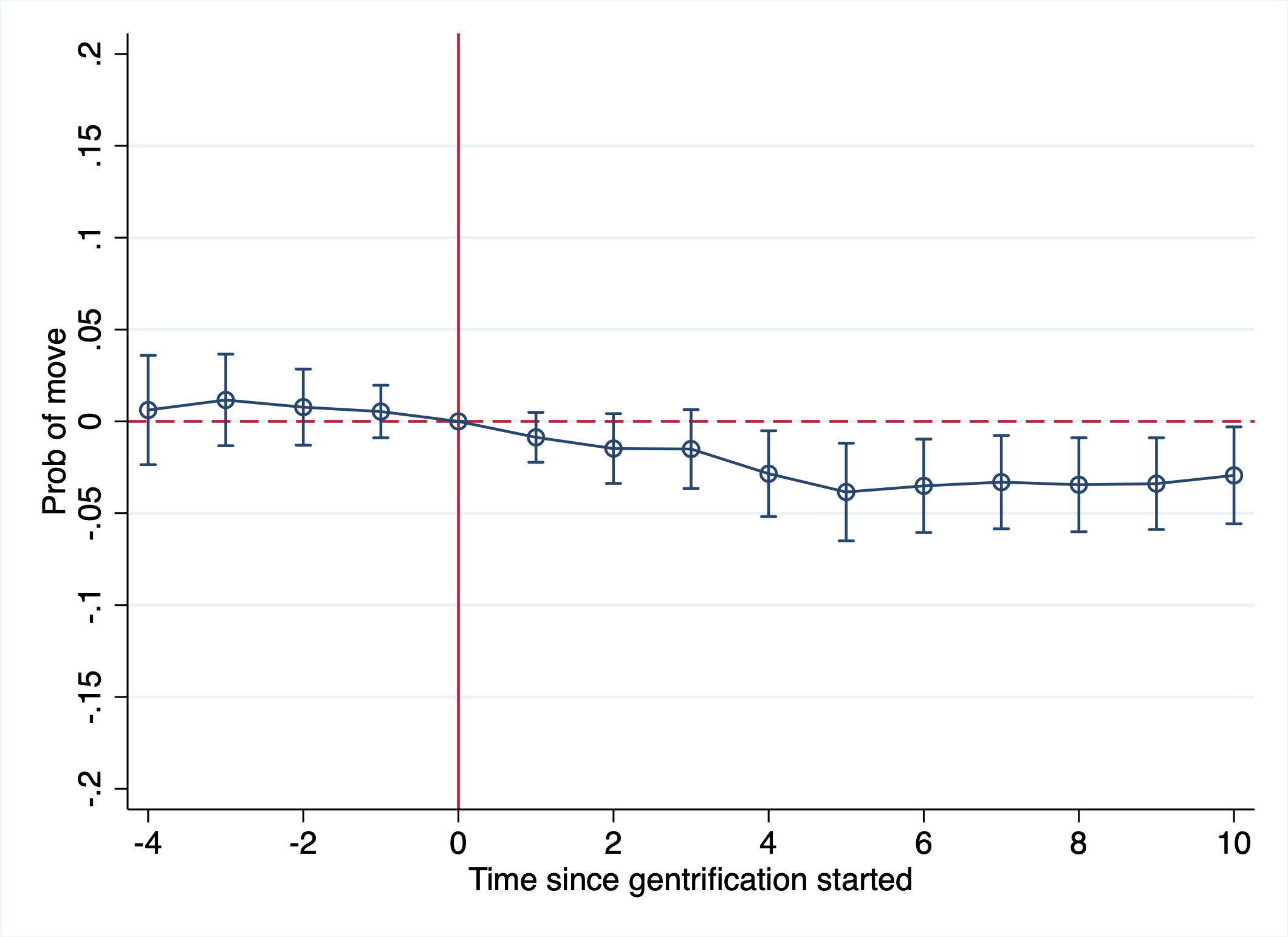}} \subfloat[Low-income]{\includegraphics[width = 3.1in]{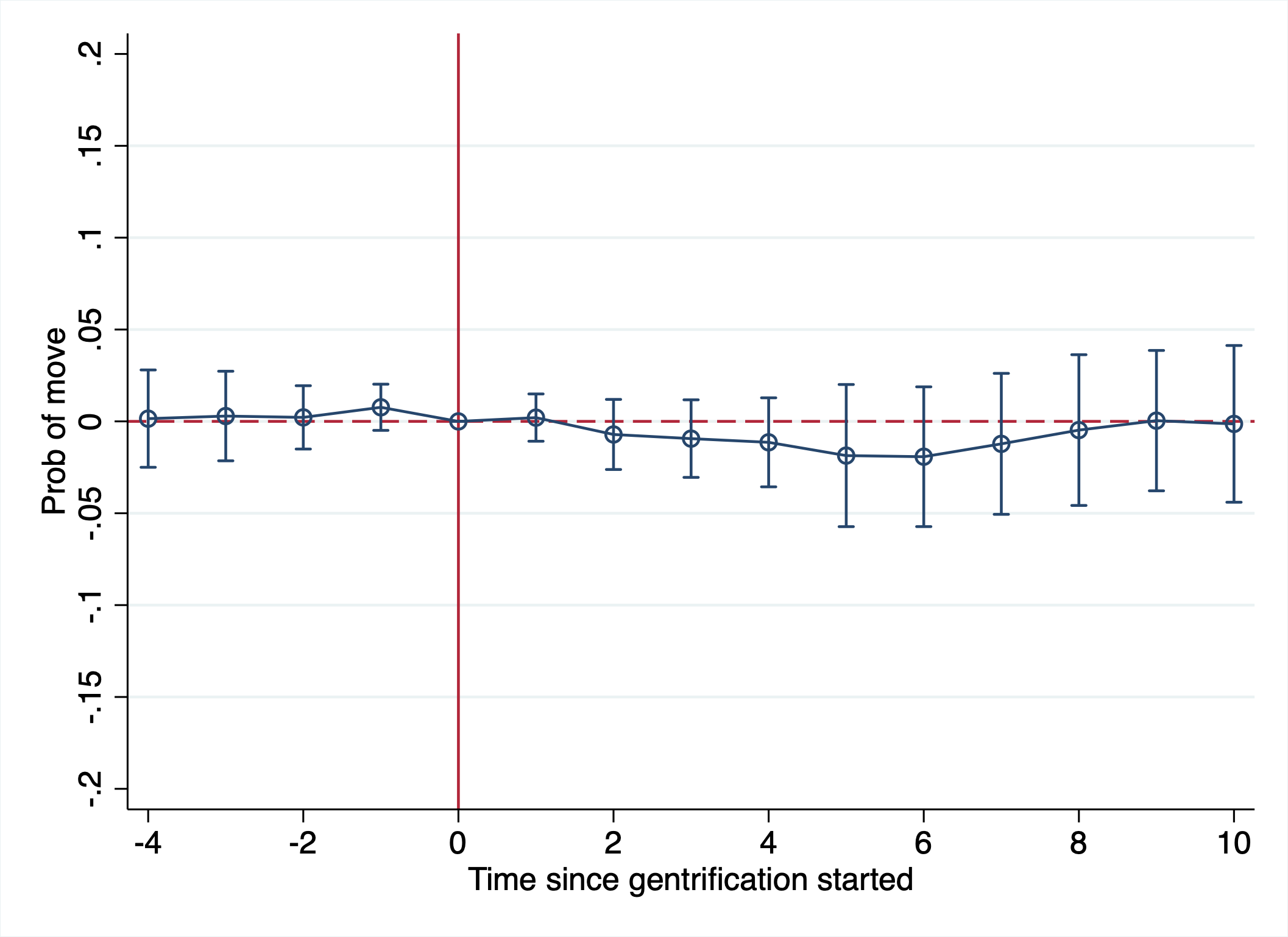}}  \\
\subfloat[High-income]{\includegraphics[width = 3.1in]{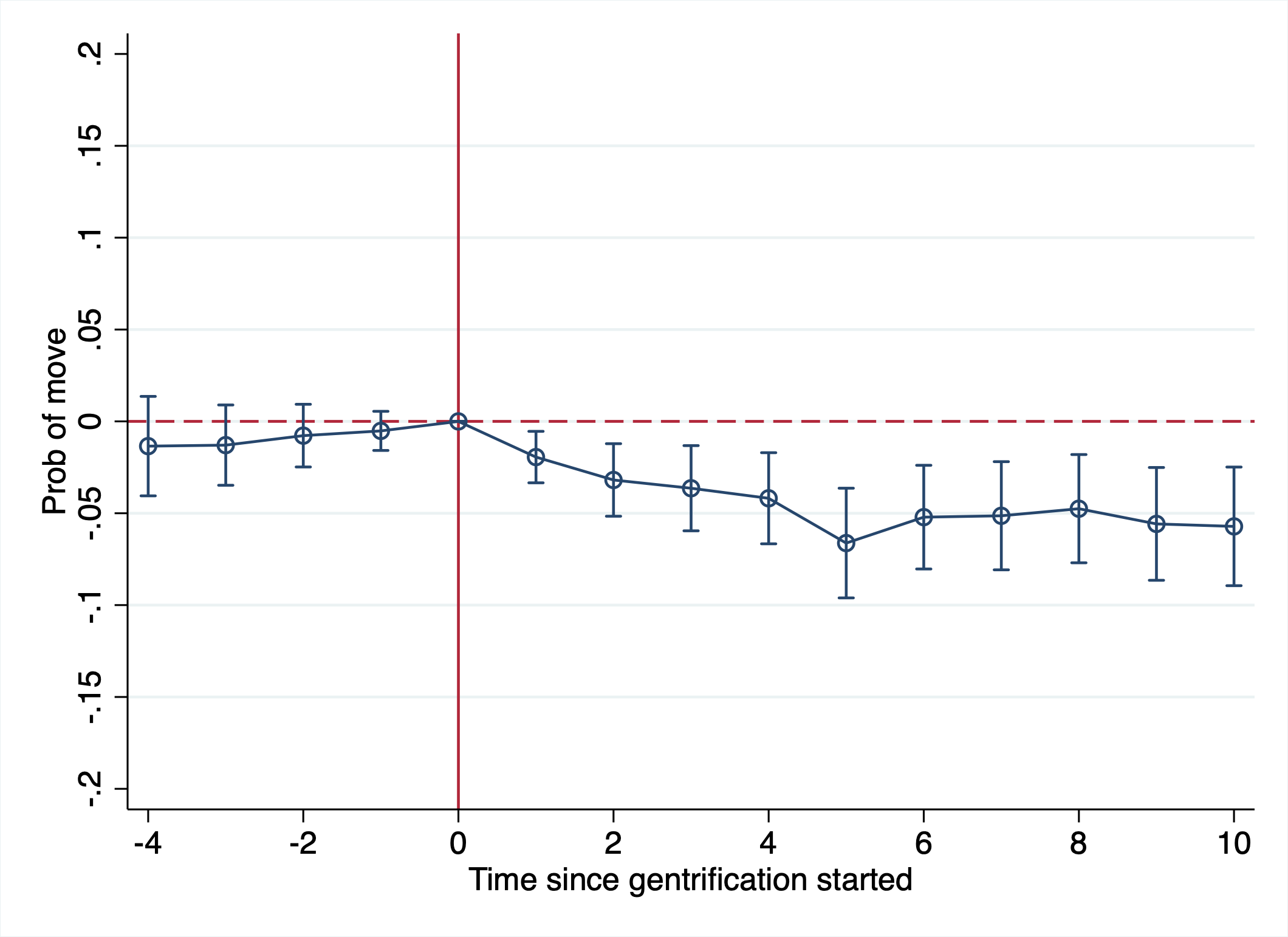}} \subfloat[High-income]{\includegraphics[width = 3.1in]{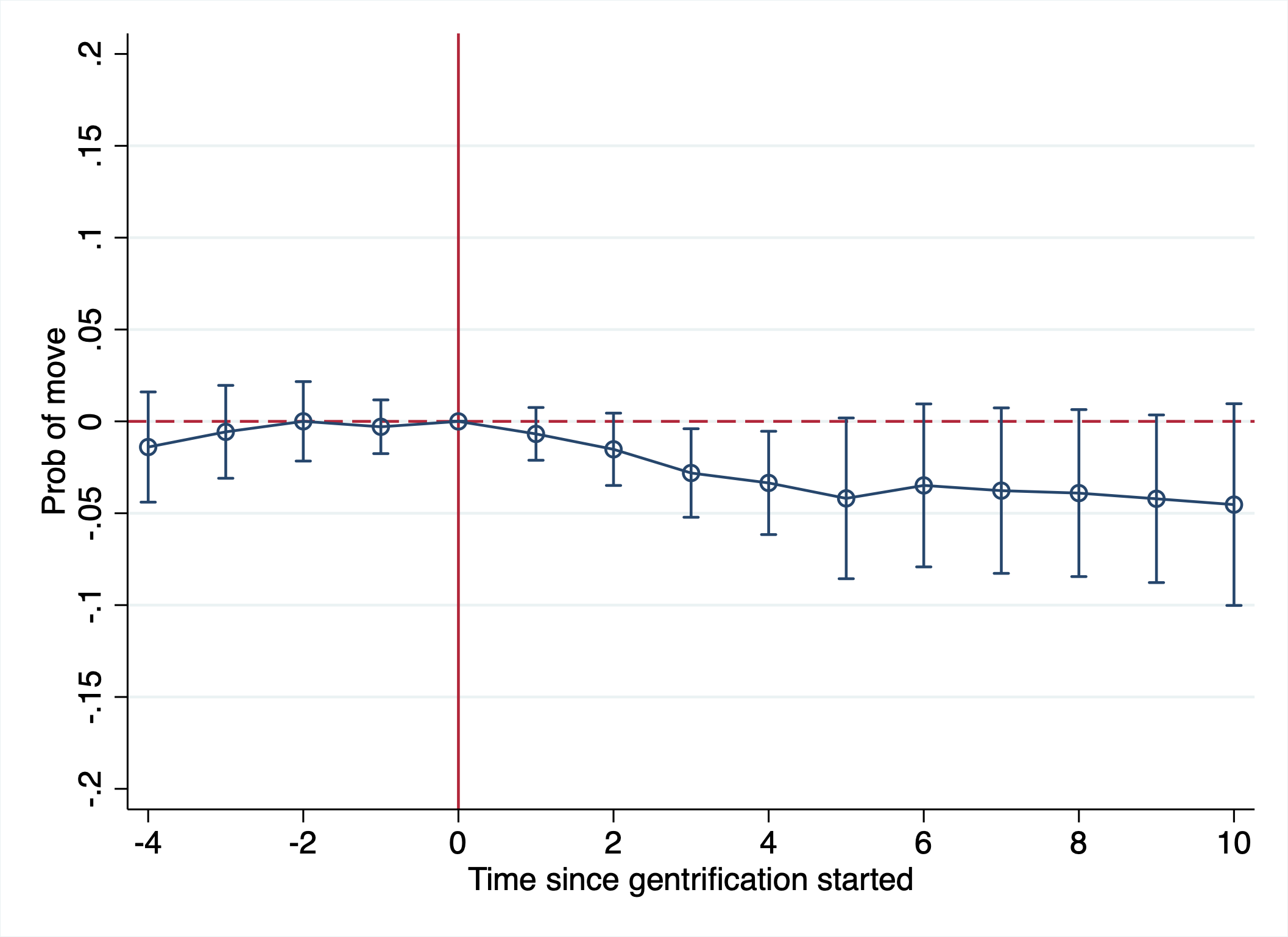}} \\
\begin{minipage}{1.2\textwidth} 
{\footnotesize \vspace{0.1cm} Source: Author's calculations using the Census and LAD.\\
Note: Dependant variables: dummy equal to one if the household lives in a different tract than at baseline. The sample is incumbent households living in gentrifiable neighborhoods (initially low-income and central city) in one of the baseline years. The control group is the matched sample discussed in section \ref{sec:Matched_Samples}. The regressions also include individual-level control variables (age, age squared, gender, family composition,  number of children, and immigrant indicator), baseline Census Tract controls (college-educated share, median income, share for low income, average rent, employment rate, visible minority share, the share of immigrant, distance from CBD), and pre-period variation controls (changes of college-educated share, median income, average rent, employment rate). \par}
\end{minipage}
\end{figure}
\end{landscape}

Figure \ref{fig:ES_by_mtl_vs_rest} shows the event-study by city\footnote{Figure \ref{fig:ES_by_city} shows the event-study for Vancouver and Toronto separately.}. In all cities, high-income households are more likely to stay longer when their neighborhood gentrifies.\footnote{The estimates for Toronto and Vancouver are imprecise, but the coefficients are systematically negative and similar to those for Montréal.} However, the effect of gentrification on low-income incumbents' mobility is negative (and significant) only in Montréal. While this is only a suggestive answer to the no-displacement paradox, it motivates further research on how rent control and tenant protection laws mitigate the potential adverse effects of gentrification.


\section{Conclusion} \label{sec:conclusion}

In this paper, I exploit administrative data on Canadian tax filers to study how neighborhood changes affect incumbent workers. In the last couple of decades, what has attracted policymakers' attention and guided the popular narrative about gentrification is the idea that it pushes poor people away from their neighborhoods. My results suggest that this isn't the primary vector that negatively affects low-income incumbent workers: incumbent residents are actually more likely to stay longer in their home neighborhoods than similar households in non-gentrifying neighborhoods. This effect is larger for high-income, who are more likely to be homeowners. This result is not incompatible with previous studies studying displacement effects. While some also find negative effects on turnover \parencite{freeman2004gentrification}, most find no effect \parencite{ellen2011low, ding2016gentrification, dragan2020does}. I argue that this contrast from the results presented in this paper comes from differences in rent control and tenant protection laws.

When incumbent households endogenously decide to leave, they do not relocate into worse neighborhoods compared to their counterparts from non-gentrifying neighborhoods. Finally, I find no effect on the total income for incumbent workers. Decomposing income into employment income and non-employment income does not show effects either. Gentrification does not change incumbents' labor market outcomes, nor does it change their non-employment income.  

I must point out two main drawbacks of my analysis. First, while I use income level to explore heterogeneous effects, it would be interesting to explore heterogeneity regarding education level and home ownership as in \citet{brummet2021effects}. This would also improve the interpretation of the role of rent control. Unfortunately, the data used here does not allow me to make such an analysis. Second, the only labor force outcome I observe is earnings. Ideally, one would observe job transition, hours worked, and commuting time to draw a complete picture of how gentrification changes the employment outcomes of incumbent workers. 

Examining the interaction between tenant protection laws and gentrification is a promising research agenda. A natural question is \textit{Can tenant protection laws limit the potential adverse effect of neighborhood changes on low socioeconomic status households?}

\newpage
\doublespacing
\printbibliography

\pagebreak
\newgeometry{top=2cm,left=2cm,right=2cm}

\section{Figures}

\vspace*{-1cm}
\begin{figure}[ht!]
\centering
\caption{Gentrification in Montréal\label{fig:gen_mtl_1}}
\includegraphics[scale=0.6]{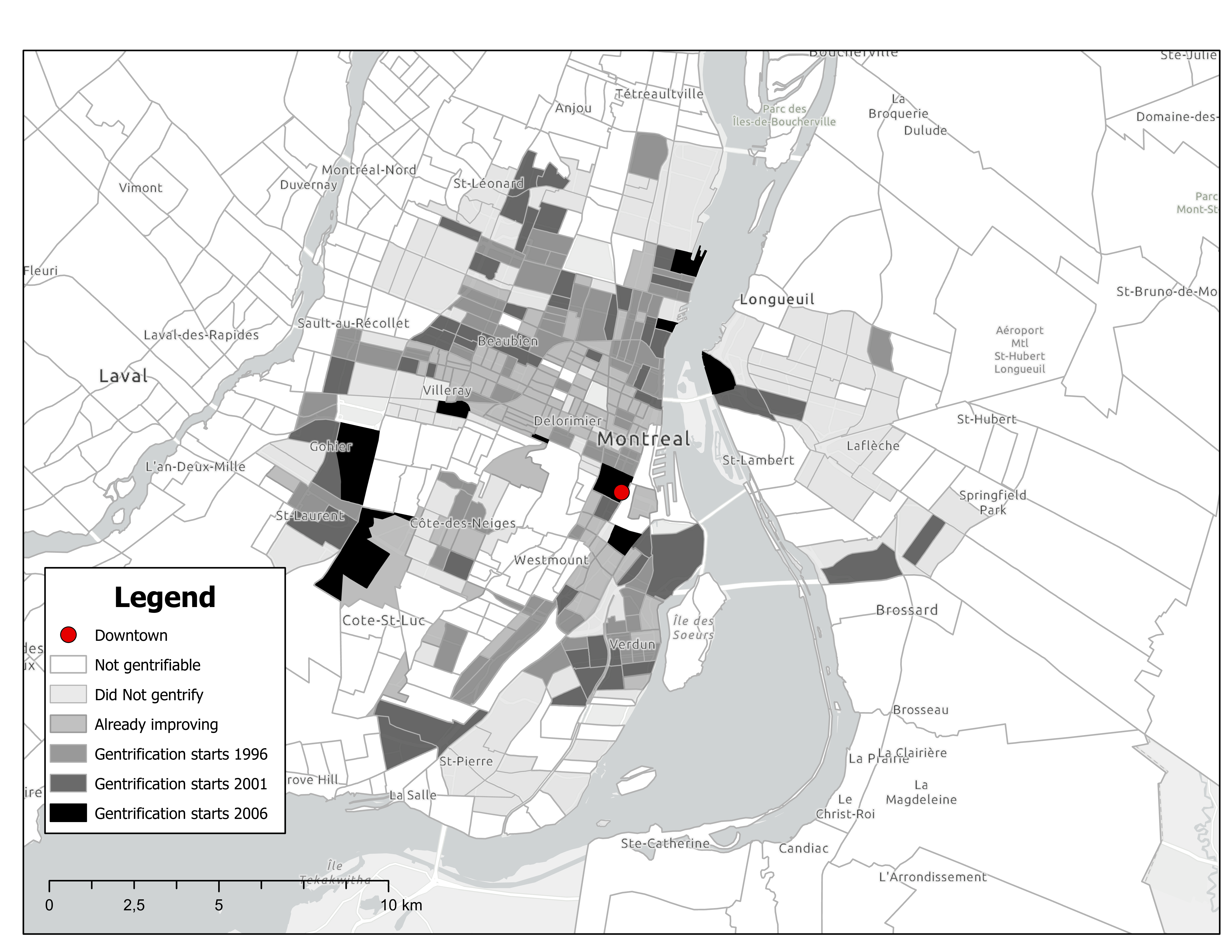}\\
\begin{minipage}{0.9\textwidth} 
{\footnotesize \vspace{0.1cm} Source: Author's calculations using the Census.\\
Note: White tracts are non-gentrifiable (outer-city) tracts, gray tracts are non-gentrifiable (high-income) tracts. The various shades of grey indicate the time at which a tract starts gentrifying. \par}
\end{minipage}
\end{figure}

\vspace*{-1cm}
\begin{figure}[ht!]
\centering
\caption{Gentrification in Toronto\label{fig:gen_tor_1}}
\includegraphics[scale=0.6]{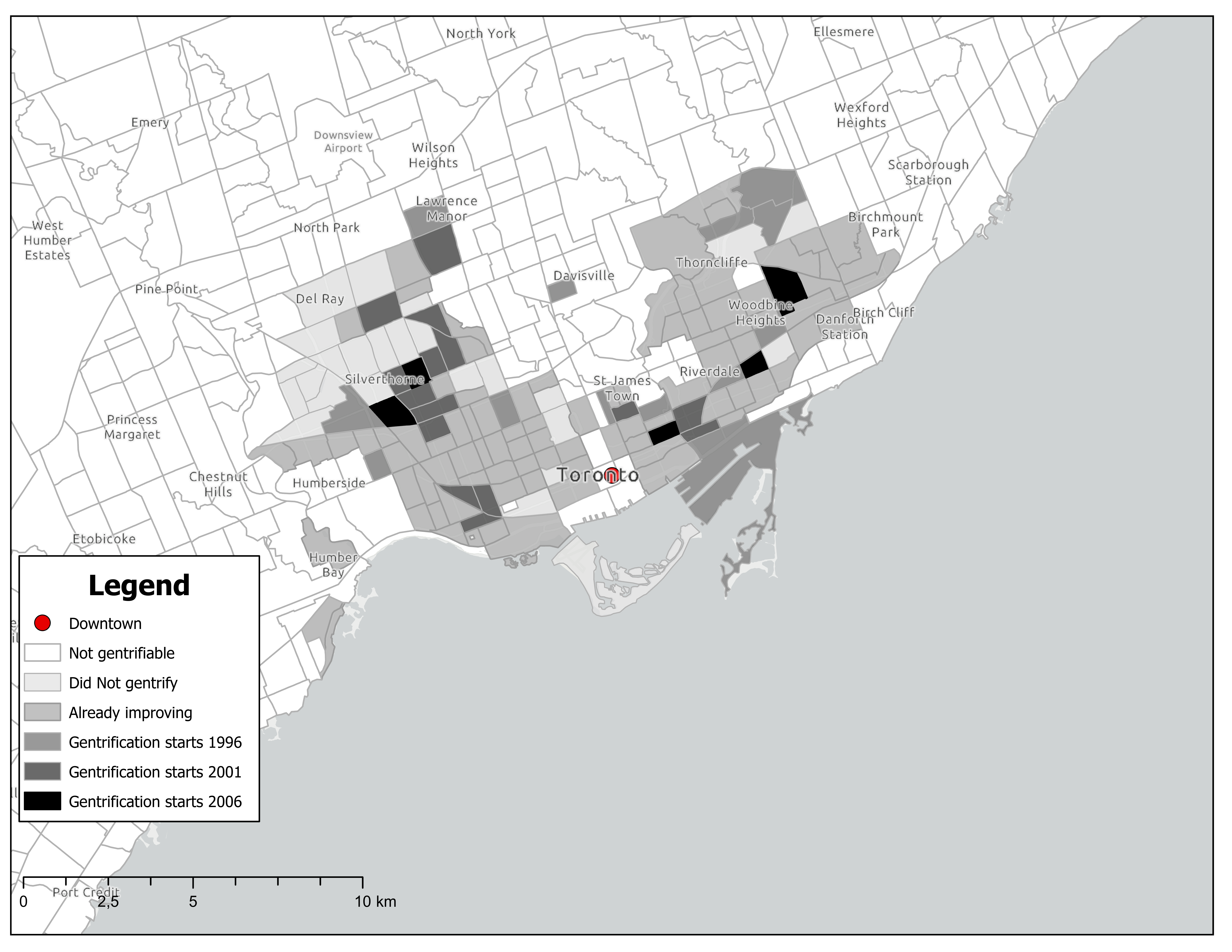}\\
\begin{minipage}{0.9\textwidth} 
{\footnotesize \vspace{0.1cm} Source: Author's calculations using the Census.\\
Note: White tracts are non-gentrifiable (outer-city) tracts, gray tracts are non-gentrifiable (high-income) tracts. The various shades of grey indicate the time at which a tract starts gentrifying. \par}
\end{minipage}
\end{figure}

\vspace*{-1cm}
\begin{figure}[ht!]
\centering
\caption{Gentrification in Vancouver\label{fig:gen_van_1}}
\includegraphics[scale=0.6]{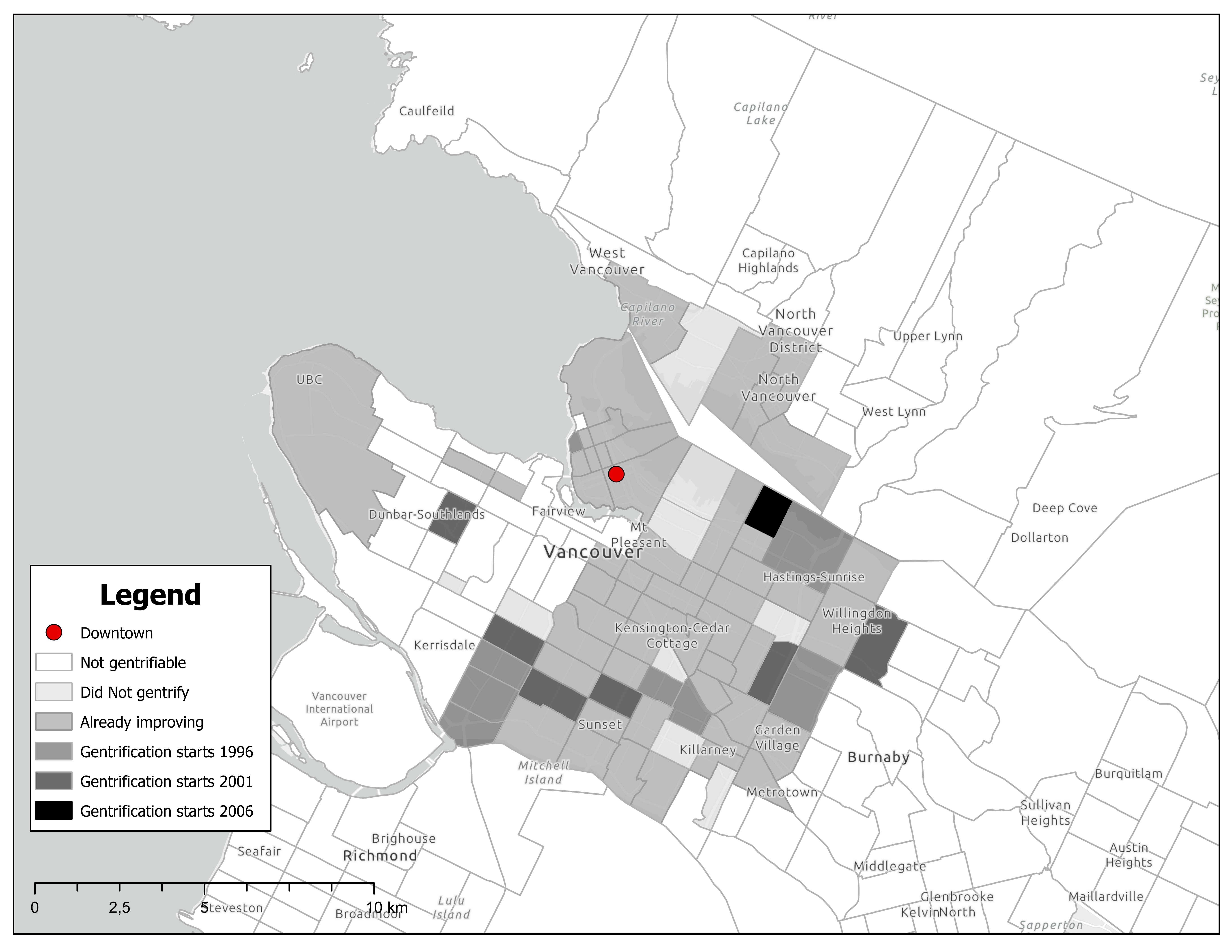}\\
\begin{minipage}{0.9\textwidth} 
{\footnotesize \vspace{0.1cm} Source: Author's calculations using the Census.\\
Note: White tracts are non-gentrifiable (outer-city) tracts, gray tracts are non-gentrifiable (high-income) tracts. The various shades of grey indicate the time at which a tract starts gentrifying. \par}
\end{minipage}
\end{figure}


\newgeometry{top=1cm,left=2cm,right=2cm}

\begin{figure}[ht!]
\caption{Effect of gentrification on location choice: Share of immigrant}
\centering
\subfloat[Low-income]{\includegraphics[width = 5in]{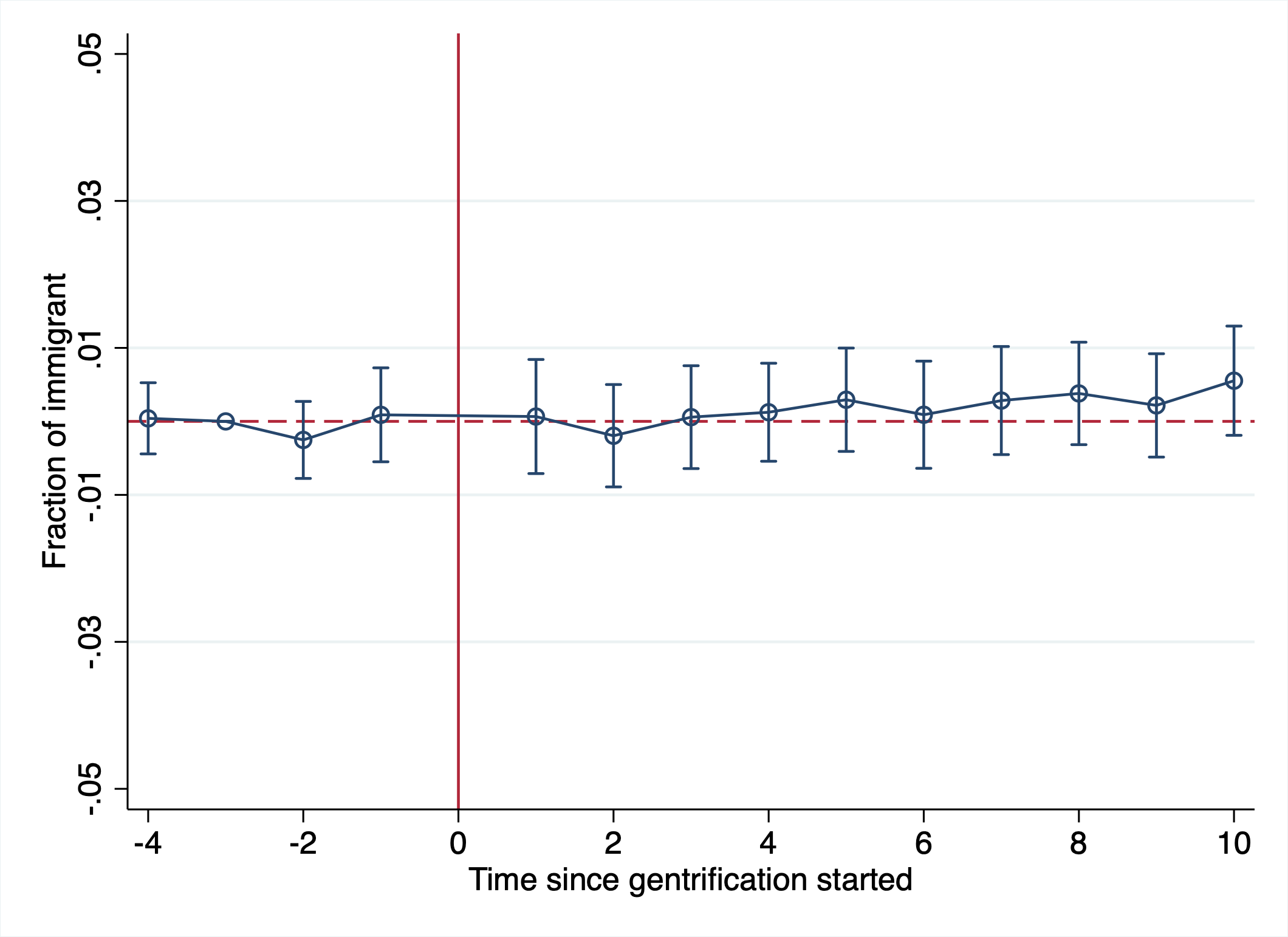}} \\
\subfloat[High-income]{\includegraphics[width = 5in]{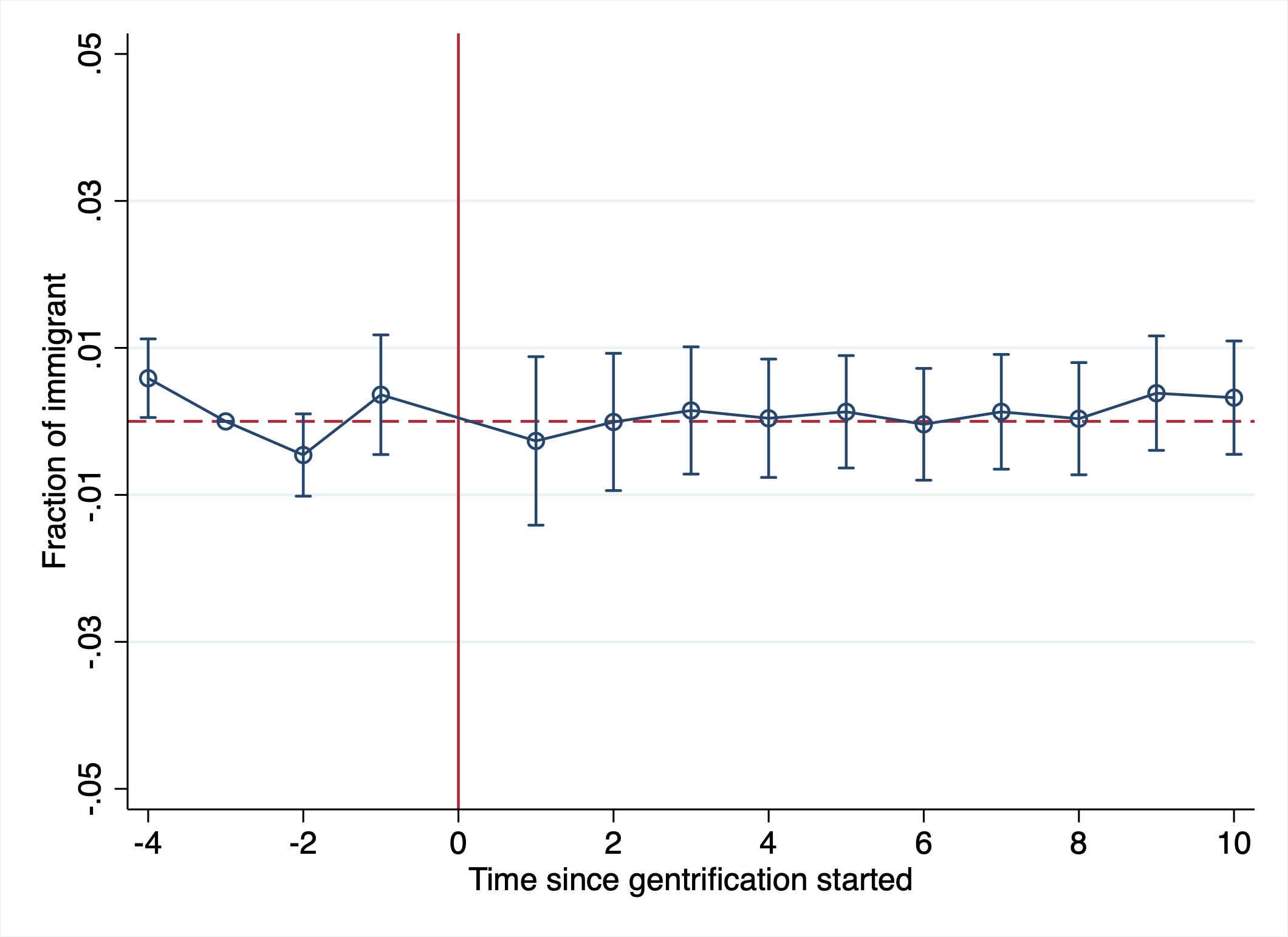}}\\
\label{fig:exp_immigrant}
\begin{minipage}{0.9\textwidth} 
{\footnotesize \vspace{0.1cm} Source: Author's calculations using the Census and LAD.\\
Note: Dependant variables: share of immigrants in the CT. The sample is incumbent households living in gentrifiable neighborhoods (initially low-income and central city) in one of the baseline years, but do not currently live in that census tract. The control group is the matched sample discussed in section \ref{sec:Matched_Samples}. The regressions also include family-level socioeconomic control variables (age, family composition,  number of children and immigrant indicator), baseline Census Tract controls (college-educated share, median income, share for low income, average rent, employment rate, visible minority share, the share of immigrant, distance from CBD), and pre-period variation controls (changes of college-educated share, median income, average rent, employment rate). Low-income number of observations: 356,900. High-income number of observations: 202,530. \par}
\end{minipage}
\end{figure}

\begin{figure}[ht!]
\caption{Effect of gentrification on location choice: Share of visible minority}
\centering
\subfloat[Low-income]{\includegraphics[width = 5in]{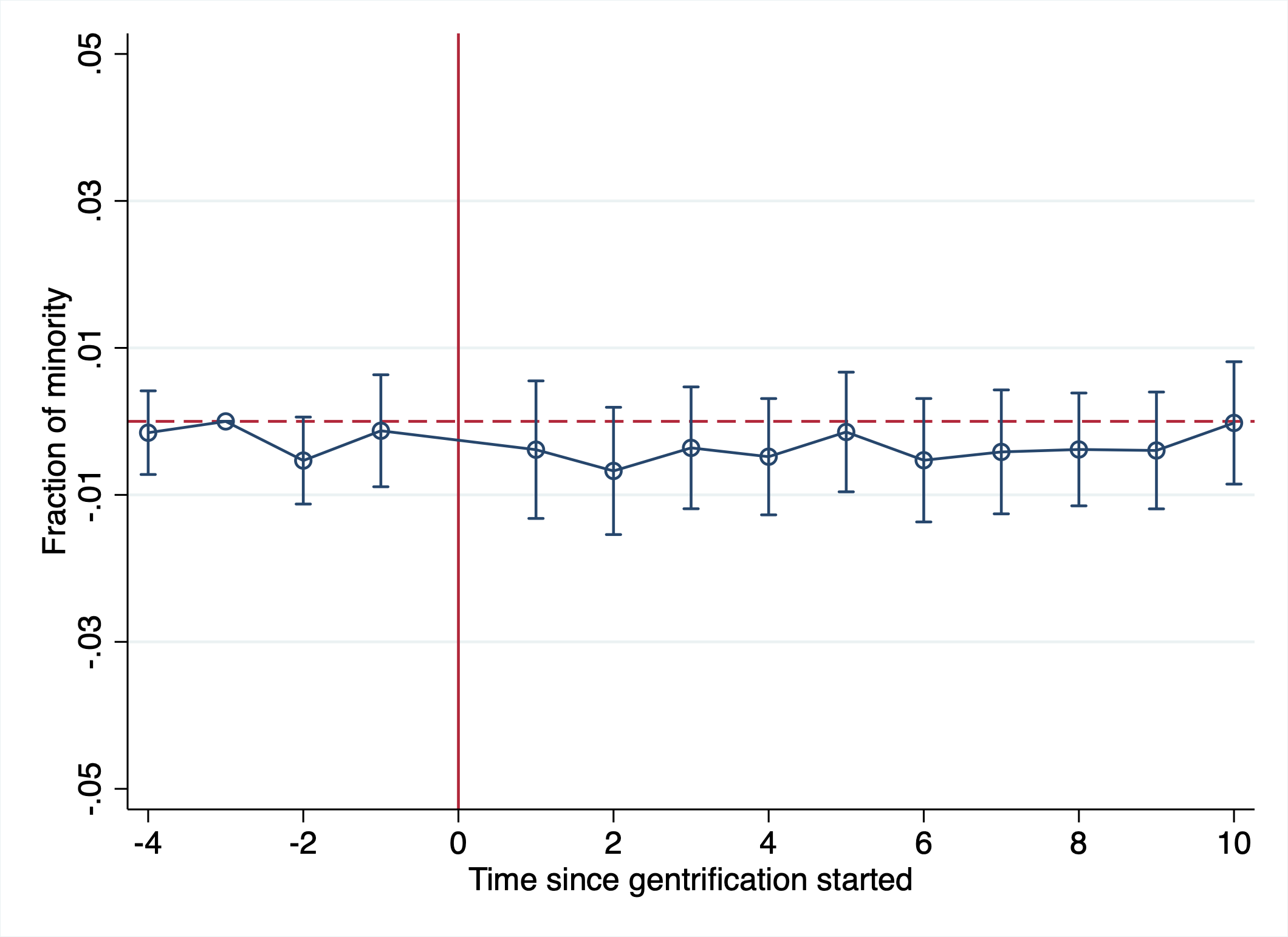}} \\
\subfloat[High-income]{\includegraphics[width = 5in]{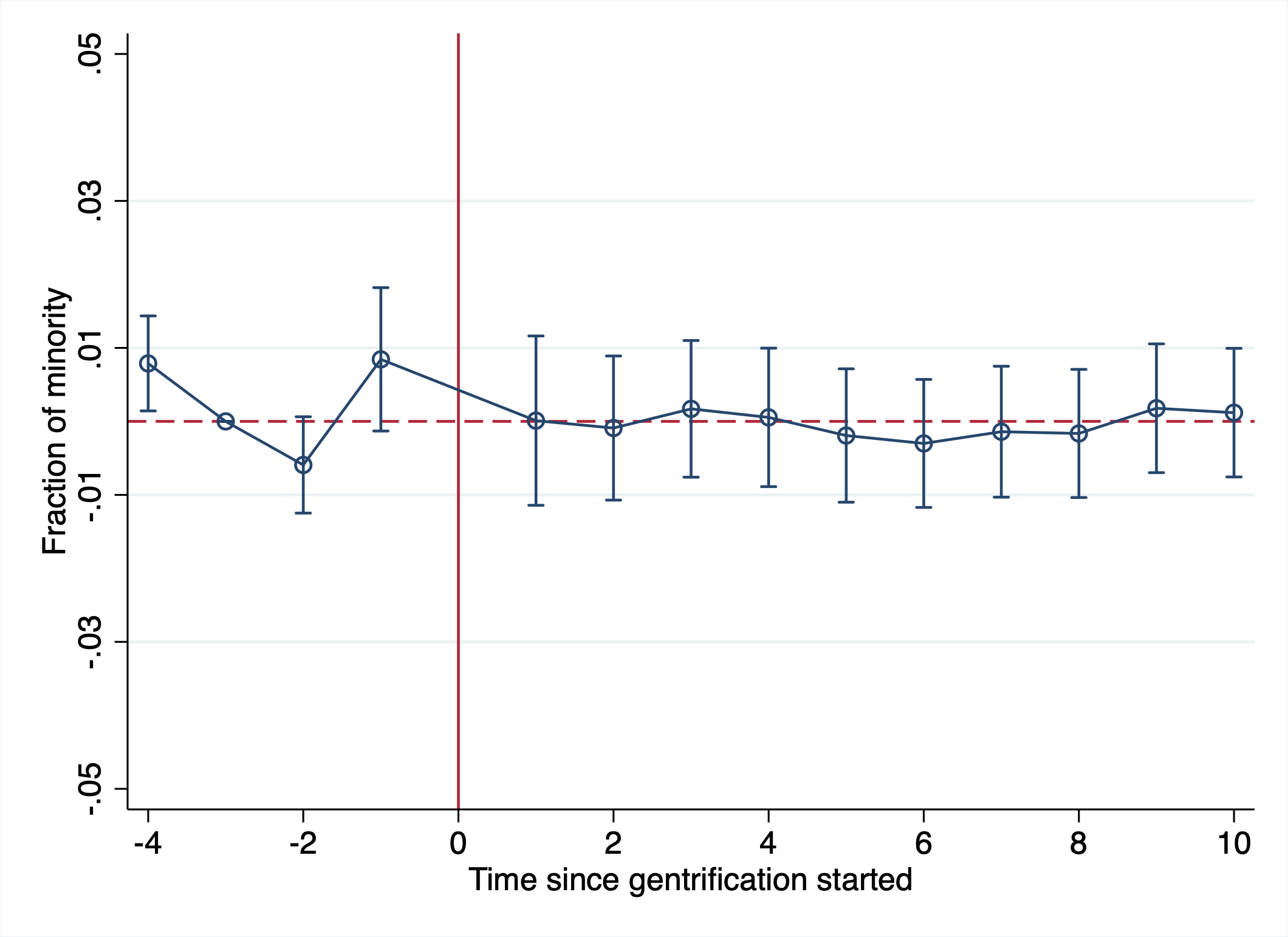}}\\
\label{fig:exp_minority}
\begin{minipage}{0.9\textwidth} 
{\footnotesize \vspace{0.1cm} Source: Author's calculations using the Census and LAD.\\
Note: Dependant variables: share of visible minorities in the CT. The sample is incumbent households living in gentrifiable neighborhoods (initially low-income and central city) in one of the baseline years, but do not currently live in that census tract. The control group is the matched sample discussed in section \ref{sec:Matched_Samples}. The regressions also include family-level socioeconomic control variables (age, family composition,  number of children and immigrant indicator), baseline Census Tract controls (college-educated share, median income, share for low income, average rent, employment rate, visible minority share, the share of immigrant, distance from CBD), and pre-period variation controls (changes of college-educated share, median income, average rent, employment rate). Low-income number of observations: 356,900. High-income number of observations: 202,530. \par}
\end{minipage}
\end{figure}

\pagebreak
 
\newgeometry{top=2cm,left=2cm,right=2cm}

\section{Tables}
 \quad \\
 
\begin{table}[h]
    \caption{Number of Census tract by gentrification status \label{tab:CTdist}}
\footnotesize
\onehalfspacing
\begin{center}
\begin{tabular}{lccc} \hline \hline
                              & Montréal & Toronto & Vancouver \\ \hline 
Not gentrifiable             & 464      & 662     & 223       \\
Improving before 1996         & 99       & 89      & 55        \\[0.2cm]
Gentrifiable        &        &       &        \\
\hspace{0.15cm} Did not gentrify        &    67    &     21  &    5    \\
\hspace{0.15cm} Gentrification starts in 1996 & 68       & 13      & 8         \\
\hspace{0.15cm} Gentrification starts in 2001 & 50       & 15      & 6         \\
\hspace{0.15cm} Gentrification starts in 2006 & 9        & 6       & 1         \\[0.2cm]
Total                         & 757      & 809     & 298      \\ \hline \hline
\end{tabular}
\end{center}
\footnotesize Source: Author's calculations using the Census. \\
Note: Number of tracts by gentrification start date and CMA. The start date is defined as the first time a CT is in the top half of the gentrification measure distribution defined in Equation \ref{equation:gent_measure}. Only gentrifiable neighbourhood can have a gentrification start date. Gentrifiable neighbourhoods are those less than 10 km away from the CBD and below the median in the median income distribution of their CMA. Not gentrifiable include outer-city neighbourhoods (those more than 10 km away from the CBD) and rich inner-city neighbourhoods (less than 10 km away from the CBD but above the median in the median income distribution of their CMA). 

\end{table}

\pagebreak

\begin{landscape}
    \begin{table}
        \caption{Summary of Neighborhood Characteristics by Gentrification status, 1996
\label{tab:descstat_1996}}
\footnotesize
\onehalfspacing
\begin{center}
\begin{tabular}{lccccccccccccc}
\hline  \hline  
&  \multicolumn{4}{c}{Not gentrifiable} & & \multicolumn{8}{c}{Gentrification quantile}  \\   \cline{2-5} \cline{7-14} 
 & \multicolumn{2}{c}{Outer-city} & \multicolumn{2}{c}{Inner-city} & & \multicolumn{2}{c}{Q1}  & \multicolumn{2}{c}{Q2} & \multicolumn{2}{c}{Q3} & \multicolumn{2}{c}{Q4}    \\   \hline  
 & Mean & SD & Mean & SD  & & Mean & SD & Mean & SD & Mean & SD & Mean & SD \\ \hline 
\multicolumn{14}{l}{\textbf{Panel A: Baseline characteristics}}   \\
Distance   from CBD (m)                      & 23156  & (11356) & 5978   & (2534) &  & 6245   & (2530)  & 5792   & (2219)  & 5393   & (2093)  & 4477   & (2459)  \\
Average Rent                                 & 740    & (235)   & 800    & (282) &   & 550    & (147)   & 580    & (149)   & 570    & (151)   & 570    & (164)   \\
Average Median income                        & 24150  & (5961)  & 28735  & (7526) &  & 15395  & (3243)  & 15565  & (3377)  & 15680  & (3049)  & 15005  & (3591)  \\
Fraction of minority                         & 0.247  & (0.212) & 0.177  & (0.111) & & 0.315  & (0.195) & 0.348  & (0.213) & 0.311  & (0.210) & 0.297  & (0.191) \\
Fraction of Immigrant                        & 0.307  & (0.187) & 0.287  & (0.113) & & 0.380  & (0.200) & 0.402  & (0.189) & 0.352  & (0.192) & 0.344  & (0.178) \\
Fraction of low income                       & 0.243  & (0.140) & 0.225  & (0.095) & & 0.473  & (0.154) & 0.493  & (0.177) & 0.485  & (0.139) & 0.515  & (0.139) \\
Employment rate                              & 0.610  & (0.081) & 0.623  & (0.083) & & 0.505  & (0.071) & 0.506  & (0.092) & 0.520  & (0.078) & 0.531  & (0.090) \\
Fraction of university graduates             & 0.167  & (0.093) & 0.377  & (0.155) & & 0.171  & (0.140) & 0.165  & (0.099) & 0.178  & (0.093) & 0.222  & (0.108) \\
Population                                   & 5585   & (2626)  & 4195   & (1996) &  & 3940   & (1722)  & 4465   & (2063)  & 4365   & (2096)  & 3555   & (2511)  \\
\multicolumn{14}{l}{\textbf{Panel B: 10 years changes}}   \\
$\Delta$ Fraction of minority               & 0.073  & (0.079) & 0.033  & (0.052) & & 0.038  & (0.062) & 0.033  & (0.068) & 0.034  & (0.067) & 0.041  & (0.078) \\
$\Delta$ Fraction of Immigrant              & 0.048  & (0.052) & 0.023  & (0.049) & & 0.020  & (0.050) & 0.012  & (0.060) & 0.018  & (0.063) & 0.036  & (0.071) \\
$\Delta$ Fraction of low income             & -0.033 & (0.067) & -0.017 & (0.053) & & -0.086 & (0.076) & -0.089 & (0.072) & -0.110 & (0.076) & -0.113 & (0.105) \\
$\Delta$ Employment rate                    & 0.017  & (0.045) & 0.006  & (0.050) & & 0.070  & (0.049) & 0.075  & (0.053) & 0.091  & (0.059) & 0.097  & (0.068) \\
$\Delta$ Fraction of university   graduates & 0.069  & (0.048) & 0.086  & (0.048) & & 0.046  & (0.032) & 0.089  & (0.027) & 0.119  & (0.019) & 0.164  & (0.047) \\
$\Delta$ Population                         & 0.193  & (0.896) & 0.081  & (0.325) & & -0.010 & (0.059) & 0.007  & (0.083) & 0.023  & (0.069) & 0.207  & (0.549) \\
Number of tracts                             & 1136   &       & 246    &   &    & 120    &       & 119    &       & 120    &       & 119    &      \\  \hline \hline
\end{tabular}
\end{center}
\footnotesize Source: Author's calculations using the Census. \\
Note: Means and standard deviations of each variable by gentrification status. Not gentrifiable (outer-city) neighbourhoods are those more than 10 km away from the CBD, Not gentrifiable (inner-city) neighbourhoods are those less than 10 km away from the CBD but above the median in the median income distribution of their CMA. Gentrifiable neighbourhoods are divided into quartiles of the measure defined in Equation \ref{equation:gent_measure}.

    \end{table}
\end{landscape} 

\begin{landscape}
    \begin{table}
        \caption{Summary of Neighborhood Characteristics by Gentrification status, 2001
\label{tab:descstat_2001}}
\footnotesize
\onehalfspacing
\begin{center}
\begin{tabular}{lccccccccccccc}
\hline  \hline  
&  \multicolumn{4}{c}{Not gentrifiable} & & \multicolumn{8}{c}{Gentrification quantile}  \\   \cline{2-5} \cline{7-14} 
 & \multicolumn{2}{c}{Outer-city} & \multicolumn{2}{c}{Inner-city} & & \multicolumn{2}{c}{Q1}  & \multicolumn{2}{c}{Q2} & \multicolumn{2}{c}{Q3} & \multicolumn{2}{c}{Q4}    \\   \hline  
 & Mean & SD & Mean & SD  & & Mean & SD & Mean & SD & Mean & SD & Mean & SD \\ \hline 
\multicolumn{14}{l}{\textbf{Panel A: Baseline characteristics}}   \\
Distance   from CBD (m)                      & 23156  & (11350)  & 5501   & (2561) & & 6354   & (2666)  & 6251   & (2182)  & 5701   & (2055)  & 4621   & (2274)  \\
Average Rent                                 & 800    & (271)    & 890    & (348)  & & 615    & (182)   & 630    & (164)   & 610    & (183)   & 575    & (185)   \\
Average Median income                        & 28365  & (7037)   & 34890  & (9497) & & 18770  & (3896)  & 19760  & (3630)  & 19650  & (2912)  & 18825  & (3518)  \\
Fraction of minority                         & 0.271  & (0.233)  & 0.176  & (0.112) & & 0.386  & (0.222) & 0.369  & (0.222) & 0.309  & (0.209) & 0.273  & (0.195) \\
Fraction of Immigrant                        & 0.326  & (0.201)  & 0.291  & (0.118) & & 0.437  & (0.193) & 0.409  & (0.185) & 0.361  & (0.197) & 0.313  & (0.178) \\
Fraction of low income                       & 0.202  & (0.128)  & 0.192  & (0.089) & & 0.436  & (0.183) & 0.390  & (0.113) & 0.386  & (0.123) & 0.434  & (0.149) \\
Employment rate                              & 0.634  & (0.077)  & 0.648  & (0.080) & & 0.558  & (0.088) & 0.570  & (0.062) & 0.570  & (0.064) & 0.574  & (0.084) \\
Fraction of university graduates             & 0.199  & (0.108)  & 0.435  & (0.147) & & 0.210  & (0.133) & 0.213  & (0.121) & 0.206  & (0.102) & 0.240  & (0.112) \\
Population                                   & 6080   & (3617)   & 4245   & (2222)  & & 4525   & (2215)  & 4630   & (2174)  & 4365   & (2146)  & 3475   & (2359)  \\
\multicolumn{14}{l}{\textbf{Panel B: 10 years changes}}   \\
$\Delta$ Fraction of minority               & 0.094  & (0.076)  & 0.063  & (0.053) & & 0.058  & (0.074) & 0.062  & (0.073) & 0.054  & (0.074) & 0.051  & (0.081) \\
$\Delta$ Fraction of Immigrant              & 0.044  & (0.058)  & 0.030  & (0.057) & & 0.010  & (0.070) & 0.019  & (0.069) & 0.017  & (0.077) & 0.044  & (0.082) \\
$\Delta$ Fraction of low income             & 0.005  & (0.061)  & 0.029  & (0.060) & & -0.017 & (0.082) & -0.026 & (0.070) & -0.025 & (0.070) & -0.059 & (0.087) \\
$\Delta$ Employment rate                    & -0.036 & (0.053)  & -0.016 & (0.046) & & -0.012 & (0.048) & 0.007  & (0.043) & 0.022  & (0.048) & 0.054  & (0.053) \\
$\Delta$ Fraction of university   graduates & 0.061  & (0.048)  & 0.086  & (0.050) & & 0.033  & (0.028) & 0.081  & (0.025) & 0.113  & (0.026) & 0.166  & (0.048) \\
$\Delta$ Population                         & 0.548  & (12.843) & 0.153  & (0.611) & & -0.052 & (0.080) & -0.017 & (0.077) & 0.007  & (0.086) & 0.115  & (0.234) \\
Number of tracts                             & 1137   &   & 277    &    &   & 113    &       & 112    &       & 112    &       & 112    &       \\   \hline \hline
\end{tabular}
\end{center}
\footnotesize Source: Author's calculations using the Census. \\
Note: Means and standard deviations of each variable by gentrification status. Not gentrifiable (outer-city) neighbourhoods are those more than 10 km away from the CBD, Not gentrifiable (inner-city) neighbourhoods are those less than 10 km away from the CBD but above the median in the median income distribution of their CMA. Gentrifiable neighbourhoods are divided into quartiles of the measure defined in Equation \ref{equation:gent_measure}.

    \end{table}
\end{landscape} 

\begin{landscape}
    \begin{table}
        \caption{Summary of Neighborhood Characteristics by Gentrification status, 2006
\label{tab:descstat_2006}}
\footnotesize
\onehalfspacing
\begin{center}
\begin{tabular}{lccccccccccccc}
\hline  \hline  
&  \multicolumn{4}{c}{Not gentrifiable} & & \multicolumn{8}{c}{Gentrification quantile}  \\   \cline{2-5} \cline{7-14} 
 & \multicolumn{2}{c}{Outer-city} & \multicolumn{2}{c}{Inner-city} & & \multicolumn{2}{c}{Q1}  & \multicolumn{2}{c}{Q2} & \multicolumn{2}{c}{Q3} & \multicolumn{2}{c}{Q4}    \\   \hline  
 & Mean & SD & Mean & SD  & & Mean & SD & Mean & SD & Mean & SD & Mean & SD \\ \hline 
\multicolumn{14}{l}{\textbf{Panel A: Baseline characteristics}}   \\
Distance from CBD (m)                     & 23152  & (11359) & 5598   & (2496) & & 6600   & (2671)  & 6544   & (2243)  & 5393   & (1895)  & 4181   & (2075)  \\
Average Rent                              & 855    & (258)   & 975    & (334)  & & 680    & (187)   & 700    & (193)   & 715    & (170)   & 685    & (178)   \\
Average Median income                     & 31015  & (8007)  & 38145  & (10189) & & 21125  & (4081)  & 21875  & (3581)  & 22685  & (3410)  & 21275  & (4246)  \\
Fraction of minority                      & 0.321  & (0.252) & 0.197  & (0.115) & & 0.402  & (0.213) & 0.432  & (0.221) & 0.347  & (0.216) & 0.313  & (0.164) \\
Fraction of Immigrant                     & 0.355  & (0.208) & 0.293  & (0.114) & & 0.449  & (0.171) & 0.444  & (0.182) & 0.389  & (0.168) & 0.347  & (0.161) \\
Fraction of low income                    & 0.210  & (0.121) & 0.209  & (0.081) & & 0.434  & (0.166) & 0.385  & (0.122) & 0.370  & (0.104) & 0.428  & (0.149) \\
Employment rate                           & 0.626  & (0.072) & 0.647  & (0.076) & & 0.556  & (0.081) & 0.573  & (0.059) & 0.612  & (0.067) & 0.597  & (0.080) \\
Fraction of university graduates          & 0.236  & (0.115) & 0.464  & (0.153) & & 0.266  & (0.158) & 0.240  & (0.125) & 0.291  & (0.107) & 0.296  & (0.112) \\
Population                                & 6680   & (5560)  & 4380   & (2465)  & & 4445   & (1834)  & 4900   & (2558)  & 4290   & (2386)  & 3535   & (2402)  \\
\multicolumn{14}{l}{\textbf{Panel B: 10 years changes}}   \\
$\Delta$ Fraction of minority             & 0.091  & (0.066) & 0.064  & (0.049) & & 0.069  & (0.064) & 0.045  & (0.066) & 0.021  & (0.066) & 0.027  & (0.079) \\
$\Delta$ Fraction of Immigrant            & 0.031  & (0.050) & 0.026  & (0.049) & & 0.014  & (0.057) & 0.002  & (0.058) & -0.009 & (0.070) & 0.012  & (0.090) \\
$\Delta$ Fraction of low income           & -0.024 & (0.051) & -0.019 & (0.050) & & -0.062 & (0.085) & -0.063 & (0.058) & -0.060 & (0.058) & -0.107 & (0.091) \\
$\Delta$ Employment rate                  & -0.031 & (0.043) & -0.015 & (0.043) & & -0.009 & (0.043) & 0.005  & (0.038) & 0.014  & (0.041) & 0.057  & (0.060) \\
$\Delta$ Fraction of university graduates & 0.049  & (0.041) & 0.077  & (0.050) & & 0.025  & (0.035) & 0.064  & (0.025) & 0.108  & (0.025) & 0.152  & (0.047) \\
$\Delta$ Population                       & 0.117  & (0.577) & 0.124  & (0.412) & & -0.033 & (0.077) & 0.011  & (0.069) & 0.021  & (0.064) & 0.192  & (0.318) \\
Number of tracts                          & 1136   &       & 282    &      & & 111    &       & 111    &       & 111    &       & 111    &   \\   \hline \hline
\end{tabular}
\end{center}
\footnotesize Source: Author's calculations using the Census. \\
Note: Means and standard deviations of each variable by gentrification status. Not gentrifiable (outer-city) neighbourhoods are those more than 10 km away from the CBD, Not gentrifiable (inner-city) neighbourhoods are those less than 10 km away from the CBD but above the median in the median income distribution of their CMA. Gentrifiable neighbourhoods are divided into quartiles of the measure defined in Equation \ref{equation:gent_measure}.

    \end{table}
\end{landscape} 

\begin{table}[h!]
        \caption{Balance table 
\label{tab:Bal_table_gent}}
\footnotesize
\onehalfspacing
\begin{center}
\begin{tabular}{lcccc}
\hline \hline
                                & \multicolumn{2}{c}{Mean and sd}                               & \multicolumn{2}{c}{Regression}  \\ 
\textbf{}                       & \multicolumn{1}{l}{Treated} & \multicolumn{1}{l}{Non-treated} & \multicolumn{2}{c}{coefficients} \\[0.1cm] \cline{2-5} 
Median income growth (\$)       & 987.97                      & 1609.25                         & -621.287***        & -105.181              \\
                                & (3134.41)                   & (2811.90)                       & (266.186)            & (209.799)            \\
                                & \multicolumn{1}{l}{}        & \multicolumn{1}{l}{}            & {[}0.020{]}          & {[}0.616{]}          \\
Employment growth               & -0.0035                     & 0.0067                         & -0.010**                & -0.001               \\
                                & (0.0584)                    & (0.0554)                        & (0.005)              & (0.004)              \\
                                & \multicolumn{1}{l}{}        & \multicolumn{1}{l}{}            & {[}0.040{]}          & {[}0.696{]}          \\
Rent growth (\%)                & 0.0753                      & 0.0910                          & -0.016*             & -0.006               \\
                                & (0.0931)                    & (0.1034)                        & (0.008)              & (0.009)              \\
                                & \multicolumn{1}{l}{}        & \multicolumn{1}{l}{}            & {[}0.059{]}          & {[}0.494{]}          \\
Fraction of low-income growth   & 0.0320                      & 0.0123                          & 0.020**              & -0.002                \\
                                & (0.1074)                    & (0.1003)                        & (0.009)              & (0.005)              \\
                                & \multicolumn{1}{l}{}        & \multicolumn{1}{l}{}            & {[}0.032{]}          & {[}0.775{]}          \\
Fraction of college educ growth & 0.0120                      & 0.0412                          & -0.029***            & -0.026***            \\
                                & (0.0375)                    & (0.0383)                        & (0.003)              & (0.003)              \\
                                & \multicolumn{1}{l}{}        & \multicolumn{1}{l}{}            & {[}0.000{]}          & {[}0.000{]}          \\[0.2cm]
Controls                        & \multicolumn{1}{l}{}        & \multicolumn{1}{l}{}            & No                   & Yes                  \\ \hline \hline
\end{tabular}
\end{center}
\footnotesize Source: Author's calculations using the Census. \\
Note: Each panel of each column presents the coefficient from a separate regression estimate from equation (\ref{equation:reg_migration}). The sample includes the 864 census tract x census year included in the main regressions.  Standard errors are reported in parentheses and are clustered at the census tract level. *** $p<0.01$, ** $p<0.05$, * $p<0.1$. P-values are reported in brackets.
Controls include CMA and year FE and distance from CBD.

\end{table}

    \begin{table}[h!]
        \caption{Balance table of matching procedure 
\label{tab:matching_table}}
\footnotesize
\singlespacing
\begin{center}
\begin{tabular}{lcccc}
\hline \hline 
            & \multicolumn{2}{c}{}                               & \multicolumn{2}{c}{}  \\[-0.3cm] 
                                & \multicolumn{2}{c}{Baseline Sample}                               & \multicolumn{2}{c}{Matched Sample}  \\[0.1cm]  \cline{2-3}  \cline{4-5} 
                                & \multicolumn{1}{l}{Control} & \multicolumn{1}{l}{Treated}& \multicolumn{1}{l}{Control} & \multicolumn{1}{l}{Treated} \\[0.1cm] \cline{2-5} 
\textbf{Panel A: Family-level}       &          &           &           &   \\[0.1cm]
Tenure                               &          &           &           &          \\
\hspace{0.2cm} Less than 1 year      & 0.313	&	0.309	&	0.317   &	0.315	          \\            
\hspace{0.2cm} 1 year                & 0.184	&	0.168	&	0.175   &	0.172	       \\    
\hspace{0.2cm} 2 year                & 0.160	&	0.136	&	0.140   &	0.139	      \\  
\hspace{0.2cm} 3 year                & 0.128	&	0.111	&	0.114   &	0.114	          \\  
\hspace{0.2cm} 4 year                & 0.081	&	0.077	&	0.080   &	0.078	        \\  
\hspace{0.2cm} 5 year or more        & 0.134	&	0.198	&	0.173   &	0.182	            \\[0.1cm]
Income quartile                   &        &           &           &                  \\
\hspace{0.2cm} Q1      & 0.236	&	0.306	&	0.336	&	0.308          \\            
\hspace{0.2cm} Q2        & 0.245	&	0.280	&	0.279	&	0.279         \\    
\hspace{0.2cm} Q3      & 0.247	&	0.227	&	0.218	&	0.228          \\  
\hspace{0.2cm} Q4        & 0.272	&	0.187	&	0.167	&	0.185           \\[0.1cm]
Family type                   &        &           &           &                  \\
\hspace{0.2cm} Single without kid      & 0.231	&	0.205	&	0.205	&	0.207          \\         
\hspace{0.2cm} Single with kids        & 0.300	&	0.210	&	0.212	&	0.214         \\    
\hspace{0.2cm} Couple without kid      & 0.095	&	0.098	&	0.102	&	0.098          \\  
\hspace{0.2cm} Couple with kids        & 0.374	&	0.487	&	0.481	&	0.480           \\[0.1cm]
Number of kids                          & 0.880  &   0.700   &  0.700    &   0.700                \\[-0.1cm]
                                       & (1.09) &   (1.03)  &  (1.02)   &  (1.03)        \\[0.1cm]
Number of family                      & 9,896,302  &   320,210   & 312,560   &   312,560     \\[0.2cm]

\textbf{Panel B: Individual-level}     &           &              &            &              \\[0.1cm]
Employment income                      & 16,800    &   17,800     &   15,800   &  16,500                 \\[-0.1cm]
                                       & (19,635)  &   (19,580)   &  (16,915)  &  (17,295)                \\[0.1cm]
Age                                    & 35.1      &   36.7       &   36.0     &  36.5                \\[-0.1cm]
                                       & (9.6)     &   (9.9)      &  (9.7)     &  (9.9)                \\[0.1cm]
Men                                    & 0.501	   &	0.494  	  &	0.486	   &	0.490                \\
Women                                  & 0.499	   &	0.506	  &	0.514	   &    0.510            \\[0.1cm]
Marital Status                          &        &           &           &                  \\
\hspace{0.2cm}  	Married	              &	0.3714	&	0.3871	&	0.3714	&	0.3831	\\  
\hspace{0.2cm}  	Common law partner	&	0.0683	&	0.0687	&	0.0711	&	0.0683	\\  
\hspace{0.2cm}  	Widowed	           &	0.0066	&	0.0081	&	0.0078	&	0.0078	\\  
\hspace{0.2cm}  	Divorced	       &	0.0470	&	0.0558	&	0.0582	&	0.0552	\\  
\hspace{0.2cm}  	Separated	       &	0.0497	&	0.0460	&	0.0534	&	0.0453	\\  
\hspace{0.2cm}  	Single	            &	0.4571	&	0.4343	&	0.4381	&	0.4403	\\[0.1cm]
Number of individual                     &	1,767,040	&	322,040	&	311,855	&	311,855	\\[0.1cm]
\hline \hline
\end{tabular}
\end{center}
\footnotesize Source: Author's calculations using the LAD. \\
Note: The top panel presents family statistics for the control and treated groups in the baseline sample and the matched sample. The bottom panel shows individual statistics for the same sample groups. 

    \end{table}

\clearpage

\appendix
\renewcommand\thetable{\thesection.\arabic{table}}   
\renewcommand\thefigure{\thesection.\arabic{figure}}   
\setcounter{figure}{0}    
\setcounter{table}{0}   

\restoregeometry
\section{Robustness}
\subsection{Alternative measure of gentrification}
\label{sec:robustness}

One might be concerned that the choice of measure of gentrification drives the main results. To address those concerns, I repeat the analysis using an alternative measure based on the growth of the median income instead of the share of college-educated individuals. The alternative measure is:

\begin{align}\label{equation:gent_measure_alt}
    gent_{jc,t} = \frac{Medincome_{jc,t+10} - Medincome_{jc,t}}{Medincome_{jc,t}}
\end{align}

As before, I define gentrifying neighborhoods as those in the top half of the distribution in their respective city. I preserve the eligibility criteria introduced in Section \ref{sec:measure}; hence the pool of gentrifiable neighborhoods remains the same. The two measures have a correlation of 0.38, implying that they both capture general trends in neighborhood characteristics and different aspects of those changes. The results below using the alternative measure are qualitatively the same as in the main analysis. This suggests that neighborhoods that saw an increase in the share of college-educated and median income are those driving the results. 


\newgeometry{top=1cm,left=2cm,right=2cm}
\begin{figure}[!htb]
\caption{Effect of gentrification on displacement, Alternative measure}
\centering
\subfloat[Low-income]{\includegraphics[width = 5in]{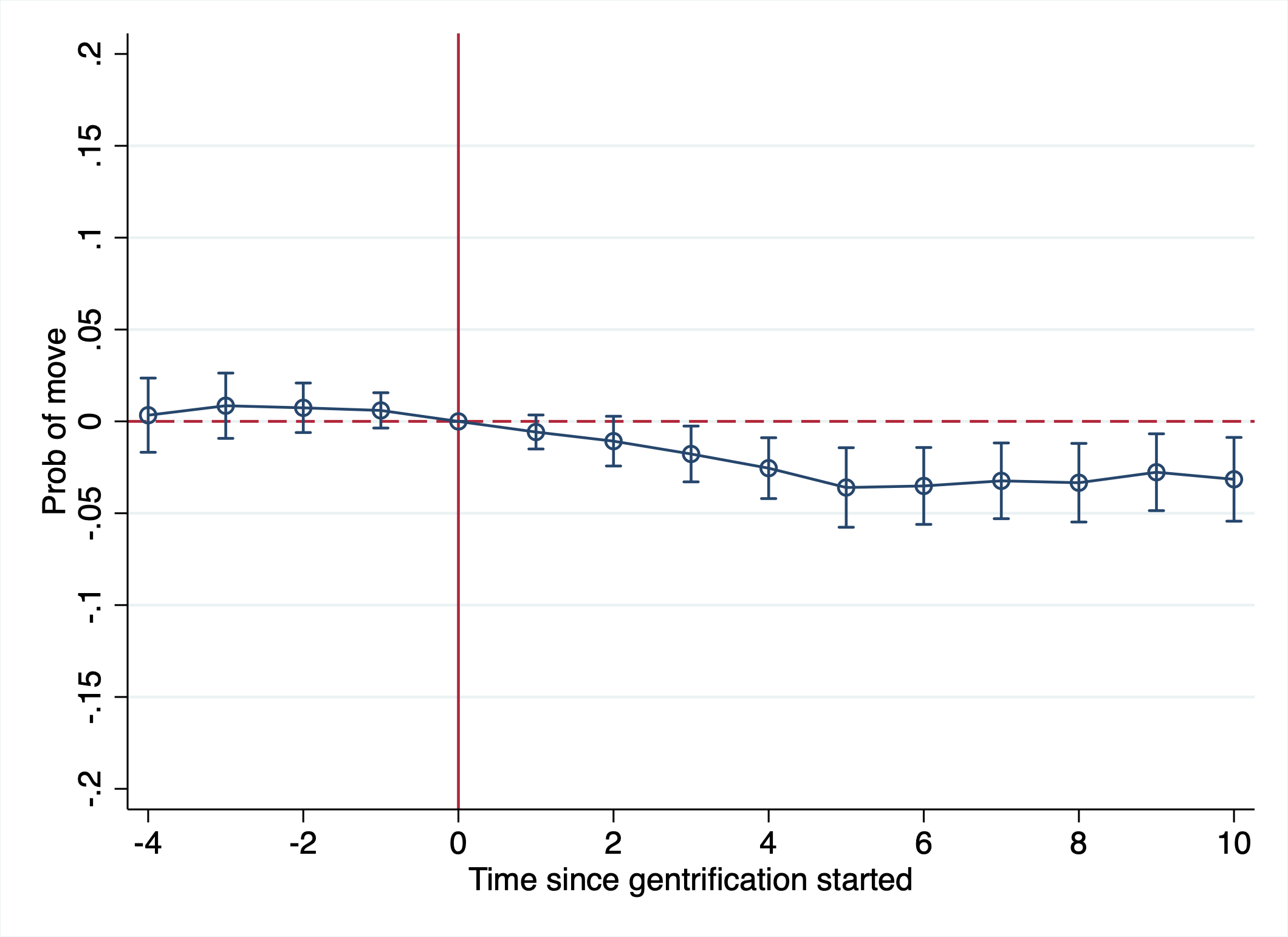}} \\
\subfloat[High-income]{\includegraphics[width = 5in]{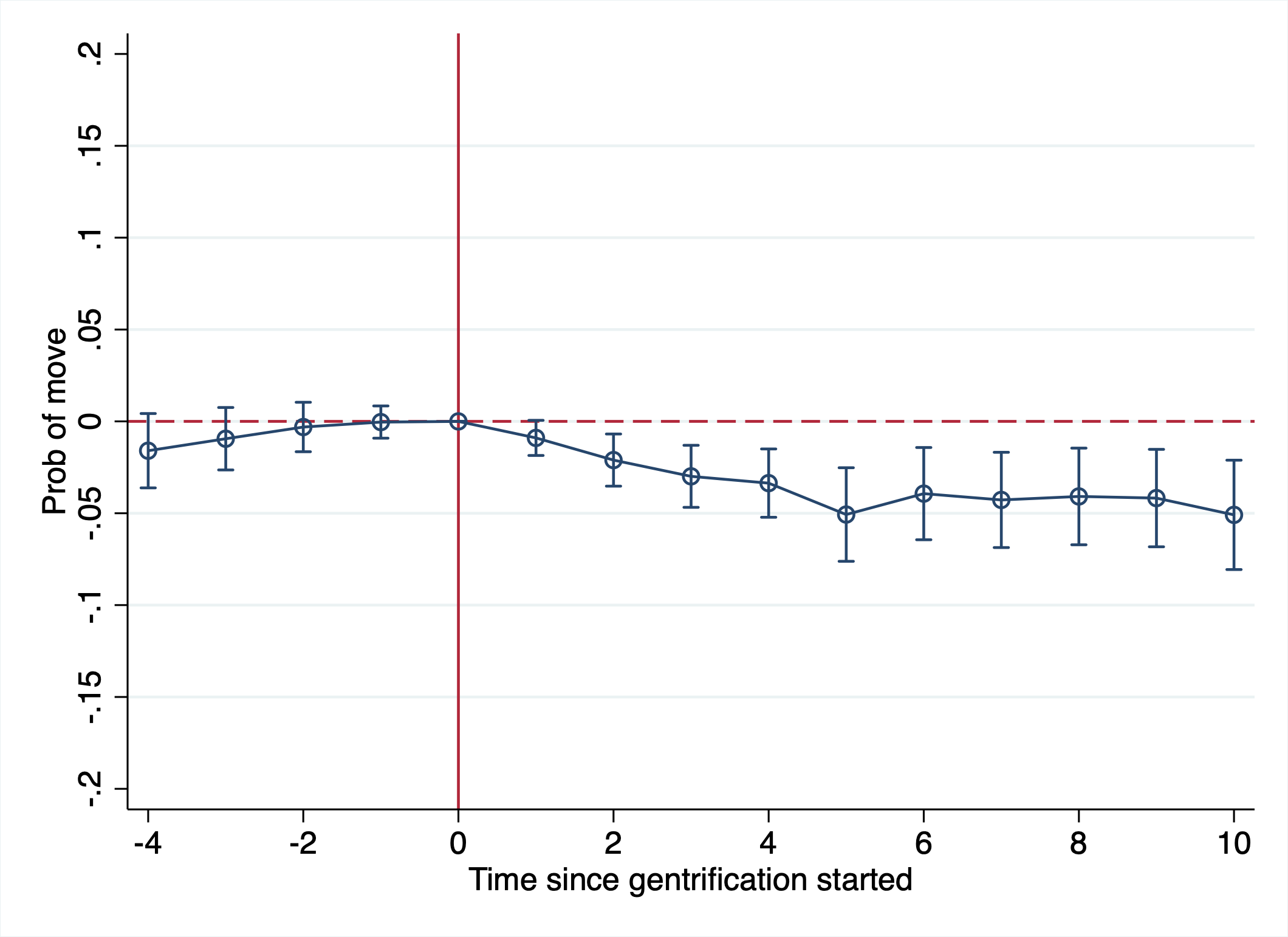}}\\
\label{fig:Displacement_2}
\begin{minipage}{0.9\textwidth} 
{\footnotesize \vspace{0.1cm} Source: Author's calculations using the Census and LAD.\\
Note: Dependant variables: dummy equal to one if the household lives in a different tract than at baseline. The sample is incumbent households living in gentrifiable neighborhoods (initially low-income and central city) prior to one of the baseline years. The control group is the matched sample discussed in section \ref{sec:Matched_Samples}. The regressions also include family-level socioeconomic control variables (age, family composition,  number of children, and immigrant indicator), baseline Census Tract controls (college-educated share, median income, share for low income, average rent, employment rate, visible minority share, the share of immigrant, distance from CBD), and pre-period variation controls (changes of college-educated share, median income, average rent, employment rate). Low-income number of observations: 991,630. High-income number of observations: 758,770. \par}
\end{minipage}
\end{figure}

\begin{figure}[!htb]
\caption{Effect of gentrification on location choice: Distance from CBD, Alternative measure}
\centering
\subfloat[Low-income]{\includegraphics[width = 5in]{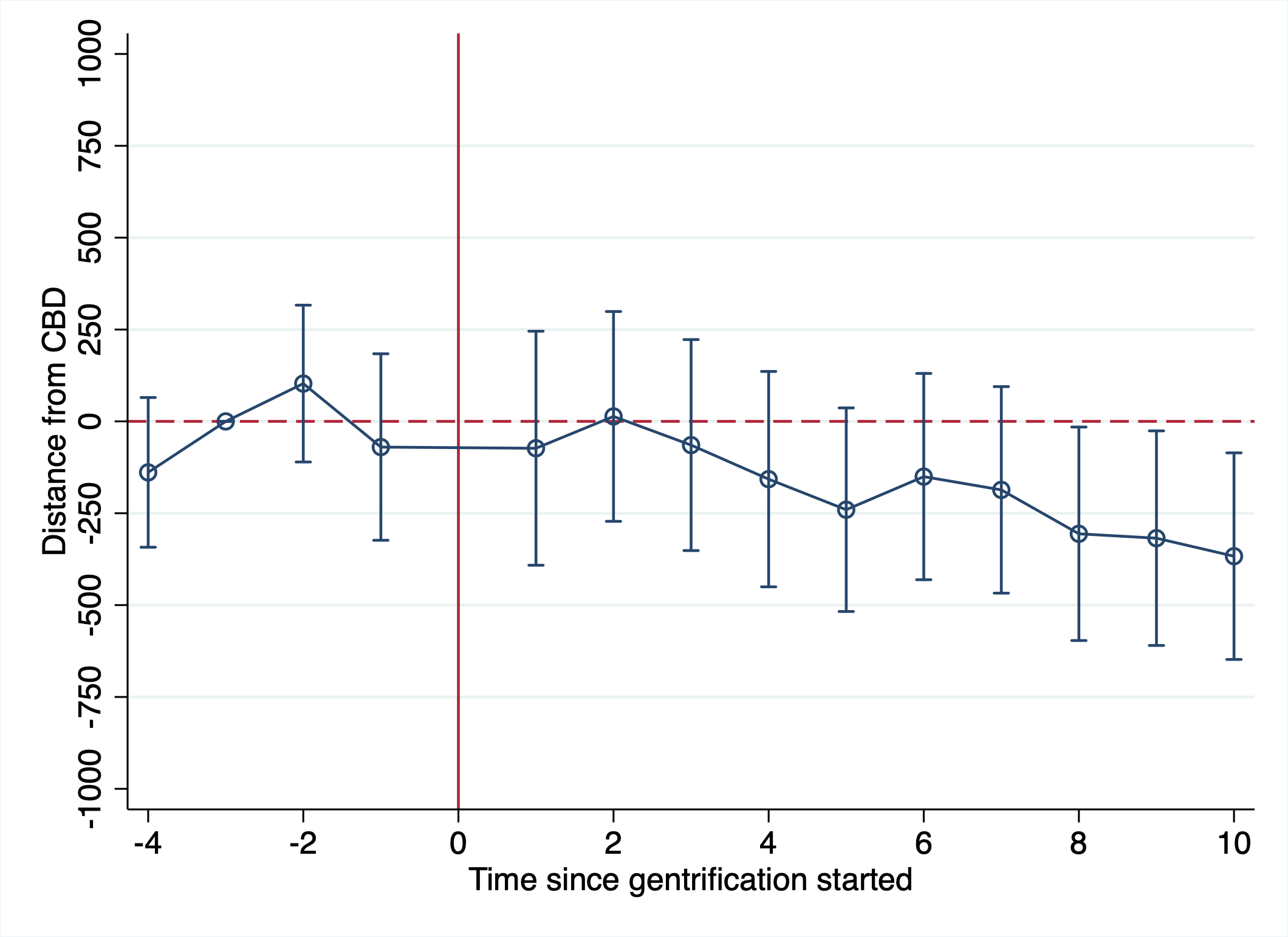}} \\
\subfloat[High-income]{\includegraphics[width = 5in]{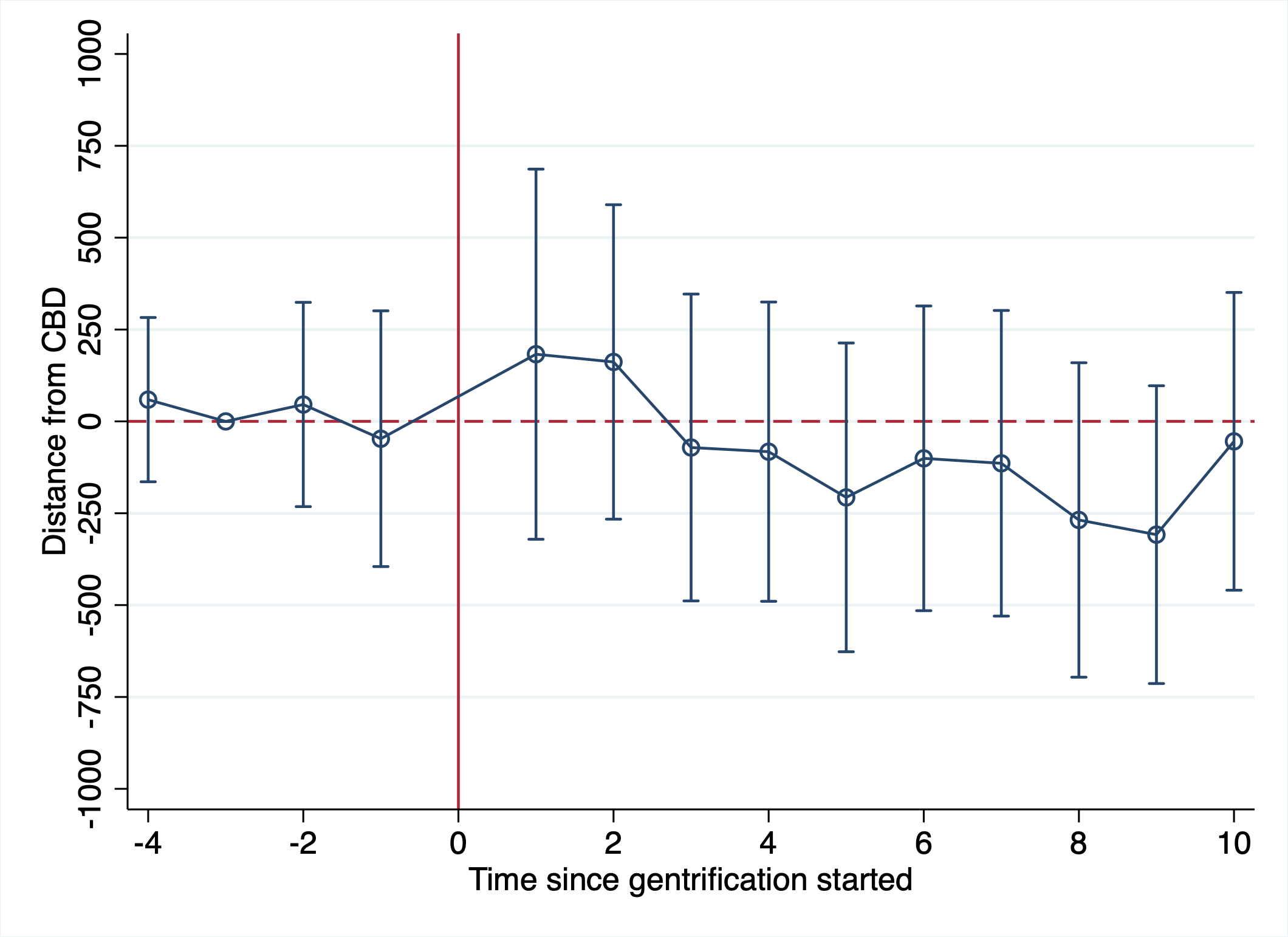}}\\
\label{fig:exp_distance_2}
\begin{minipage}{0.9\textwidth} 
{\footnotesize \vspace{0.1cm} Source: Author's calculations using the Census and LAD.\\
Note: Dependant variables: distance from central business district. The sample is incumbent households living in gentrifiable neighborhoods (initially low-income and central city) in one of the baseline years, but do not currently live in that census tract. The control group is the matched sample discussed in section \ref{sec:Matched_Samples}. The regressions also include family-level socioeconomic control variables (age, family composition,  number of children and immigrant indicator), baseline Census Tract controls (college-educated share, median income, share for low income, average rent, employment rate, visible minority share, the share of immigrant, distance from CBD), and pre-period variation controls (changes of college-educated share, median income, average rent, employment rate). Low-income number of observations: 358,070. High-income number of observations: 201,839. \par}
\end{minipage}
\end{figure}

\begin{figure}[!htb]
\caption{Effect of gentrification on location choice: Share of low-income, Alternative measure}
\centering
\subfloat[Low-income]{\includegraphics[width = 5in]{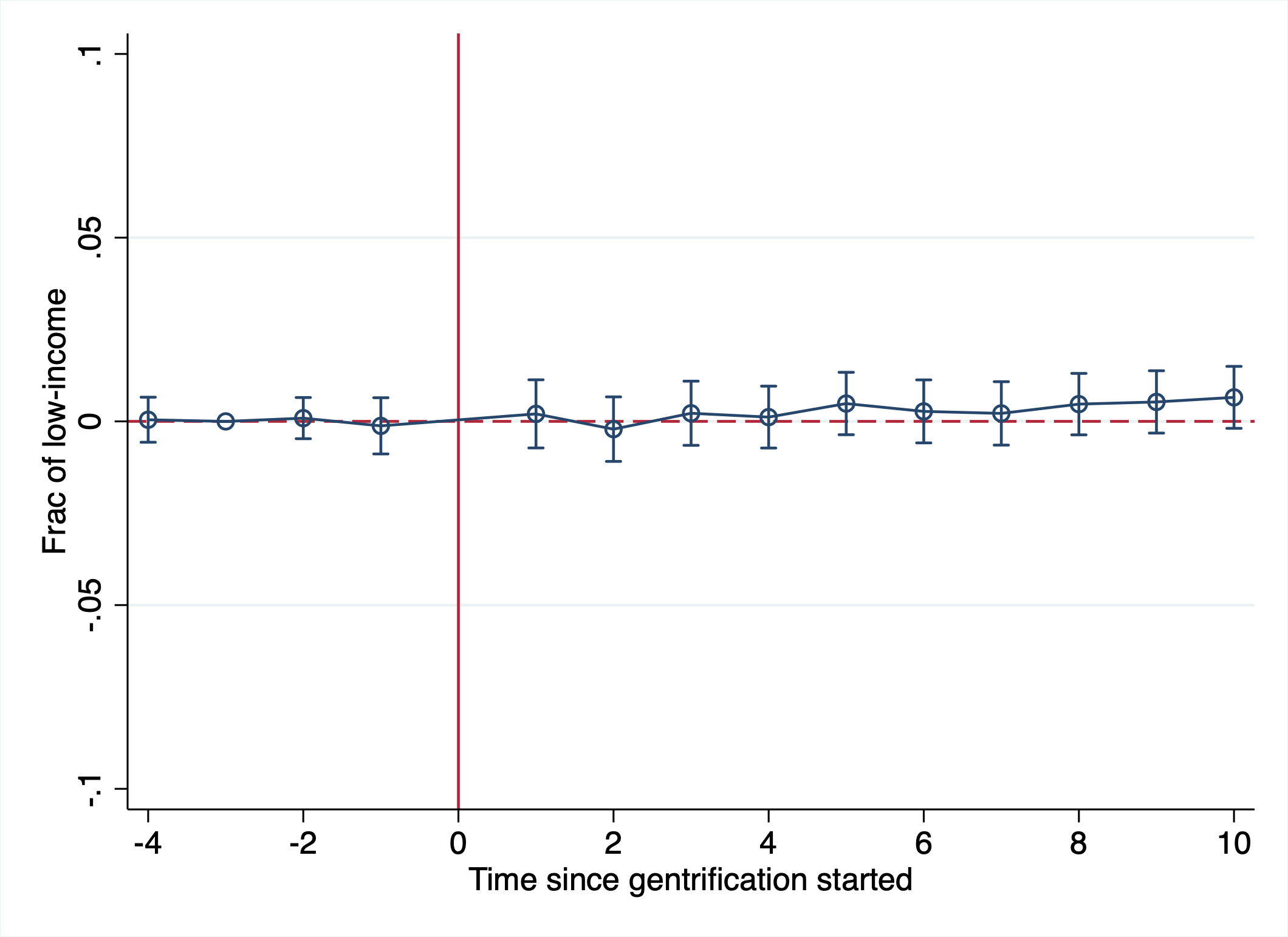}} \\
\subfloat[High-income]{\includegraphics[width = 5in]{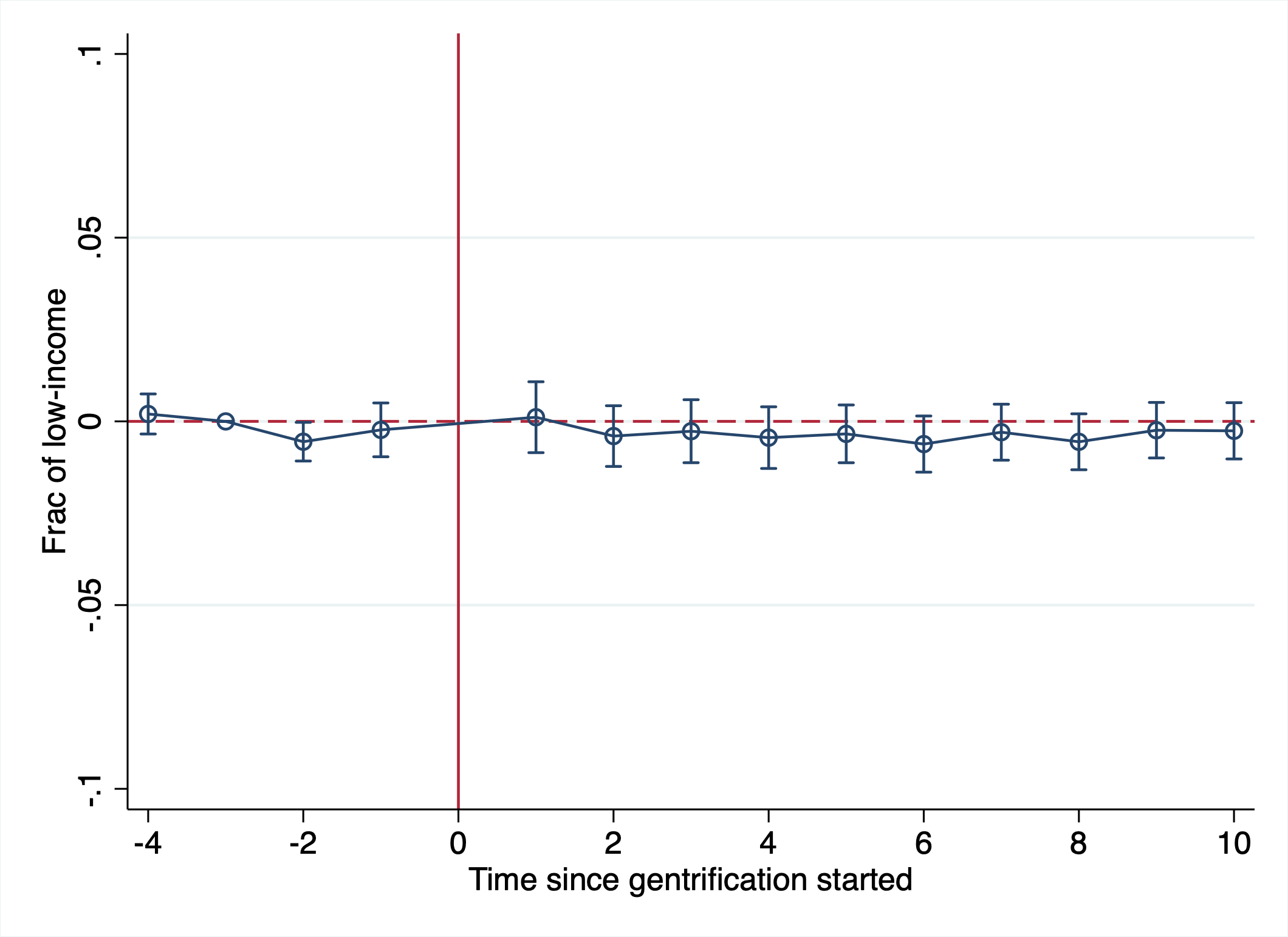}}\\
\label{fig:exp_lowinc_2}
\begin{minipage}{0.9\textwidth} 
{\footnotesize \vspace{0.1cm} Source: Author's calculations using the Census and LAD.\\
Note: Dependant variables: share of low-income households in the CT. The sample is incumbent households living in gentrifiable neighborhoods (initially low-income and central city) in one of the baseline years, but do not currently live in that census tract. The control group is the matched sample discussed in section \ref{sec:Matched_Samples}. The regressions also include family-level socioeconomic control variables (age, family composition,  number of children and immigrant indicator), baseline Census Tract controls (college-educated share, median income, share for low income, average rent, employment rate, visible minority share, the share of immigrant, distance from CBD), and pre-period variation controls (changes of college-educated share, median income, average rent, employment rate). \par}
\end{minipage}
\end{figure}

\begin{figure}[!htb]
\caption{Effect of gentrification on location choice: Share of university-educated, Alternative measure}
\centering
\subfloat[Low-income]{\includegraphics[width = 5in]{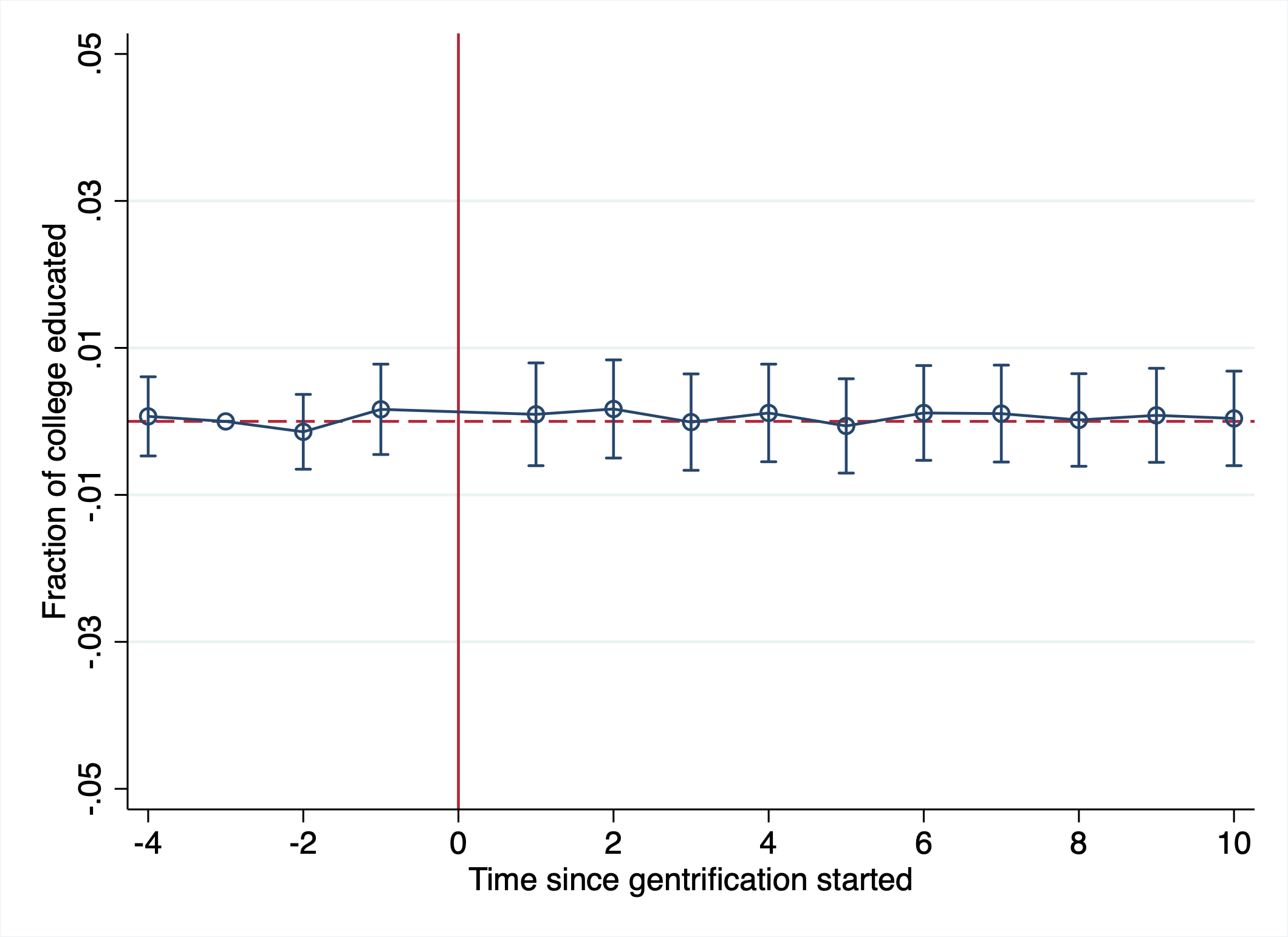}} \\
\subfloat[High-income]{\includegraphics[width = 5in]{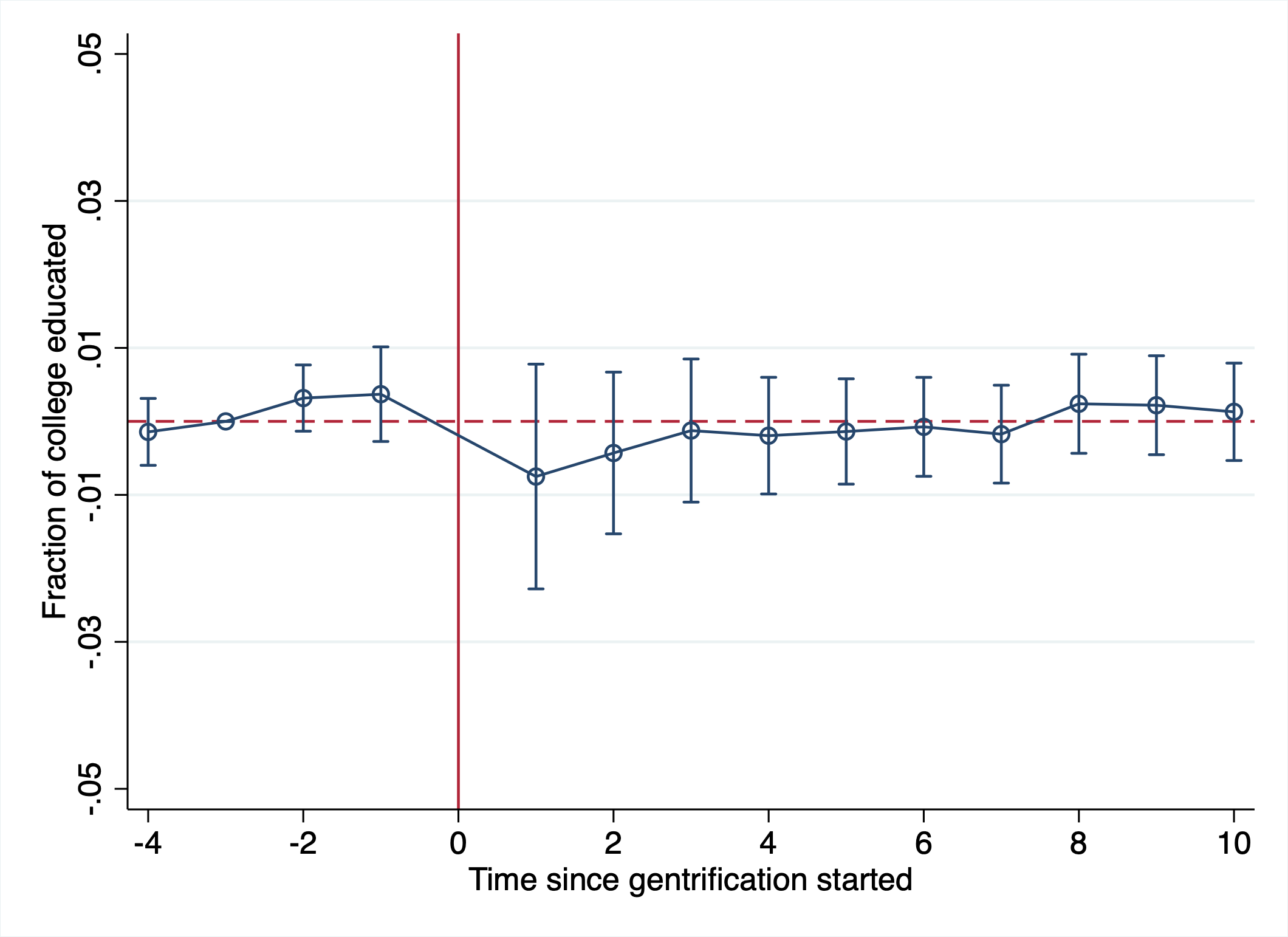}}\\
\label{fig:exp_colgrad_2}
\begin{minipage}{0.9\textwidth} 
{\footnotesize \vspace{0.1cm} Source: Author's calculations using the Census and LAD.\\
Note: Dependant variables: share of university-educated in the CT. The sample is incumbent households living in gentrifiable neighborhoods (initially low-income and central city) in one of the baseline years, but do not currently live in that census tract. The control group is the matched sample discussed in section \ref{sec:Matched_Samples}. The regressions also include family-level socioeconomic control variables (age, family composition,  number of children and immigrant indicator), baseline Census Tract controls (college-educated share, median income, share for low income, average rent, employment rate, visible minority share, the share of immigrant, distance from CBD), and pre-period variation controls (changes of college-educated share, median income, average rent, employment rate). \par}
\end{minipage}
\end{figure}

\begin{figure}[!htb]
\caption{Effect of gentrification on location choice: Employment rate, Alternative measure}
\centering
\subfloat[Low-income]{\includegraphics[width = 5in]{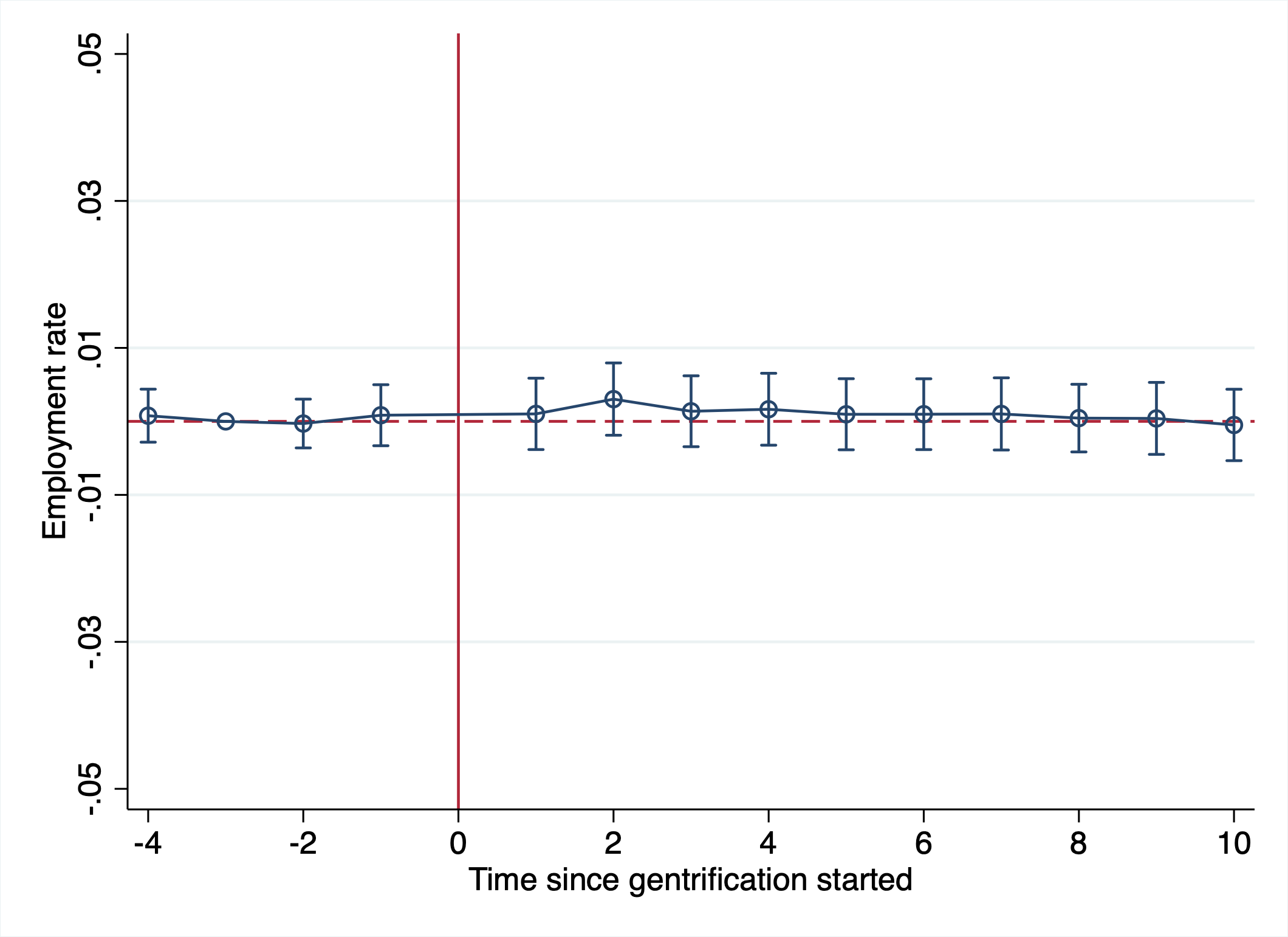}} \\
\subfloat[High-income]{\includegraphics[width = 5in]{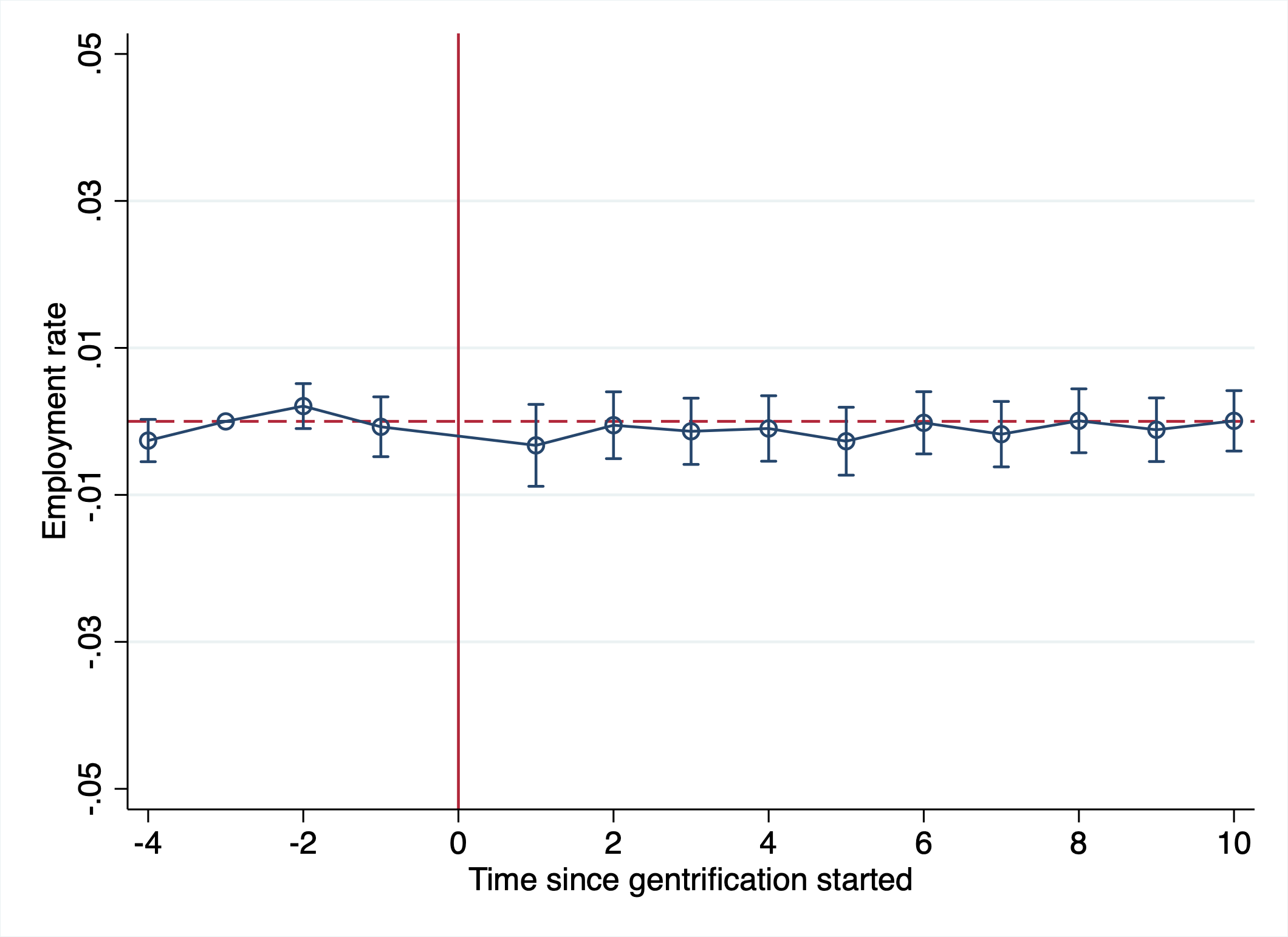}}\\
\label{fig:exp_unempl_2}
\begin{minipage}{0.9\textwidth} 
{\footnotesize \vspace{0.1cm} Source: Author's calculations using the Census and LAD.\\
Note: Dependant variables: employment rate in the CT. The sample is incumbent households living in gentrifiable neighborhoods (initially low-income and central city) in one of the baseline years, but do not currently live in that census tract. The control group is the matched sample discussed in section \ref{sec:Matched_Samples}. The regressions also include family-level socioeconomic control variables (age, family composition,  number of children and immigrant indicator), baseline Census Tract controls (college-educated share, median income, share for low income, average rent, employment rate, visible minority share, the share of immigrant, distance from CBD), and pre-period variation controls (changes of college-educated share, median income, average rent, employment rate). Low-income number of observations: 358,070. High-income number of observations: 201,839. \par}
\end{minipage}
\end{figure}

\begin{figure}[!htb]
\caption{Effect of gentrification on location choice: Median income, Alternative measure}
\centering
\subfloat[Low-income]{\includegraphics[width = 5in]{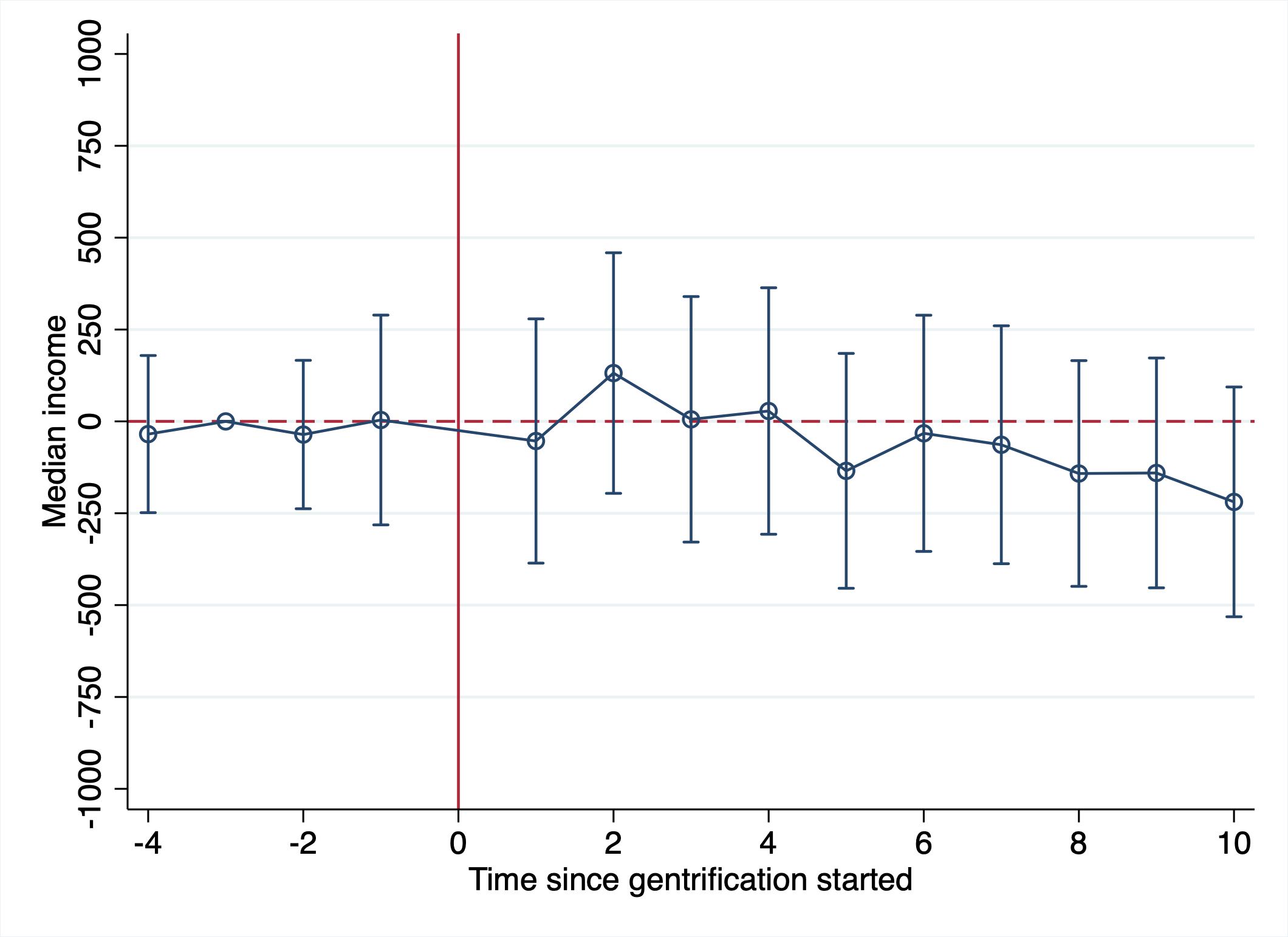}} \\
\subfloat[High-income]{\includegraphics[width = 5in]{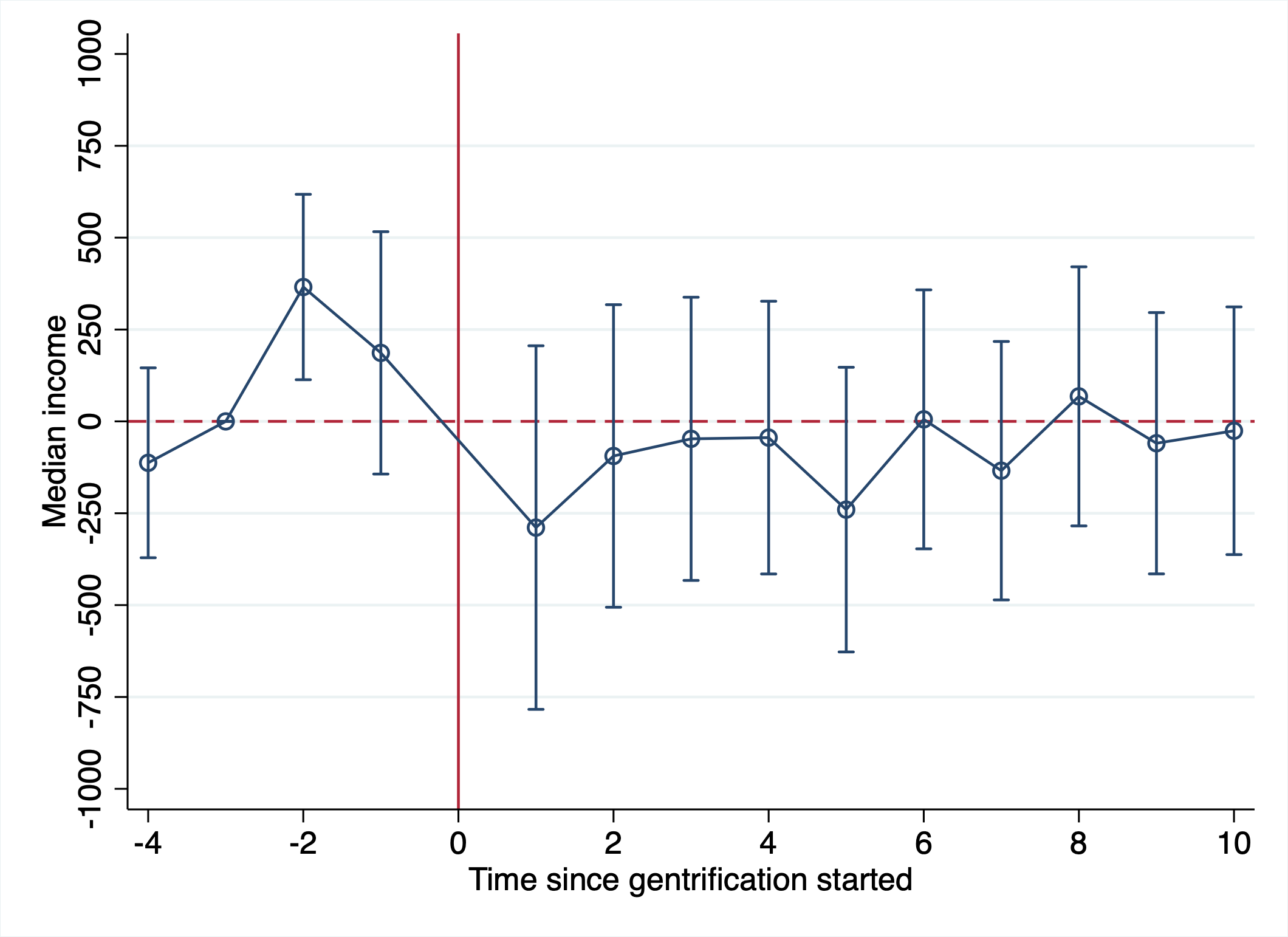}}\\
\label{fig:exp_meadian_inc_2}
\begin{minipage}{0.9\textwidth} 
{\footnotesize \vspace{0.1cm} Source: Author's calculations using the Census and LAD.\\
Note: Dependant variables: median household income in the CT. The sample is incumbent households living in gentrifiable neighborhoods (initially low-income and central city) in one of the baseline years, but do not currently live in that census tract. The control group is the matched sample discussed in section \ref{sec:Matched_Samples}. The regressions also include family-level socioeconomic control variables (age, family composition,  number of children and immigrant indicator), baseline Census Tract controls (college-educated share, median income, share for low income, average rent, employment rate, visible minority share, the share of immigrant, distance from CBD), and pre-period variation controls (changes of college-educated share, median income, average rent, employment rate). Low-income number of observations: 358,070. High-income number of observations: 201,839. \par}
\end{minipage}
\end{figure}

\begin{figure}[!htb]
\caption{Effect of gentrification on location choice: Share of immigrant, Alternative measure}
\centering
\subfloat[Low-income]{\includegraphics[width = 5in]{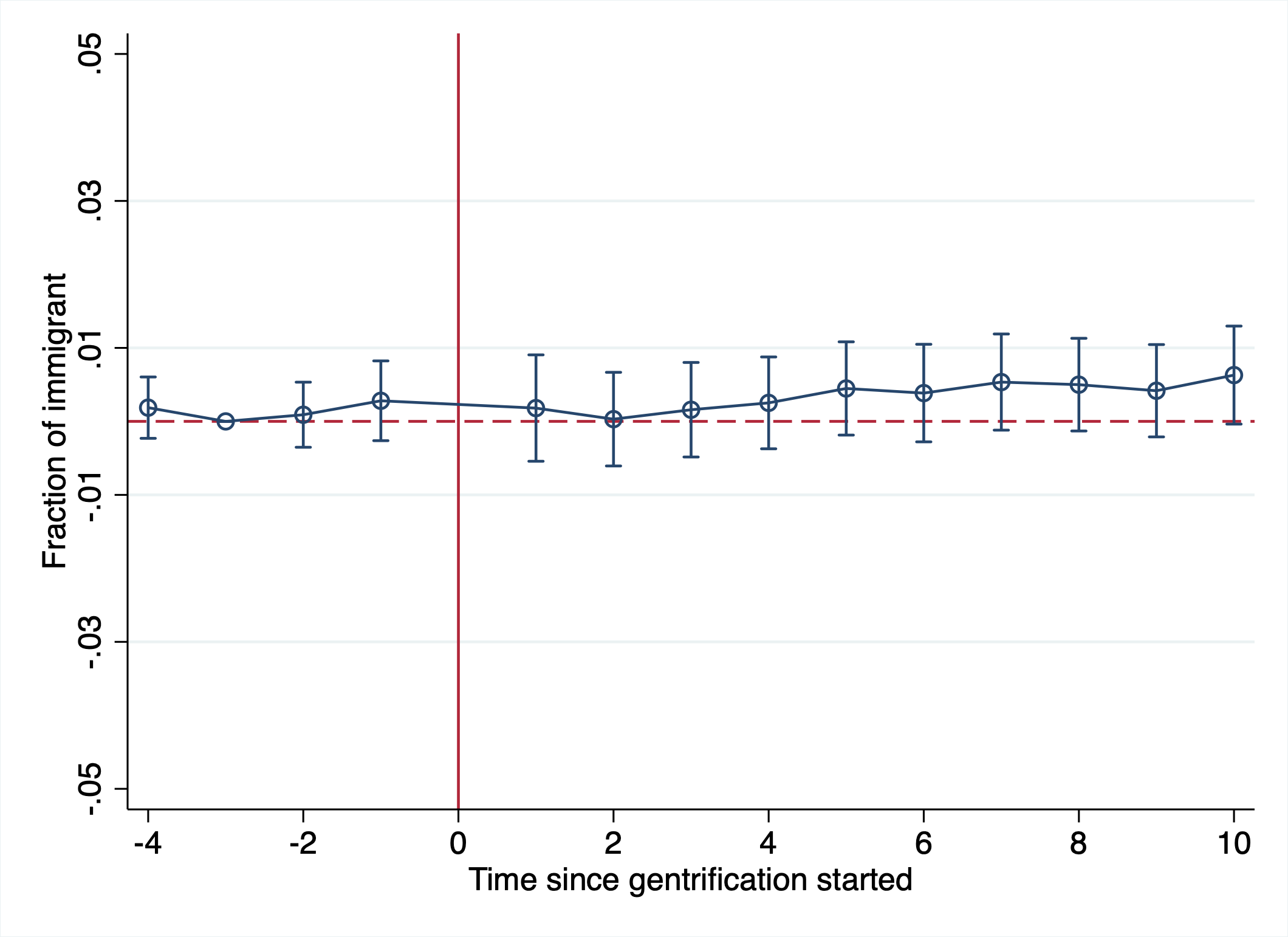}} \\
\subfloat[High-income]{\includegraphics[width = 5in]{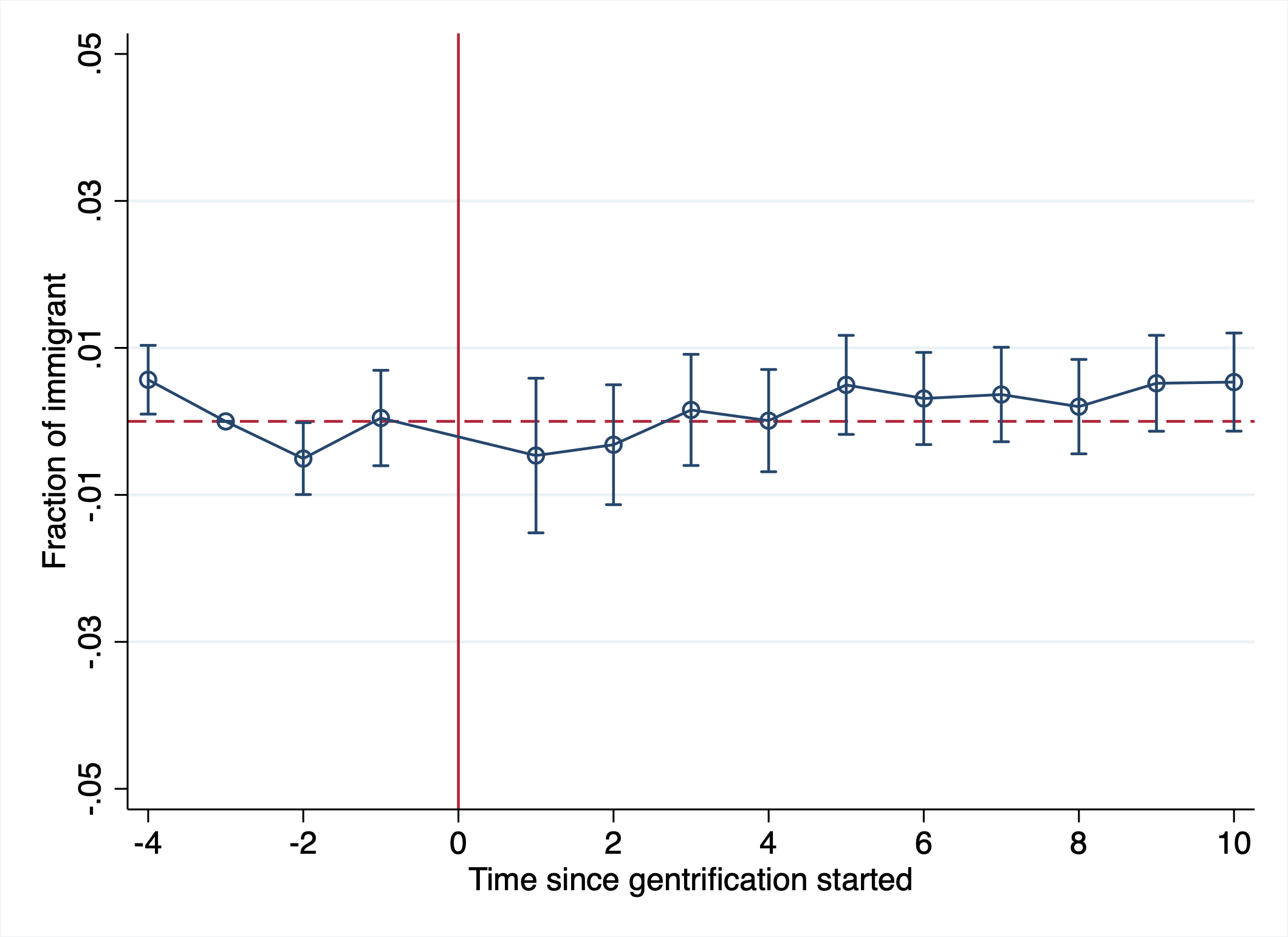}}\\
\label{fig:exp_immigrant_2}
\begin{minipage}{0.9\textwidth} 
{\footnotesize \vspace{0.1cm} Source: Author's calculations using the Census and LAD.\\
Note: Dependant variables: share of immigrants in the CT. The sample is incumbent households living in gentrifiable neighborhoods (initially low-income and central city) in one of the baseline years, but do not currently live in that census tract. The control group is the matched sample discussed in section \ref{sec:Matched_Samples}. The regressions also include family-level socioeconomic control variables (age, family composition,  number of children and immigrant indicator), baseline Census Tract controls (college-educated share, median income, share for low income, average rent, employment rate, visible minority share, the share of immigrant, distance from CBD), and pre-period variation controls (changes of college-educated share, median income, average rent, employment rate). Low-income number of observations: 358,070. High-income number of observations: 201,839. \par}
\end{minipage}
\end{figure}

\begin{figure}[!htb]
\caption{Effect of gentrification on location choice: Share of visible minority, Alternative measure}
\centering
\subfloat[Low-income]{\includegraphics[width = 5in]{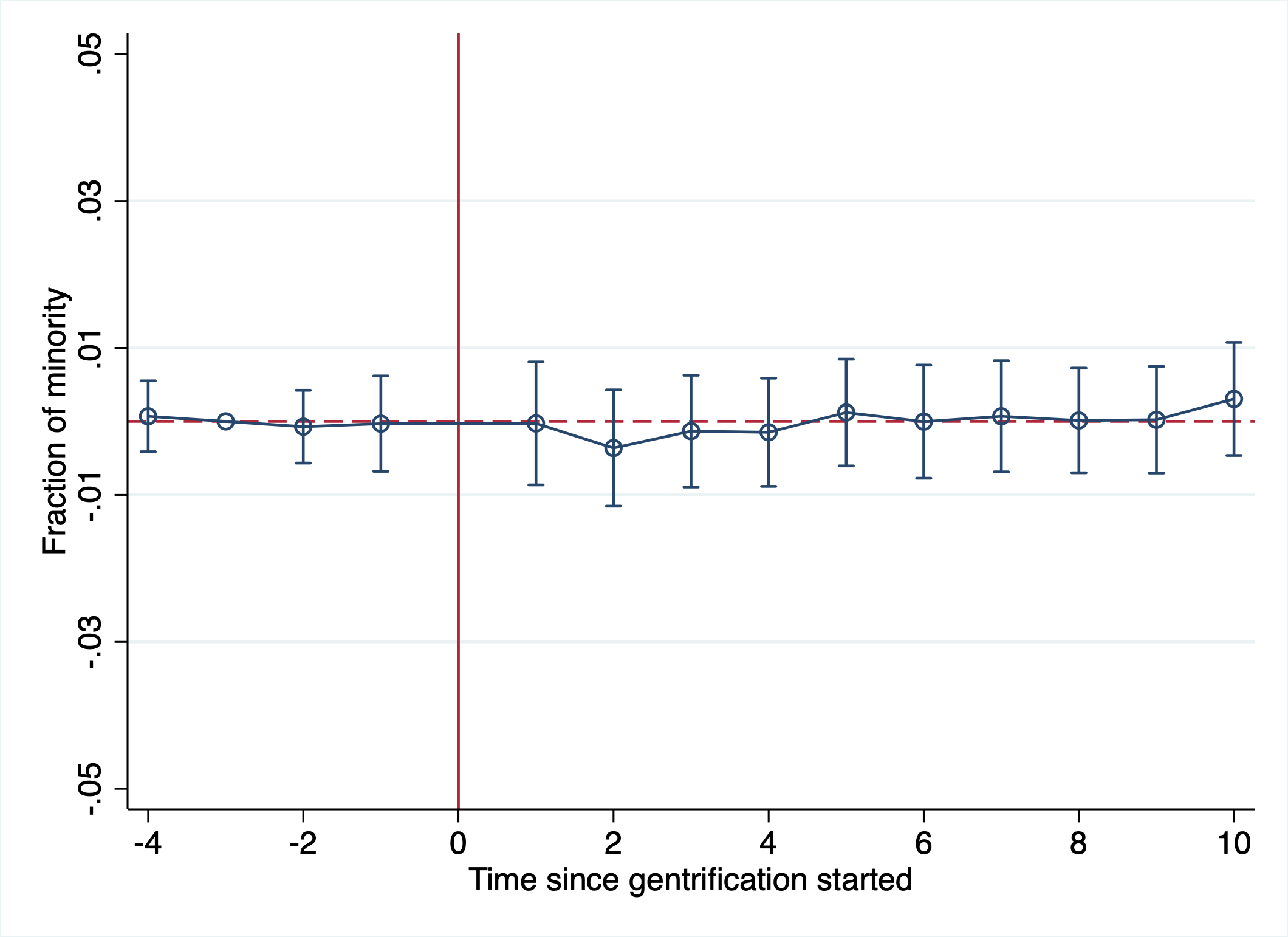}} \\
\subfloat[High-income]{\includegraphics[width = 5in]{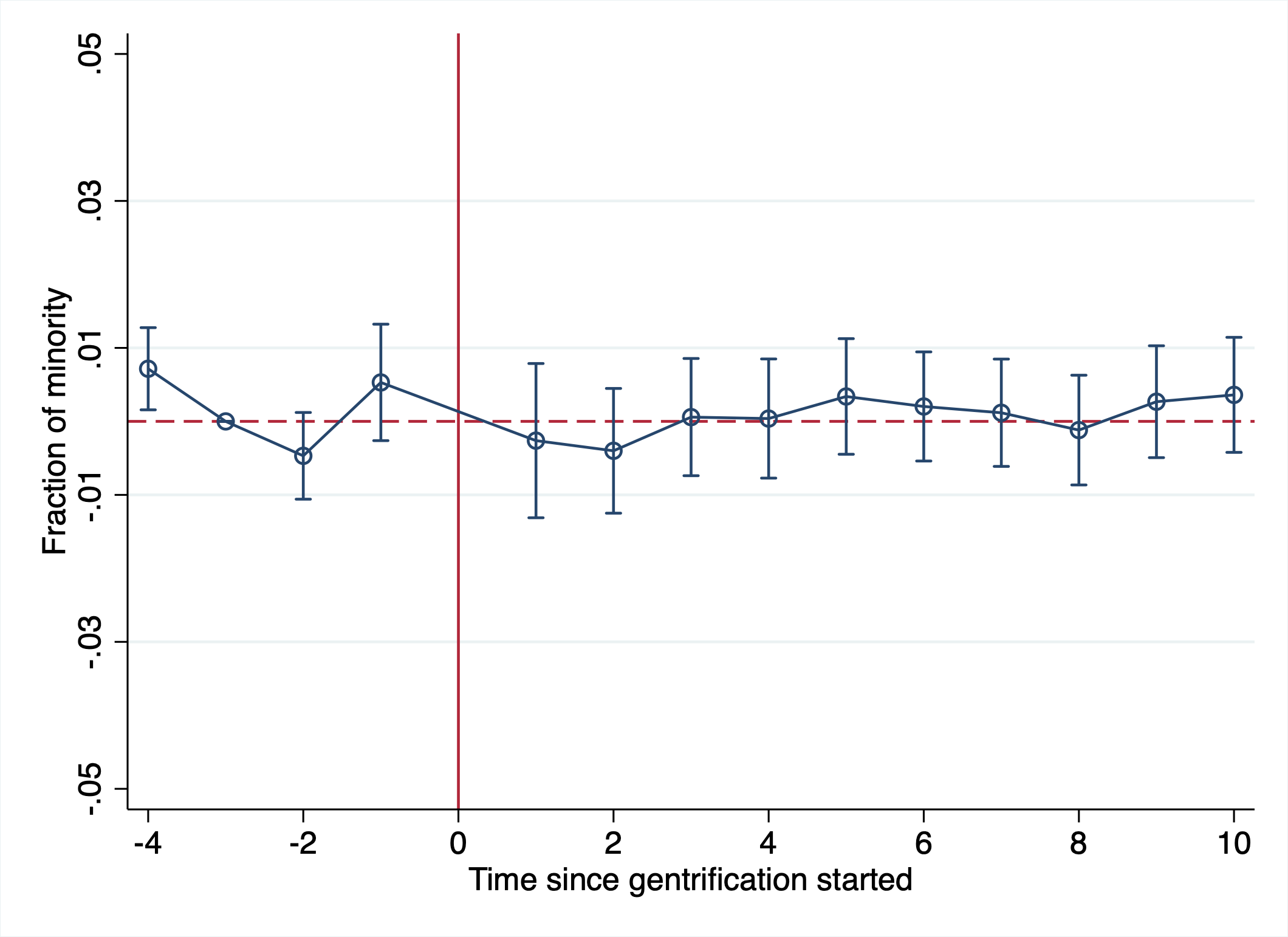}}\\
\label{fig:exp_minority_2}
\begin{minipage}{0.9\textwidth} 
{\footnotesize \vspace{0.1cm} Source: Author's calculations using the Census and LAD.\\
Note: Dependant variables: share of visible minorities in the CT. The sample is incumbent households living in gentrifiable neighborhoods (initially low-income and central city) in one of the baseline years, but do not currently live in that census tract. The control group is the matched sample discussed in section \ref{sec:Matched_Samples}. The regressions also include family-level socioeconomic control variables (age, family composition,  number of children and immigrant indicator), baseline Census Tract controls (college-educated share, median income, share for low income, average rent, employment rate, visible minority share, the share of immigrant, distance from CBD), and pre-period variation controls (changes of college-educated share, median income, average rent, employment rate). Low-income number of observations: 358,070. High-income number of observations: 201,839. \par}
\end{minipage}
\end{figure}

\begin{figure}[!htb]
\caption{Effect of gentrification on total income, Alternative measure}
\centering
\subfloat[Low-income]{\includegraphics[width = 5in]{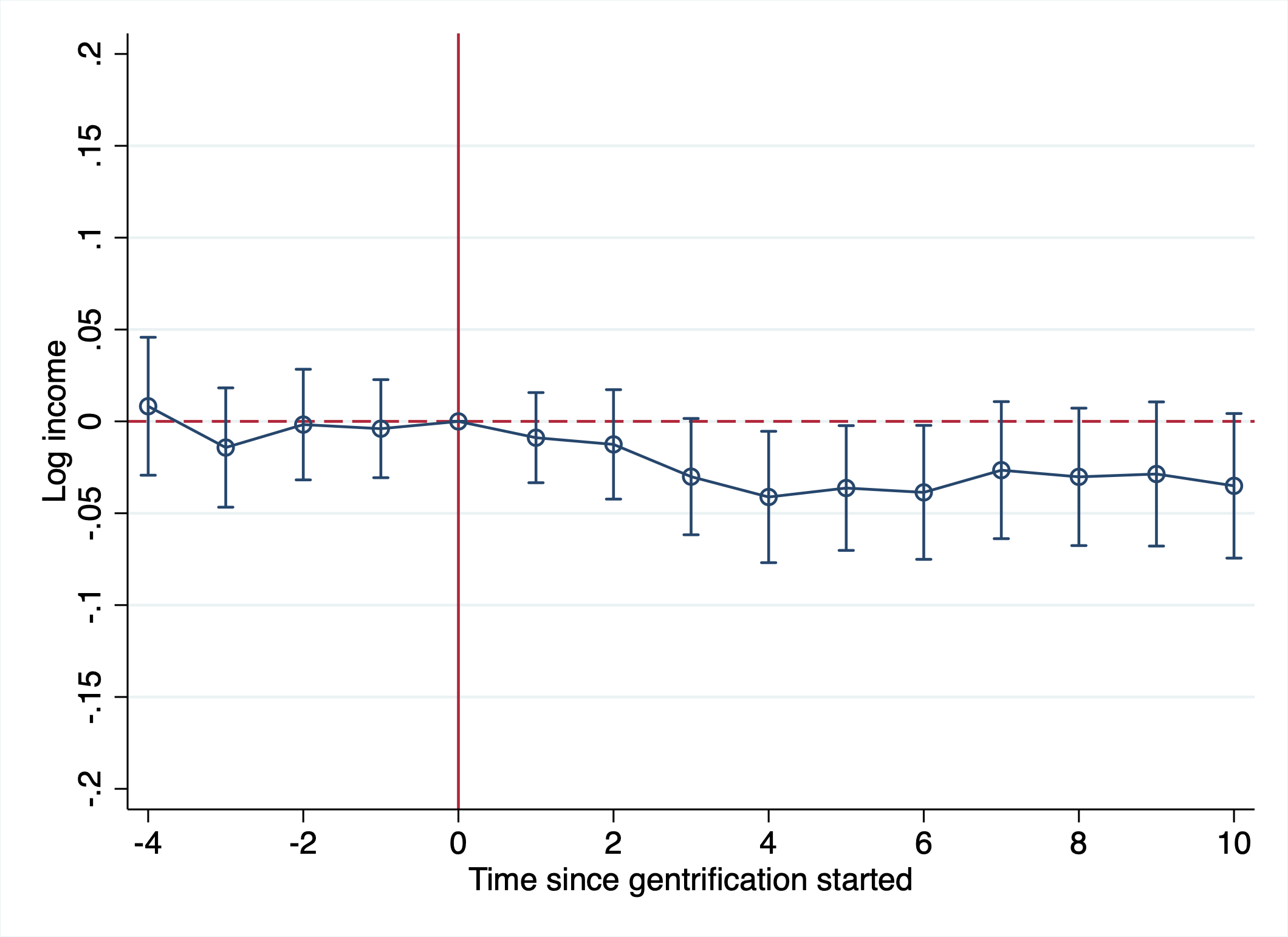}} \\
\subfloat[High-income]{\includegraphics[width = 5in]{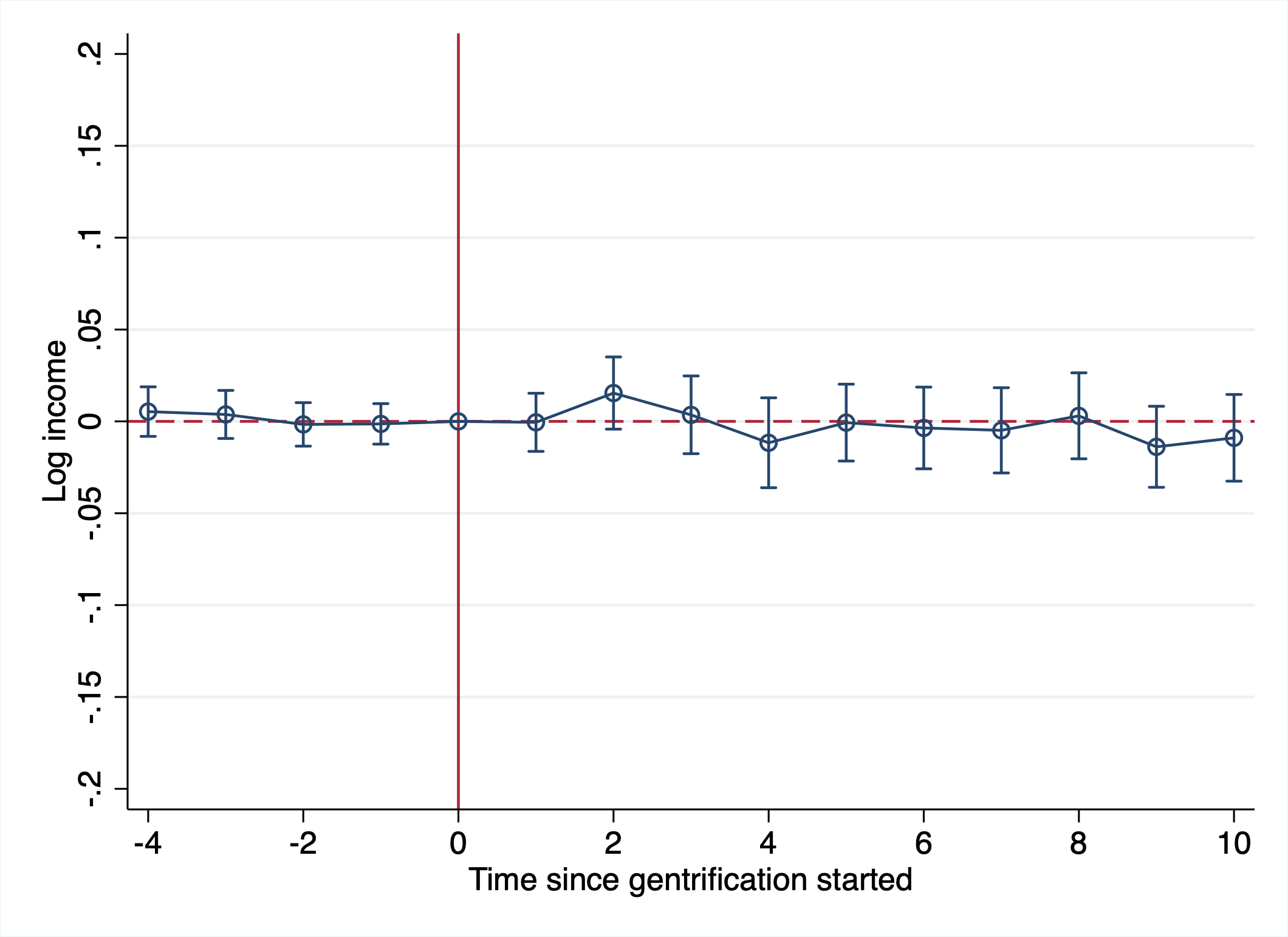}}\\
\label{fig:ES_total_earnings_2}
\begin{minipage}{0.9\textwidth} 
{\footnotesize \vspace{0.1cm} Source: Author's calculations using the Census and LAD.\\
Note: Dependant variables: total gross income. The sample is incumbent households living in gentrifiable neighborhoods (initially low-income and central city) in one of the baseline years. The control group is the matched sample discussed in section \ref{sec:Matched_Samples}.  The regressions also include individual-level control variables (age, age squared, gender, family composition,  number of children and immigrant indicator), baseline Census Tract controls (college-educated share, median income, share for low income, average rent, employment rate, visible minority share, the share of immigrant, distance from CBD), and pre-period variation controls (changes of college-educated share, median income, average rent, employment rate). Low-income number of observations: 2,385,445. High-income number of observations: 1,314,480. \par}
\end{minipage}
\end{figure}

\begin{figure}[!htb]
\caption{Effect of gentrification on employment earnings, Alternative measure}
\centering
\subfloat[Low-income]{\includegraphics[width = 5in]{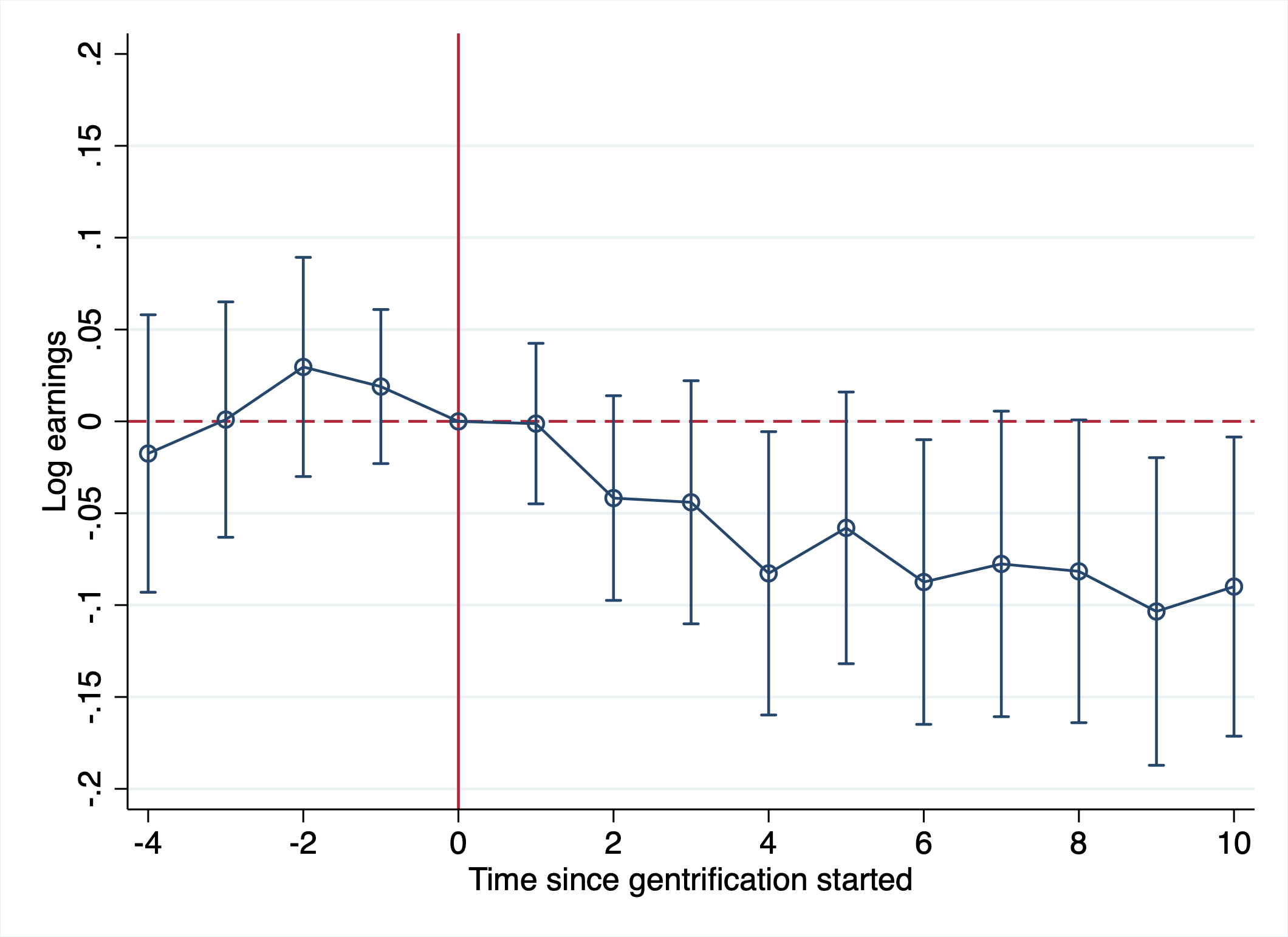}} \\
\subfloat[High-income]{\includegraphics[width = 5in]{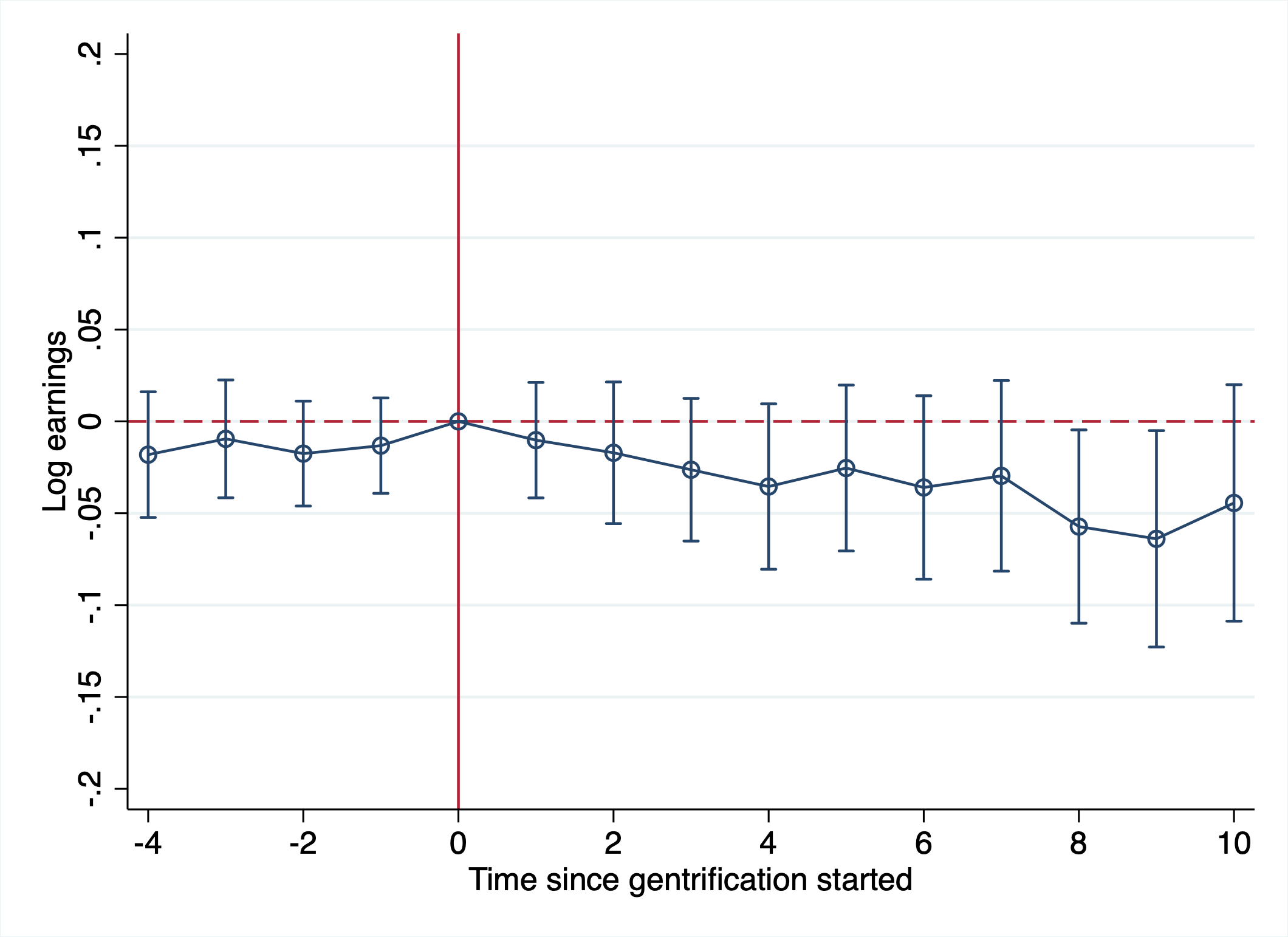}}\\
\label{fig:ES_empl_earnings_2}
\begin{minipage}{0.9\textwidth} 
{\footnotesize \vspace{0.1cm} Source: Author's calculations using the Census and LAD.\\
Note: Dependant variables: employment earnings. The sample is incumbent households living in gentrifiable neighborhoods (initially low-income and central city) in one of the baseline years. The control group is the matched sample discussed in section \ref{sec:Matched_Samples}.  The regressions also include individual-level control variables (age, age squared, gender, family composition,  number of children and immigrant indicator), baseline Census Tract controls (college-educated share, median income, share for low income, average rent, employment rate, visible minority share, the share of immigrant, distance from CBD), and pre-period variation controls (changes of college-educated share, median income, average rent, employment rate). Low-income number of observations: 2,385,445. High-income number of observations: 1,314,480. \par}
\end{minipage}
\end{figure}

\begin{figure}[!htb]
\caption{Effect of gentrification on other earnings, Alternative measure}
\centering
\subfloat[Low-income]{\includegraphics[width = 5in]{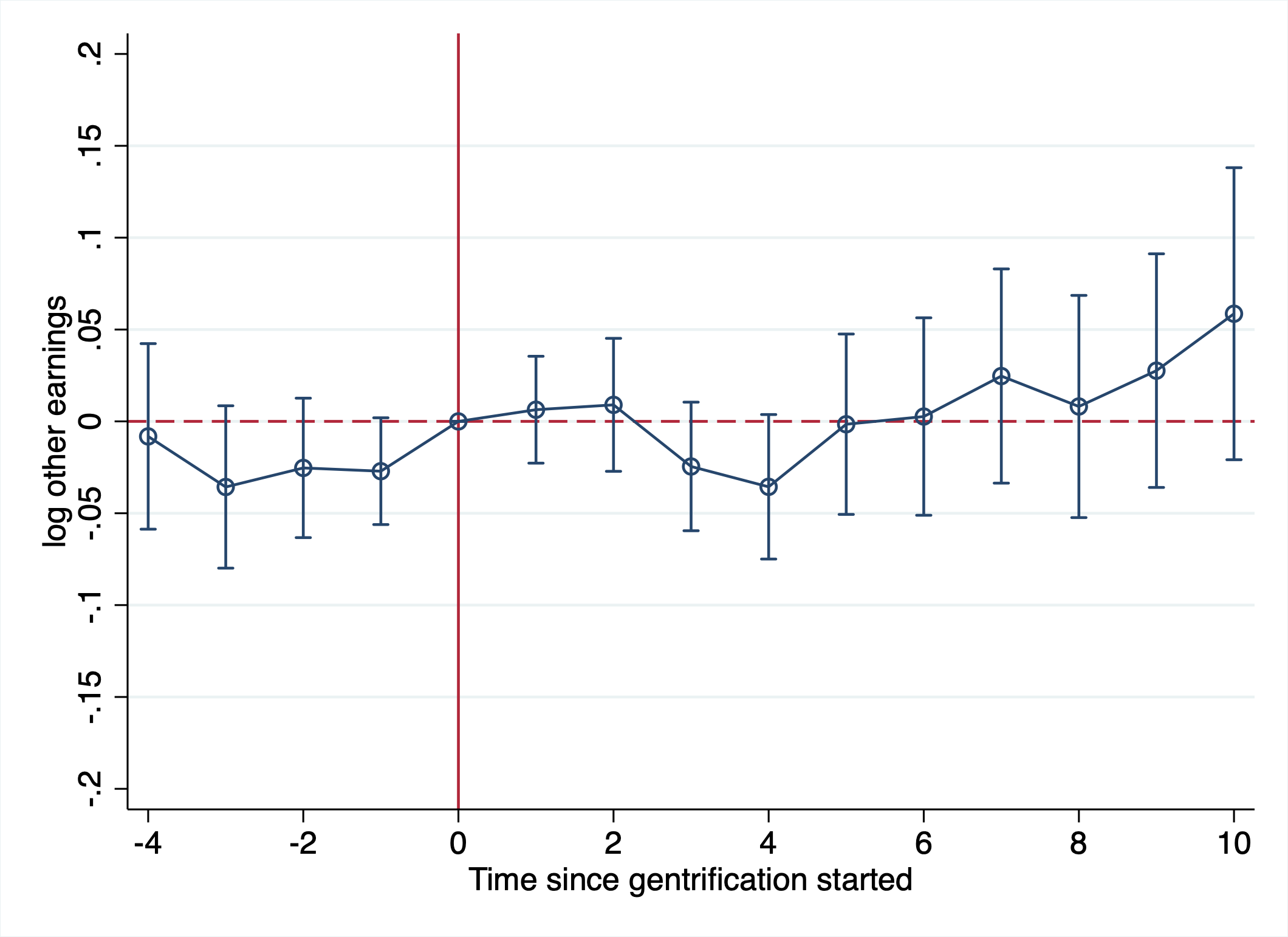}} \\
\subfloat[High-income]{\includegraphics[width = 5in]{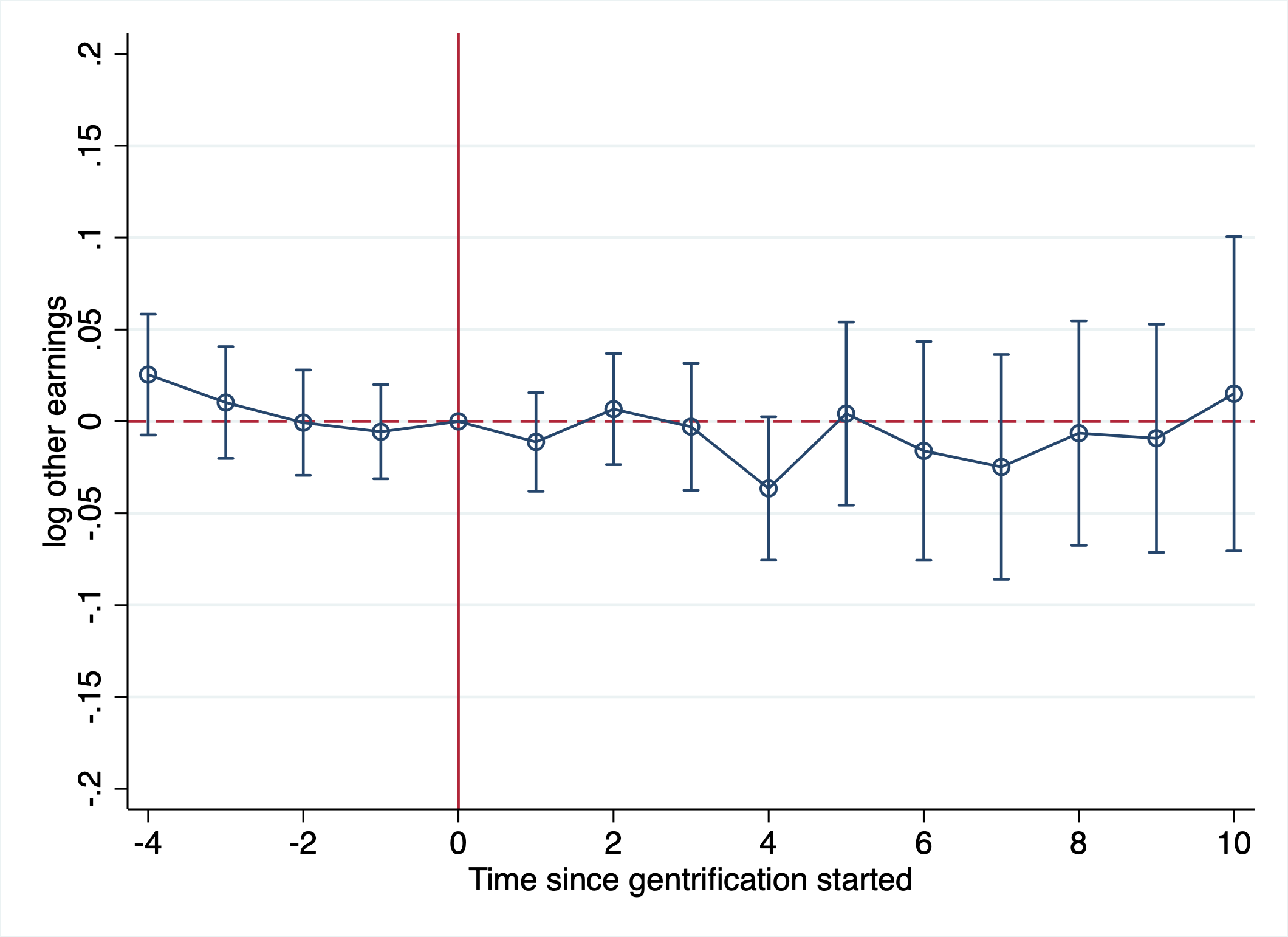}}\\
\label{fig:ES_other_earnings_2}
\begin{minipage}{0.9\textwidth} 
{\footnotesize \vspace{0.1cm} Source: Author's calculations using the Census and LAD.\\
Note: Dependant variables: employment earnings. The sample is incumbent households living in gentrifiable neighborhoods (initially low-income and central city) in one of the baseline years. The control group is the matched sample discussed in section \ref{sec:Matched_Samples}.  The regressions also include individual-level control variables (age, age squared, gender, family composition,  number of children and immigrant indicator), baseline Census Tract controls (college-educated share, median income, share for low income, average rent, employment rate, visible minority share, the share of immigrant, distance from CBD), and pre-period variation controls (changes of college-educated share, median income, average rent, employment rate). Low-income number of observations: 2,385,445. High-income number of observations: 1,314,480. \par}
\end{minipage}
\end{figure}
\begin{landscape}
\begin{figure}[ht!]
\caption{Effect of gentrification by cities}
\label{fig:ES_by_city}
\begin{sloppypar}
\hspace{3cm} \textbf{Montréal} \hspace{6cm}        
\textbf{Toronto}         \hspace{6cm}        
\textbf{Vancouver} \\
\end{sloppypar}
\centering
\subfloat[Low-income]{\includegraphics[width = 3.1in]{Figures/Event-study/move_matched_462_0.png}} \subfloat[Low-income]{\includegraphics[width = 3.1in]{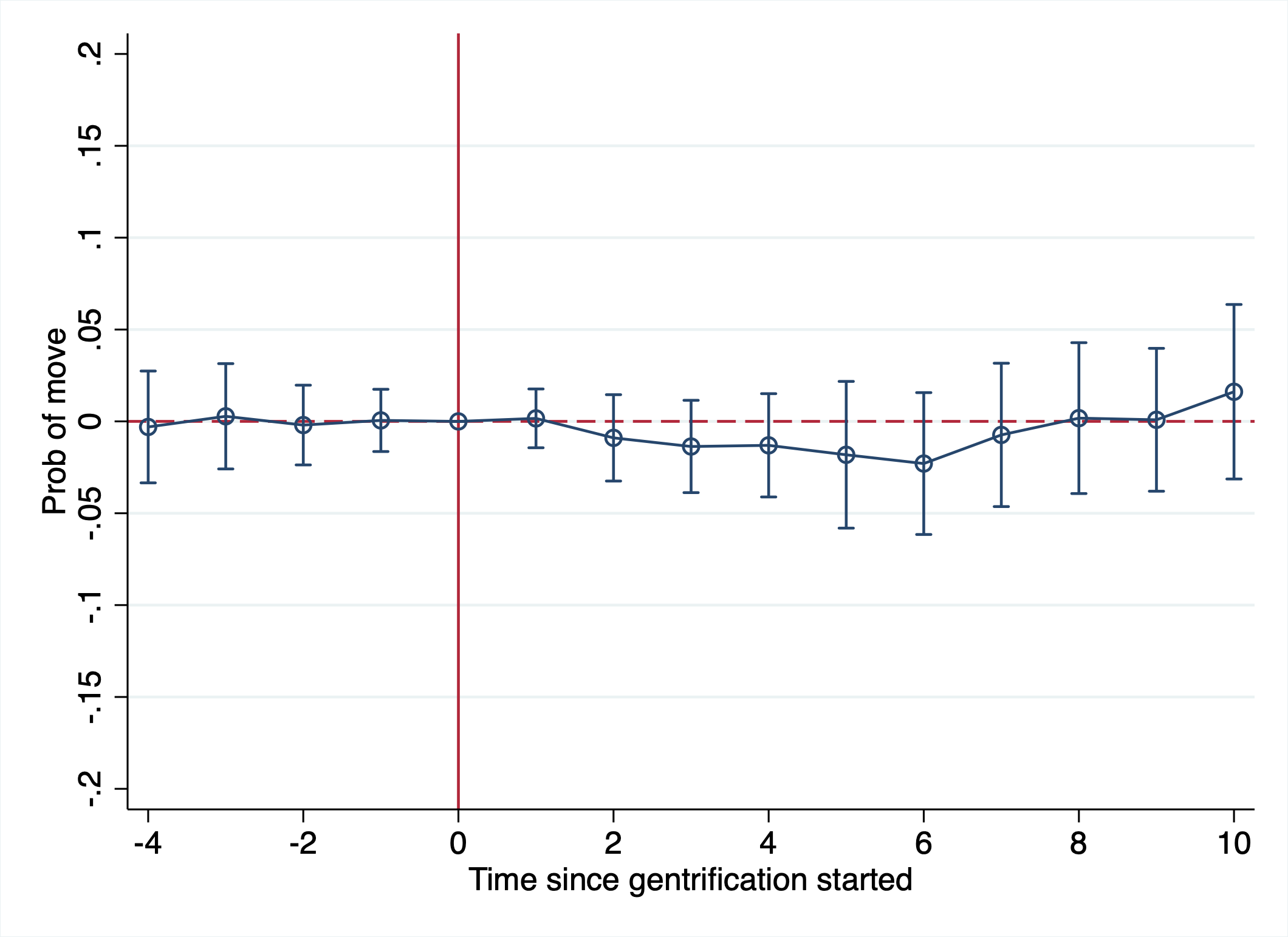}} \subfloat[Low-income]{\includegraphics[width = 3.1in]{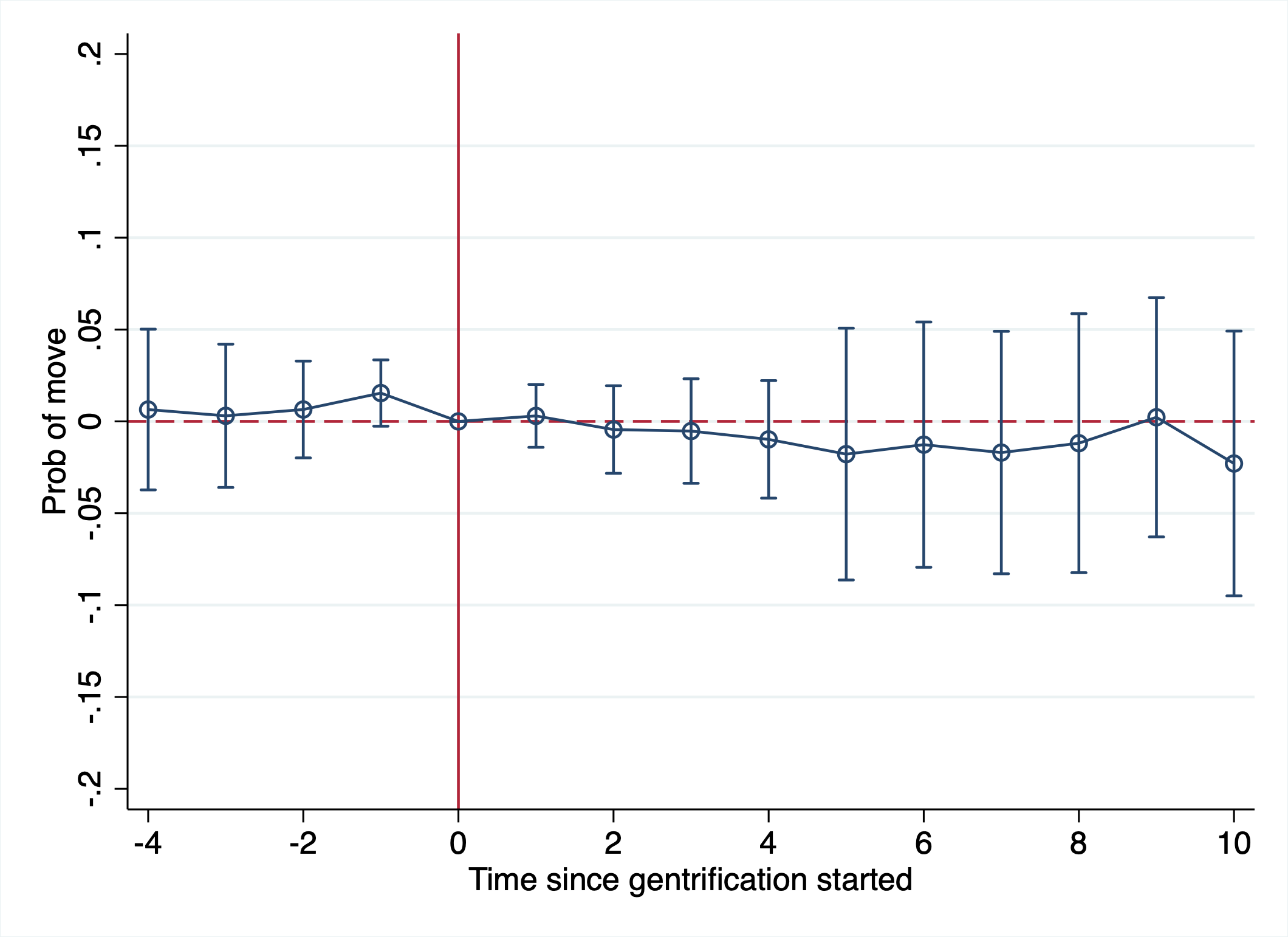}} \\
\subfloat[High-income]{\includegraphics[width = 3.1in]{Figures/Event-study/move_matched_462_1.png}} \subfloat[High-income]{\includegraphics[width = 3.1in]{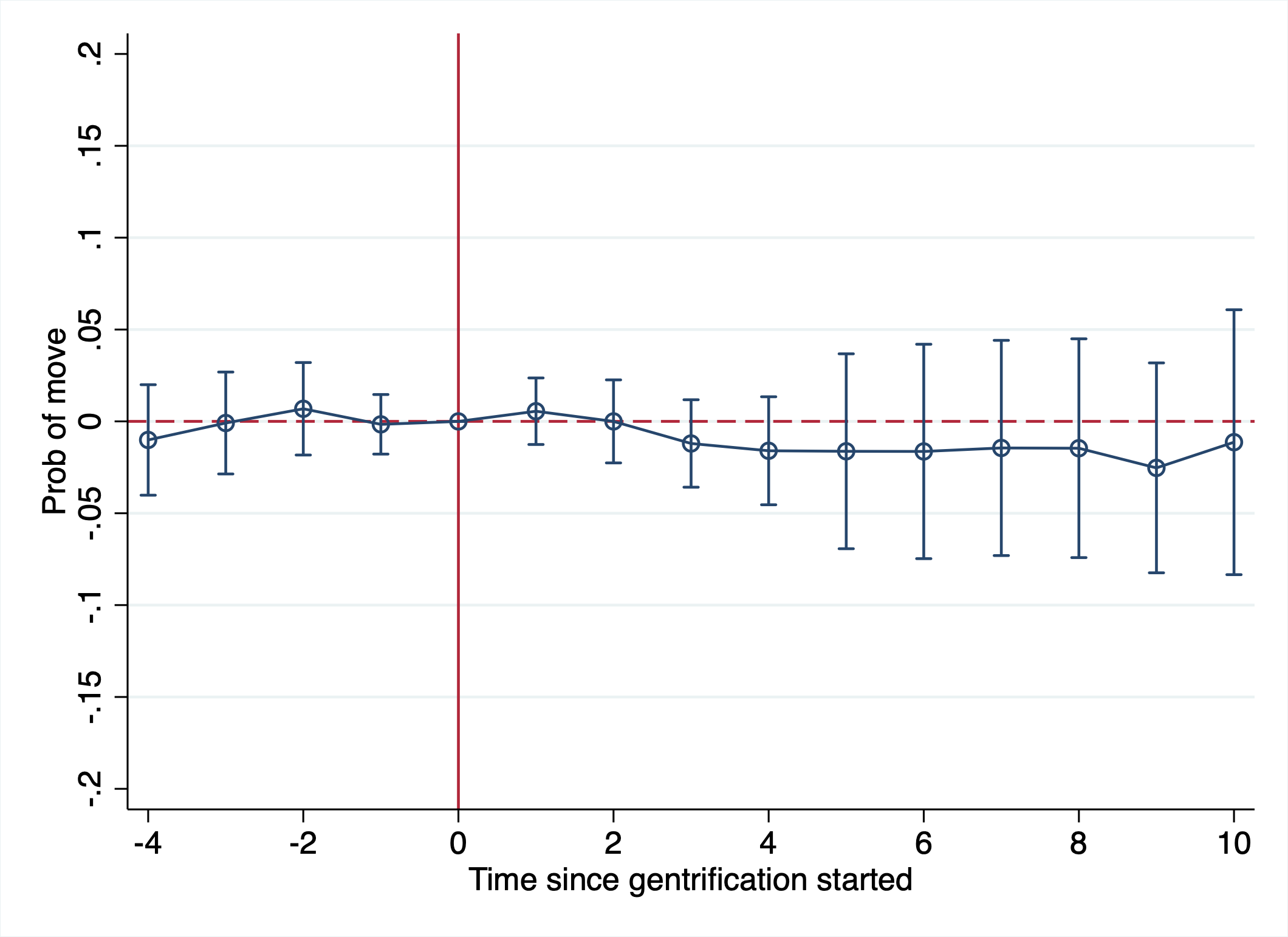}} \subfloat[High-income]{\includegraphics[width = 3.1in]{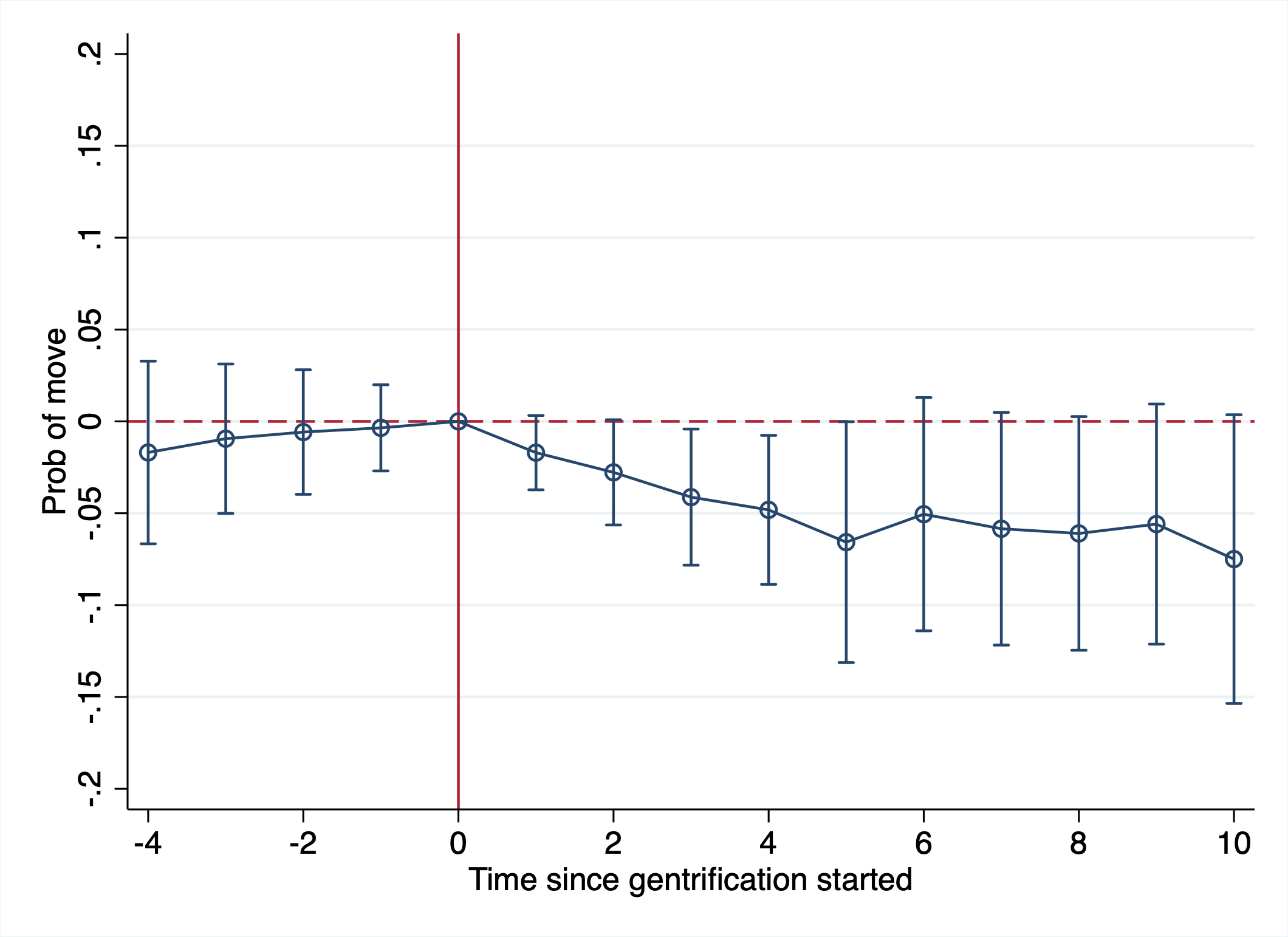}} \\
\begin{minipage}{1.2\textwidth} 
{\footnotesize \vspace{0.1cm} Source: Author's calculations using the Census and LAD.\\
Note: Dependant variables: dummy equal to one if the household lives in a different tract than at baseline. The sample is incumbent households living in gentrifiable neighborhoods (initially low-income and central city) in one of the baseline years. The control group is the matched sample discussed in section \ref{sec:Matched_Samples}. The regressions also include individual-level control variables (age, age squared, gender, family composition,  number of children and immigrant indicator), baseline Census Tract controls (college-educated share, median income, share for low income, average rent, employment rate, visible minority share, the share of immigrant, distance from CBD), and pre-period variation controls (changes of college-educated share, median income, average rent, employment rate). \par}
\end{minipage}
\end{figure}
\end{landscape}


\clearpage

\section{Historical census tract}

\doublespacing

One challenge of studying neighborhood dynamics across time is that statistical geographic units change over time. Each census year, Statistics Canada revises small geographical units such as census tracts to account for changing populations. This is usually not a problem when comparing two censuses, but it becomes problematic once we look at long time horizons. 

\citet{allen2018new} use a combination of map-matching techniques, dasymetric overlays, and population-weighted areal interpolation to create a set of a cross-walk table that link Census tract identifiers across years. This enables researchers to study more constant geographical units across longer periods. 

Significant problems arise when studying very long periods (more than 20 years) and smaller CMAs that grow quickly. In the present case, we restrict the analysis to the three largest CMAs, for which the inner-city CTs have stayed remarkably stable across recent years. Additionally, I do not study a given neighborhood for a very long period; instead, I study neighborhoods for a limited period of about 15 years. Table \ref{fig:tractchange} shows that from Census to Census since 1991, between 90\% and 98\% of census tract remains the same or undergo ``perfect" splits from which we can retrieve initial CT. 

\begin{table}[h]
\centering
\caption{Census tract changes across years  \label{fig:tractchange}}
  \includegraphics[scale=0.43]{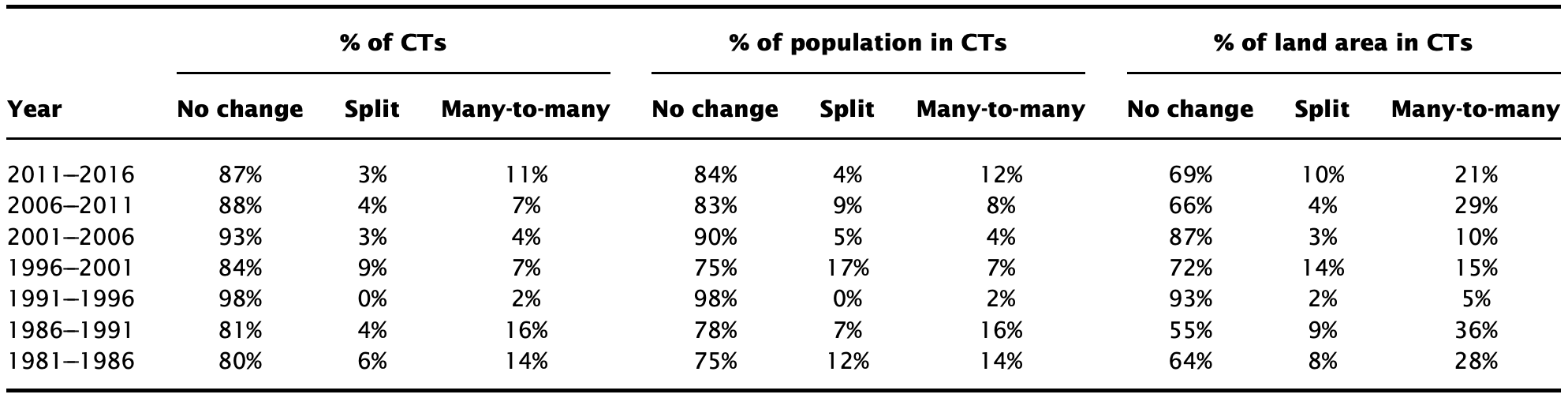}\\
\begin{minipage}{0.9\textwidth}\footnotesize \vspace{0.1cm} Source: \citet{allen2018new} \par
\end{minipage}

\end{table}
To mitigate this problem, I bring all observations to the 1996 census tract definition. For the Census of Population data, this involves dropping all areas that were not Census tracts in 1996. This is not a problem here as the focus is on inner-city neighborhoods, and most of the new CTs are outer-city neighborhoods (e.g. expansion of CMA to include the peripheral suburbs). 

For the Longitudinal Administrative Databank, I use the postal code conversion file from Statistics Canada to assign individuals to the 1996 CT boundary. 

\end{document}